\documentclass[a4paper,12pt,
]{scrbook}

\usepackage{uqthesis}

 \usepackage{makeidx}  

\usepackage[square,comma,numbers,sort&compress]{natbib}

\usepackage[nottoc]{tocbibind}  

\usepackage{amsmath,amsfonts} 

\usepackage{graphicx} 

\usepackage{epsfig} 

\copyrightpagetrue 

\examinerscopytrue   

\titlesmallcapstrue

 \makeindex  

\begin{document}
\setkomafont{caption}{\itshape}
\setkomafont{pagehead}{\itshape}

\frontmatter
\large\sffamily
\title{\sf{\LARGE\bfseries  Novel fluctuations at constrained interfaces }} 
\author{\sf{\large\bfseries Abhishek Chaudhuri}} 
\dept{\sf{\LARGE\bfseries S. N. B. N. C. B. S., Kolkata}} 

\titlepage 

\signaturepage  

\chapter{Acknowlegements}
\noindent
First, I would like to thank my supervisor, Surajit Sengupta, for his
continuous support in the Ph.D program. He was always there to listen
and to give advice. Interactions with him were always so friendly that
it gave me ample opportunity to ask questions and express my ideas. He
showed me the different ways to approach a research problem and always
insisted on staying focussed and working hard. It has really been a
wonderful experience working with him.
\vskip 1cm
\noindent
I would like to thank my collaborators, P. A. Sreeram and Debasish
Chaudhuri at S. N. Bose National Centre for Basic Sciences (SNBNCBS),
and Madan Rao at Raman Research Institute. Sreeram helped
me make my life with computers much simpler than I would have
imagined. I enjoyed working with him. Madan always impressed me with
his ideas and his encouragement to do good work. The innumerable
discussions that I had with Debasish during my Ph.D program helped me
to understand the diverse aspects of my research problem. He has been
a guiding force in my research work. He was always there to meet and
talk about my ideas, to proofread, and to ask me good questions to
help me think through my problems.
\vskip 1cm
\noindent
As for other members of our group, I would like to thank Ankush
Sengupta, with whom I have always shared a brotherly relationship. His
vibrancy and wonderful attitude always kept me going and academic
discussions with him are also acknowledged. Tamoghna and Arya always
brought with them a breath of fresh air and interacting with them was
fun. They provided the much needed boost in the final year of my Ph.D
and helped a lot while I was writing my thesis.
\vskip 1cm
\noindent
I would like to acknowledge helpful discussions with M. Barma (Tata
Institute of Fundamental Research), J. Krug (Universit{\"a}t Essen, Germany),
J. K. Bhattacharya (Indian Association for the Cultivation of
Science), S. N. Majumdar (CNRS, France), G. I. Menon (Indian Institute
of Mathematical Sciences), P.B. Sunil Kumar (I.I.T. Chennai), A. Mookerjee, 
S. S. Manna and A. K. Raychaudhuri (SNBNCBS) during the course of my research
work. A special thanks to P.B. Sunil Kumar and G. I. Menon and all the folks
at Sunil's lab for the wonderful time that I had at Chennai during my visit 
there. I would also like to acknowledge the Council of Scientific and
Industrial Research, Government of India for the CSIR (NET) research
fellowship.
\vskip 1cm
\noindent
My heartiest thanks goes to my friends, seniors and juniors who made my stay 
at S.N. Bose Centre colourful and enjoyable. Life here would not have been
the same without them. I relish the memories of the interactions that I 
had with them during my Ph.D program and it would be fair enough to say that
their support and cooperation helped me immensely to complete my research work.
I wish to thank the staff members of this centre for their sincere cooperation 
and help.
\vskip 1cm
\noindent
Last, but not the least, I thank my family : my parents, for unconditional
support and encouragement to pursue my interests and my sister, for all
the sweet memories that I share with her.


\chapter{List of Publications}
\noindent
{\bf I.} articles in journals/contributions to books/unpublished eprints:
\vskip 0.4cm

\newcounter{refno}
\begin{list}{[\arabic{refno}]}{\usecounter{refno}}
\item Abhishek Chaudhuri, Surajit Sengupta and Madan Rao, \,{\em 
  Stress relaxation in a perfect nanocrystal by coherent ejection of lattice 
 layers}, Phys. Rev. Lett. {\bf 95}, 266103 (2005).

\item Abhishek Chaudhuri, P. A. Sreeram and Surajit Sengupta, \,{\em A 
Kinetics driven commensurate - incommensurate transition},
Phase Transitions {\bf 77}, 691 (2004).

\item  Abhishek Chaudhuri, P. A. Sreeram and Surajit Sengupta, \,{\em
Growing smooth interfaces with inhomogeneous, moving external fields: dynamical 
transitions, devil's staircases and self-assembled ripples}, 
Phys. Rev. Lett. {\bf 89}, 176101 (2002).

\end{list}

\noindent
{\bf II.} published contributions in academic conferences: 
\vskip 0.4cm

\newcounter{refno2}
\begin{list}{[\arabic{refno2}]}{\usecounter{refno2}}
\item Abhishek Chaudhuri, Debasish Chaudhuri and Surajit Sengupta, \,{\em 
   Induced Interfaces at Nano-scales: Structure and Dynamics},
   International Journal of Nanoscience 2005 (in press). Proceedings
of International Conference on Nanoscience and Technology,
Hyaat Regency, Kolkata, 17-20 December 2003.

\item Abhishek Chaudhuri and Surajit Sengupta, \,{\em Profile-driven 
interfaces in 1+1 dimensions: periodic steady states, dynamical melting and 
detachment},
Physica A, {\bf 318}, 30  (2003). Proceedings of the International conference 
on Statistical Physics, Statphys - Kolkata IV, IACS, Kolkata (14-16 January,
2002) and SNBNCBS, Kolkata (17-19 January, 2002). 
 
\end{list}

\chapter{Abstract}
\noindent
In the study of condensed matter systems, a paradigm which is very
generally invoked is that of a ground state and its low energy
excitations. A classical solid, for example, is a periodic crystal
with low energy, long wavelength phonon fluctuations. Often these low
energy excitations appear due to the complete or partial breaking of a
spatial symmetry e.g. rotational or translational symmetry.  What
happens if explicit constraints are introduced such that these low
energy modes are unavailable?
\vskip 0.5cm
\noindent
This question has assumed some importance in recent years due to the
advent of nano technology and the growing use of nanometer scale
devices and structures. In a small system, the size limits the scale
of the fluctuations and makes it imperative for us to understand how
the response of the system is altered in such a situation. In this
thesis, this question is answered for the special case of interfacial
fluctuations in two dimensions (2d). The energy of an interface
between two phases in equilibrium is invariant with respect to
translations perpendicular to the plane (or line in 2d) of the
interface. We study the consequence of breaking this symmetry
explicity using an external field gradient. One expects that since low
energy excitations are suppressed, the interface would be flat and
inert at all times. We show that surprisingly there are novel
fluctuations and phenomena associated with such constrained interfaces
which have static as well as dynamic consequences.
\vskip 0.5cm
\noindent
After a couple of chapters containing introductory material on
interfaces, in Chapter 3 we investigate static and dynamic properties
of an Ising interface defined on a 2d oriented square lattice. The
interface is constrained by a non-uniform external field with a
profile which is such that the field changes sign across a (straight)
line which we call the ``edge'' of the profile. The Ising interface lies
as close to the edge as possible and the presence of the symmetry
breaking field suppresses long wavelength interfacial fluctuations as
expected. The interface therefore remains essentially flat over long
length scales. At short scales of the order of a few lattice
parameters, however, the interface begins to show ``self-assembled''
patterns depending on the orientation of the edge. These patterns may
be indexed by all possible rational fractions. The energy of the
interface depends on the index of the rational fraction and the
interface deforms locally to conform to low energy states. The local
orientation of the interface plotted as a function of the average
orientation shows a complete Devil's staircase structure. If the
external field profile is now translated perpendicular to the
interface with a fixed velocity, the interface follows the profile,
but we now have transitions among the patterns which occur in a way so
as to render the Devil's staircase incomplete. Low index patterns are
stable while patterns indexed by higher order rationals either
transform to low index ones or remain ``amorphous'' i.e. fluctuate
randomly between nearby stable states. At very high velocities the
interface detaches from the field edge and coarsens dynamically. We
believe that our results have some relevance on real crystal growing
techniques apart from being of theoretical interest due to the
presence of an infinite hierarchy of driven dynamical transitions. 
\vskip 0.5cm
\noindent
The later part of the thesis deals with a more realistic interface
(also in 2d) namely that between a classical liquid and a
solid. Similar to the Ising interface, this is created using a
chemical potential field in the form of an atomic trap. The
presence of the trap allows the atoms to condense into a solid which
is separated from the surrounding liquid by a liquid solid
interface. For reasons similar to the Ising case, the interface is
flat and ``crystallization waves'' which normally tend to roughen a
solid liquid interface are suppressed. The small size solid thus
formed can exchange particles with the liquid through the
interface. We show that this transfer of particles happens
predominantly by a coherent addition and removal of entire atomic
layers as the depth of the potential well is changed. The elastic
energy cost of flaking off incomplete layers is prohibitively large if
the thickness of the crystal is below a certain critical amount. This
ensures that only whole, single, layers are exchanged.  This provides
a novel mechanism of stress relaxation in a nano-sized single
crystal. Topological defects like dislocations tend to destroy this
coherence but are rapidly eliminated once they appear on the interface
as steps.
\vskip 0.5cm
\noindent
Finally, we study momentum and energy transfer across the liquid solid
interface in the presence of this ``layering'' transition using
molecular dynamics simulations. With the depth of the potential held
close to the value required for a layering transition, a tiny momentum
pulse is generated, near a free surface of the crystal. The pulse
travels as a shock wave through the solid and emerges on the other
side taking with it an entire crystal layer; the ejected layer
carrying away a definite amount of momentum and energy behaving, as it
were, like a coherent ``quasi particle'' in a completely classical
context. This results in a prominent peak in the absorption
spectrum. We believe that this driven layer ejection may have
practical applications apart from its interest from a fundamental
viewpoint. The transfer of energy through the interface is
characterized by the Kapitza or contact resistance which we measure
using non-equilibrium molecular dynamics where a temperature gradient
is setup using proper boundary conditions. We observe that the Kapitza
resistance shows a discontinuity at the layering transition. This may
enable one in future to design interfaces with controllable thermal
properties

\mainmatter

\tableofcontents




\chapter{Introduction to interfaces and summary of the thesis}

To a large extent, the properties of most everyday materials are 
determined by the properties of their internal interfaces \cite{sutton,safran}. 
This is equally
true, for example for a block of steel, where grain boundaries interact with
dislocations to produce hardness and resistance to plastic deformation, 
or a catalytic convertor in an automobile exhaust where interface properties
control crucial chemical reactions, or in a mammalian lung tissue where
lung surfactants residing in the inner lining of alveoli reduce the amount
of work required in breathing and thereby contribute directly to the
survival of the organism. 

It is therefore obvious that any means that we discover, which alters the 
properties of interfaces will have a profound effect on the overall structure 
and properties of many materials of technological importance. Indeed, a 
large amount of work exists \cite{sze,rhoderick,margaritondo} which attempts 
precisely that. Interfacial properties may be altered using alloying elements 
\cite{mullins} which either segregate or prevents segregation of
specific elements at interfaces, surfactants \cite{surfactant} which 
reduce the surface tension, stress etc.

In this thesis, we explore the possibility of tuning interfacial structure
and properties using non-uniform external fields. We show that large 
potential gradients at interfaces reduces the dominant low-energy, long 
wavelength fluctuations. Instead of such interfaces becoming inert, 
however we show that novel fluctuations are generated which lead to 
interesting behaviour which may have some technological applications. Such
fine tuning of interfacial fluctuations may be of use in the emerging area 
of nanotechnology \cite{balzani} where the properties of interfaces at the 
length and time scales of interest in this thesis, is crucial. 

We begin this chapter by providing a brief introduction to the thermodynamics
of interfaces followed by a discussion of the dominant low
energy equilibrium fluctuations. We then introduce
the basic paradigms for understanding the dynamics of driven interfaces. 
We end this chapter with a systematic overview and summary of the rest of
the chapters of this thesis.

\section{Definition of interfaces and interfacial region}
By definition, bulk phases are regions of configuration space associated 
with a global minimum of an appropriate free energy. Often, instead of
a single phase, one obtains a mixture of two or more phases which coexist.
The region in-between two coexisting bulk phases is an interface, a surface
being simply an interface where one of the bulk phases happens to be a gas
or vacuum. An interface can also be composed of a material that is 
different from the two bulk phases such as surface coatings, epitaxial layers 
and membranes composed of amphiphilic molecules that may lie between
bulk domains of water and oil. Although one may think of an
interface as being of negligible thickness, in fact the thickness of the 
interfacial region is significant and definitely non-zero, when we consider 
phenomena occurring at nano-meter scales. 

Any intensive property, such as density, varying from, say, its value 
in a liquid phase through the interface to a gas phase gives a plot
such as that in Fig. 1.1. 
\begin{figure}[h]
\begin{center}
\includegraphics[width=5.0cm]{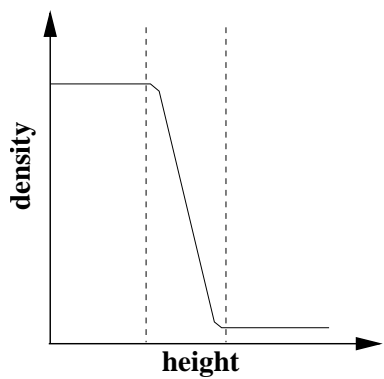}
\end{center}
\caption{ 
Variation of density from the high density liquid to the lower density gas 
phase. Dashed lines indicate the interfacial region}
\label{}
\end{figure}
In this case the density shows a smooth 
transition from the high density of the liquid to the much lower density
of the gas. The bulk phases can be separated from the interface by two
surfaces parallel to one another and positioned so that the bulk phases
are homogeneous and uniform (uniform density in this case) while the 
inhomogeneity and non-uniformity are connected entirely within the 
interfacial region lying between the two surfaces. The dashed lines in 
Fig. 1.1 illustrate this point. However we should note that for some
properties the transition from one bulk phase to another does not follow
a smooth monotonic form such as that in Fig. 1.1. Instead we
may have a behavior as shown in Fig. 1.2. 
 \begin{figure}[h]
\begin{center}
\includegraphics[width=15.0cm]{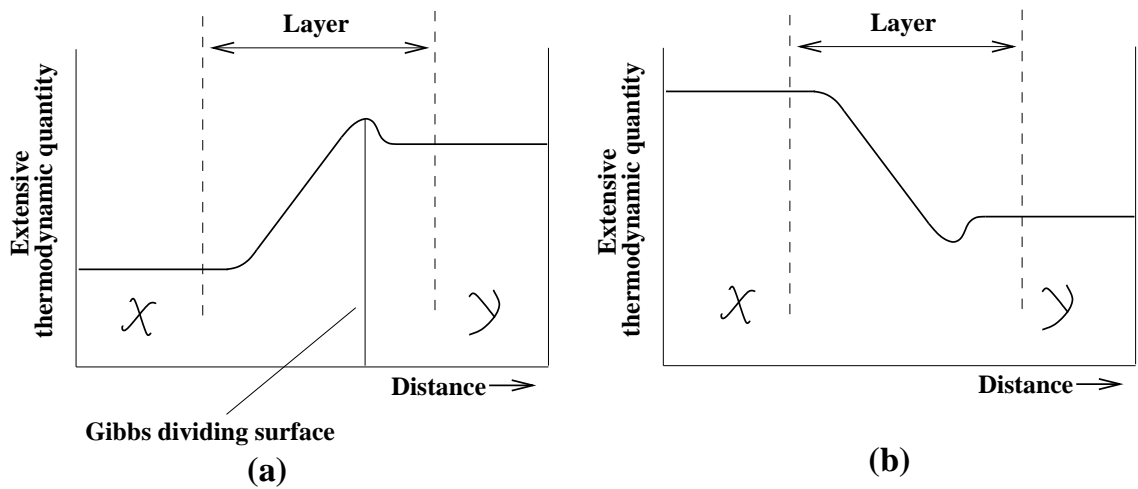}
\end{center}
\caption{(a) Variation of an extensive thermodynamic quantity with distance 
across an interface between ${\cal X}$ and ${\cal Y}$ phases. The layer
between the two dashed lines includes the interfacial region. Also shown is
the ``dividing surface'' of Gibbs (b) Variation of the same except that
here there is a relative depletion as opposed to interface excess in (a).
}
\label{}
\end{figure}
As an example, the concentration
of some solutes at the gas-solution interface may reach values much
higher than those in either bulk phase and exhibit a profile such as that
in Fig. 1.2(a). Such quantities are said to show an interfacial excess.
Conversely, there may be relative depletion of the concentration of a 
solute at an interface (Fig. 1.2(b))

\section{The thermodynamics of interfaces}
Given the above definition of an interface it is possible to define 
interfacial analogues of bulk thermodynamic quantities. Formally, this
may be achieved in several equivalent ways. Below, we follow the method
of Cahn \cite{cahn}. An alternative derivation was formulated by Gibbs more than
a century ago \cite{sutton}.

\subsection{The interface free energy : Cahn's method}
Consider a system consisting of two bulk phases, ${\cal X}$ and ${\cal Y}$, 
each containing ${\cal C}$ components, which are in contact along a flat 
interface. The interface is growing in equilibrium within a container in 
contact with suitable reservoirs. The entire system, including the reservoirs, 
is maintained at constant temperature, ${\cal T}$, hydrostatic pressure, 
${\cal P}$, and chemical potential, $\mu_i$, for each of the components. 
The increase in internal energy, $d{\cal E}$, due to the accretion of $d{\cal 
N}_i$ particles of component leading to corresponding changes in the entropy 
${\cal S}$, volume ${\cal V}$ and area ${\cal A}$ is given by,
\begin{eqnarray}
d{\cal E} &=& {\cal T}d{\cal S} - {\cal P}d{\cal V} + \sum_{i=1}^{\cal C}
 \mu_i d{\cal N}_i + \sigma_f d{\cal A}
\end{eqnarray}
Eqn. 1.1 contains the usual bulk term plus the added term 
$\sigma_f d{\cal A}$
which is required to account for the increase in internal energy of the
system which is associated with the increase in the area of the interface.
From Eqn. 1.1, we have
\begin{eqnarray}
\sigma_f &=& [\frac {\partial {\cal E}}{\partial {\cal A}}]_{{\cal S},{\cal V},
{\cal N}_i}, 
\end{eqnarray}  
and $\sigma_f$, the interfacial free energy per unit area, is therefore 
defined as the increase in internal energy of the 
entire system per unit increase in interface area at constant ${\cal S}$
and ${\cal V}$ and ${\cal N}_i$ are held constant. We may as well define 
$\sigma_f$ in terms of the Gibbs free energy, ${\cal G}$, the Helmholtz free 
energy, ${\cal F}$, and the grand potential, $\Omega$, as
\begin{eqnarray}
\sigma_f =  \left[\frac{\partial {\cal G}}{\partial {\cal A}}\right]_
{{\cal T,P,N_i}}  
=  \left[\frac{\partial {\cal F}}{\partial {\cal A}}\right]_{{\cal T},{\cal V},
{\cal N}_i} 
=  \left[\frac{\partial \Omega}{\partial {\cal A}}\right]_{{\cal T,V},\mu_i}.
\end{eqnarray}   

Further, on integrating Eqn 1.1, we obtain finally
\begin{eqnarray}
\sigma_f = \frac{1}{\cal A}\left[ {\cal G} - \mu_i {\cal N}_i \right].
\end{eqnarray}
where ${\cal G = E + PV - TS}$
Since the quantity $\sum_{i=1}^C \mu_i {\cal N}_i$ is the total Gibbs free 
energy which the homogeneous ${\cal X}$ and ${\cal Y}$ phases would possess 
together, we may identify $\sigma_f$ simply as the excess Gibbs
free energy of the entire system per unit interface area due to the presence 
of the interface. Note that the entire contribution to $\sigma_f$ arises
from the excess Gibbs free energy of the interfacial region. There is no
contribution from the bulk. This ensures that $\sigma_f$ is independent of
our choice for the width of the interface as long as it is large enough 
to contain the full variation of all intensive interfacial quantities 
e.g., the region between the dashed lines in Fig. 1.1.
 
\section{Equilibrium fluctuations of interfaces}
Interfaces in equilibrium rarely remain flat.
As can be easily seen, translating a planar interface between two 
coexisting phases in equilibrium in a direction perpendicular to the interface
costs no energy. It therefore follows that long wavelength fluctuations 
of the interface would be energetically cheap and would therefore dominate and
roughen the regions between two materials. How strong
these effects are depends on dimensionality (i.e., if the interface is a
line or a plane) and on temperature. For instance, for fluid interfaces, 
two-dimensional
systems with one-dimensional interfaces have a mean-square roughness of 
the interface that varies linearly with system size; the corresponding
quantity in three-dimensional systems is logarithmic with the system size.
To understand the role of fluctuations in interfaces, we consider first the
simplest case where the fluctuations are thermal in origin. For fluid 
interfaces, in the limit of zero viscosity, these fluctuations are known
as capillary waves.

\subsection{Thermal fluctuations}
Consider the fluctuation \cite{Barabasi,safran} of a two-dimensional surface defined in the Monge 
representation as $z = h(x,y)$. The area of the flat surface is denoted by $A$.
For slowly varying fluctuations of this surface about a flat shape ($h = h_0$,
where $h_0$ is a constant) the additional surface free energy of the
undulated interface over that of the flat one ($\Delta {\cal F}_s = {\cal F}_s - \gamma {\cal A}$)
is approximately
\begin{eqnarray}
\Delta {\cal F}_s = \frac{1}{2} \gamma \int dxdy(
(\frac{\partial h}{\partial x})^2 + (\frac{\partial h}{\partial y})^2)
\end{eqnarray}
Note that the free energy $\Delta {\cal F}_s$ is independent of $h$ itself and 
depends only on the derivatives, due to translational invariance. 
In Fourier space,
\begin{eqnarray}
\Delta {\cal F}_s = \frac{1}{2} \gamma \sum_{\bf q} q^2 |h({\bf q})|^2
\end{eqnarray}
The mean-square value of the fluctuating variable $h({\bf q})$, using
equipartition, can be written as
\begin{eqnarray}
<|h({\bf q})|^2> = \frac{{\cal T}}{\gamma q^2}
\end{eqnarray}
This gives the mean-square value of each Fourier mode in thermal equilibrium.

The mean-squared, real-space fluctuations of the surface about a flat profile,
are given by
\begin{eqnarray}
<h^2({\bf r)}> = \frac{1}{A}\sum_{\bf q}<|h({\bf q})|^2> =
\frac{1}{(2\pi)^2}\int d{\bf q} <|h({\bf q})|^2>
\end{eqnarray}
To avoid logarithmic divergences (when ${\bf q}$ is two-dimensional) while 
performing the integral, we introduce a lower limit to the integral, 
$\pi/L$, related to the
finite size of the system ($L \propto \sqrt A$) and an upper limit, $\pi/a$
due to the finite molecular size (proportional to $a$, the intermolecular
separation). Then we have \cite{safran},
\begin{eqnarray}
<h^2({\bf r})> = \frac{1}{2\pi \gamma}log{\frac{L}{a}}
\end{eqnarray}
By symmetry, $<h^2({\bf r})>$ is independent of the point ${\bf r}$
in the $x - y$ plane at which the fluctuation is calculated.
Note that the mean-square fluctuation diverges as the logarithm of
the system size. For a one-dimensional interface (i.e., a line instead of
a surface) this divergence is even more severe and increases linearly with
the size of the system. Thus, due to reduced dimensionality of these 
interfaces, the thermal fluctuations can have drastic effects on 
the ``flatness'' of the interfaces.

\subsection{Capillary waves}
For fluid interfaces, in the limit of negligible viscosity, the dominant
fluctuations are capillary waves. We first write down the potential
energy of an incompressible fluid surface acting under gravity and surface
tension. The free energy, ${\cal F}_s$, is given by a constant term that
comes from the bulk (where density variations are not allowed) free energy,
${\cal F}_0$, 
and a contribution from the surface tension and gravitational terms :
\begin{eqnarray}
{\cal F}_s = {\cal F_0} + \frac{1}{2}\int dA[\rho_0gh^2 + \gamma|\nabla h|^2]
\end{eqnarray}
where $A$ is the area, $\rho_0$ is the density, and $\gamma$ is the surface
tension. Solving the equation of motion for the interfacial position
$h(x,y)$ resulting from the above free energy in the undamped, zero viscosity,
inertial limit, we can obtain the dispersion relation \cite{safran}
\begin{eqnarray}
\omega = \sqrt{q\left(g + \frac{\gamma q^2}{\rho_0} \right)}
\end{eqnarray}
which shows that the frequency of such capillary fluctuations, 
$\omega \rightarrow 0$, as $q \rightarrow 0$. For a liquid solid interface,
these capillary fluctuations are known as ``crystallization waves'' 
\cite{balibar}. For most ordinary solids, such waves are highly
damped, however, though they have been experimentally studied in great detail
for liq helium at temperatures 0.03 to 0.5 K \cite{wang,helium}
\section{Morphology and dynamics of growth of interfaces}
In most situations, interfaces are actually not in equilibrium and it is 
difficult to define equilibrium structural and thermodynamic quantities. 
In such cases however, certain scaling laws, which determine how 
non-equilibrium interfacial fluctuations behave with time and system size come
in useful in characterizing the structure of an interface. The scaling 
exponents behave similar to the exponents well known from studies of
equilibrium critical point phenomena such that they can be used to define
distinct "universality classes" of non-equilibrium, driven interfaces.
In this section, we discuss briefly two such universality classes, viz. 
the Edwards-Wilkinson and the Kardar-Parisi-Zhang.

\subsection{Scaling concepts}
Consider a one-dimensional interface between two phases in two dimensions. 
Let us consider a situation where one of the phases is growing at the 
expense of the other. To describe the growth of this one-dimensional 
interface quantitatively, we introduce two functions \cite{Barabasi}. 
\begin{itemize}
\item The {\em mean height} of the surface, $\overline h$, is defined by
\begin{eqnarray}
\overline h(t) \equiv \frac{1}{L}\int_0^Ldx h(x,t)
\end{eqnarray}
where $h(x,t)$ is the position of the interface at $x$ at time $t$.  
The mean height increases linearly with time,
\begin{eqnarray}
\overline h(t) \sim t
\end{eqnarray}
if the interface moves at a constant rate.
\item The {\em interface width}, which characterizes the {\em roughness} 
of the interface, is defined by the rms fluctuation in the height,
\begin{eqnarray}
w(L,t) \equiv \sqrt{\frac{1}{L}\int_0^L[h(x,t) - \overline h(t)]^2}
\end{eqnarray}
\end{itemize} 
The time evolution of the surface width has two regions 
separated by a "crossover" time $t_x$ \cite{Barabasi}

(i) Initially, the width increases as a power of time,
\begin{eqnarray}
w(L,t) \sim t^\beta \hskip 2cm [t << t_x]
\end{eqnarray} 
The exponent $\beta$, which we call the {\em growth exponent}, characterizes
the time-dependent dynamics of the roughening process.

(ii) This is followed by a saturation regime during
which the width reaches a {\em saturation value}, $w_{sat}$. As $L$
increases, the saturation width, $w_{sat}$, increases as a power law,
\begin{eqnarray}
w_{sat}(L) \sim L^\alpha \hskip 2cm [t >> t_x]
\end{eqnarray}
The exponent $\alpha$, called the {\em roughness exponent}, characterizes the 
roughness of the saturated interface.

(iii) The crossover time $t_x$ (or {\em saturation} time) at
which the interface crosses over from the behavior of (i) to that of (ii) 
depends on the system size,
\begin{eqnarray}
t_x \sim L^z
\end{eqnarray}
where $z$ is called the {\em dynamic exponent}.

The exponents $\alpha$, $\beta$ and $z$ are not independent. If we
approach the crossover point $(t_x, w(t_x))$ from the left, we find, 
according to Eqn. 1.15, that $w(t_x) \sim t_x^\beta$. However, approaching
the same point from the right, we have from Eqn. 1.16, 
$w(t_x) \sim L^\alpha$. From these two relations follows 
$t_x^\beta \sim L^\alpha$ which according to Eqn. 1.17, implies that
\begin{eqnarray}
z = \frac{\alpha}{\beta}
\end{eqnarray}
Eqn 1.18, a scaling law linking the three exponents, is valid for any growth
process that obeys the {\em Family-Vicsek} scaling relation \cite{Barabasi} 
\begin{eqnarray}
w(L,t) \sim L^\alpha f\left(\frac{t}{L^z}\right)
\end{eqnarray}

\subsection{The Edwards-Wilkinson equation}
Consider an interface, defined by a height function $h({\bf x},t)$ which 
obeys the following stochastic growth equation,
\begin{eqnarray}
\frac{\partial h({\bf x},t)}{\partial t} = G(h,{\bf x},t) + \eta({\bf x},t)
\label{stoch}
\end{eqnarray}
where $\eta({\bf x},t)$ is a random noise and $G(h,{\bf x},t)$ contains the
deterministic part. Consider also that the equation of motion Eqn.~\ref{stoch} 
obeys the following symmetries \cite{Barabasi}: 

(i) {\em Invariance under translation in time}. The growth equation is 
invariant under the transformation $t \rightarrow t + \delta_t$. This
symmetry rules out an explicit time dependence of $G$.

(ii) {\em Translation invariance along the growth direction}. The growth
rule is independent of where we define $h = 0$, so the growth equation
should be invariant under the translation $h \rightarrow h + \delta_h$. 
This symmetry rules out the explicit $h$ dependence of $G$, so that the 
equation must be constructed from combinations of $\nabla h, \nabla^2 h,....,
\nabla^n h$, etc.

(iii) {\em Translation invariance in the direction perpendicular to 
the growth direction}. The equation is independent of the actual value 
of ${\bf x}$, having the symmetry ${\bf x} \rightarrow {\bf x} + \delta_x$.
This exclude explicit ${\bf x}$ dependence of $G$.

(iv) {\em Rotation and inversion symmetry about the growth direction 
${\bf n}$}. This rules out odd order derivatives in the coordinates, 
excluding vectors such as $\nabla h, \nabla(\nabla^2 h)$, etc.

(v) {\em Up/down symmetry for h}. The interface fluctuations are similar 
with respect to the mean interface height. This rules out even powers of
$h$, terms such as $(\nabla h)^2, (\nabla h)^4$, etc. A moments reflection
tells us that this symmetry implies that $h({\bf x},t)$ represents in 
this case an interface between two equilibrium coexisting phases.

The final form of the growth equation, consistent with all  
the symmetries listed above is, 
\begin{eqnarray}
\nonumber
\frac{\partial h({\bf x},t)}{\partial t} = (\nabla^2 h) + (\nabla^4 h) + ... 
+ (\nabla^{2n} h) + (\nabla^2 h)(\nabla h)^2 + ... \\
+ (\nabla^{2k}h)(\nabla h)^{2j} + \eta({\bf x},t),
\end{eqnarray}
where $n,k,j$ can take any positive integer value. For simplicity of 
notation we do not explicitly indicate the coefficients in front of the terms.

Since we are interested in the scaling properties, we focus on the long-time
$(t \rightarrow \infty)$, long-distance $(x \rightarrow \infty)$ behavior
of functions that characterize the surface. In this {\em hydrodynamic limit},
higher order derivatives should be less important compared to the lowest
order derivatives, as can be confirmed using scaling arguments. 

The noise term, $\eta({\bf x},t)$ in Eqn. 1.21 incorporates the stochastic
character of the fluctuation process and is assumed to
be uncorrelated, with the properties 
\begin{eqnarray}
\langle \eta({\bf x},t) \rangle &=& 0 \\
<\eta({\bf x},t)\eta({\bf x^{\prime}},t^{\prime})> &=& 2D\delta^d({\bf x} - 
{\bf x^{\prime}})\delta(t - t^{\prime})
\end{eqnarray}
Thus, the simplest equation describing the fluctuations of an equilibrium
interface, called the Edwards-Wilkinson (EW) equation \cite{EW} has the form
\begin{eqnarray}
\frac{\partial h({\bf x},t)}{\partial t} = 
\nu \nabla^2 h + \eta({\bf x},t)
\end{eqnarray}
Here $\nu$ is sometimes called a "surface tension", for the $\nu \nabla^2 h$
term tends to smoothen the interface. Eqn 1.24 is valid in the small 
gradient approximation, i.e., in the limit $(\nabla h) << 1$.

The average velocity of the interface is zero, which can be seen by using
Eqn 1.24,
\begin{eqnarray}
v \equiv \int_0^L d^d{\bf x} \left< \frac{\partial h}{\partial t} \right> = 0.
\end{eqnarray}
The contribution from the Laplacian term is zero, due to the periodic 
boundary conditions, and since the noise has zero average according to 
Eqn 1.22. 

The scaling exponents of the EW equation can be obtained
either by using scaling arguments or by solving the equation. The exponents
in d-dimensions are given as follows :
\begin{eqnarray}
\alpha = \frac{2 - d}{2},\,\,\,\,\,\beta = \frac{2 - d}{4},\,\,\,\,\,z = 2
\end{eqnarray}
Thus for $d = 2$, we find that $\alpha = \beta = 0$, i.e., the width scales
logarithmically with time at early times, and the saturation width depends 
on the logarithm of the system size. For $d > 2$, the roughness exponent
$\alpha$ becomes negative, which means that the interface is flat,
because the surface tension suppresses any noise-induced irregularity.

\subsection{Non-equilibrium interface : Kardar-Parisi-Zhang equation}
The first extension of the EW equation to include non-linear terms was 
proposed by Kardar, Parisi and Zhang (KPZ) \cite{KPZ,Barabasi}. 
The KPZ equation is the simplest growth equation which has the symmetries
(i)-(iv) of the linear theory discussed in Eqn. 1.21, but the up-down 
symmetry of the interface height $h({\bf x},t)$ is broken. The source of 
the symmetry breaking is, of course, the existence of a driving force, $F$, 
perpendicular to the interface, and uniform in space and time which selects
a particular growth direction for the interface. In the
linear theory this up-down symmetry excludes terms such as $(\nabla h)^{2n}$
from the growth equation. The lowest order term of this sort is $(\nabla h)^2$
which, if added to the EW Eqn. 1.26, results in the KPZ equation 
given as 
\begin{eqnarray}
\frac{\partial h({\bf x},t)}{\partial t} = 
\nu \nabla^2 h + \frac{\lambda}{2}(\nabla h)^2 + \eta({\bf x},t)
\end{eqnarray} 

In general, the existence of a driving force is a {\em necessary}, but not
{\em sufficient} condition for the broken up-down symmetry in $h$, and hence
for the appearance of the nonlinear term. As an example, consider the 
random deposition (RD) model with surface relaxation \cite{RD}. In the RD model, each
particle falls along a single column toward the surface until it reaches 
the top of the interface, whereupon it sticks irreversibly. To include 
surface relaxation, the deposited particle is allowed to diffuse along the
surface up to a finite distance, stopping when it finds the position with
the {\em lowest} height. In this model, particle deposition generates a 
driving force that makes the interface grow. Deposition apparently breaks
the up-down symmetry of the growth, but the model is nevertheless described
by the EW equation, so in fact no symmetry breaking occurs. If we transform
to a system of coordinates moving together with the interface, the model
generates an interface which is invariant under the transformation 
$h \rightarrow -h$.

Scaling exponents for the KPZ equation can be obtained exactly for $d = 1$
via a dynamical renormalization group calculation, to be $\alpha = 1/2$,
$\beta = 1/3$ and $z = 3/2$. However for $d > 1$, dynamic RG analysis does
not give the scaling exponents. There are several models that predict the
scaling exponents in higher dimensions but none of them are exact. For
$d = 2$, large scale simulations predict that 
$\beta = \alpha/z = 0.240\pm0.0001$ (see \cite{Barabasi} and references 
therein). There is however a scaling relation that
is true in any dimensions : 
\begin{eqnarray}
\alpha + z = 2
\end{eqnarray}

\section{Plan of the thesis}
In the previous sections we have tried to define and describe equilibrium
and driven interfaces in general terms as briefly as possible. There are
extensive review articles \cite{helium} and books 
\cite{Barabasi,sutton,safran} on this subject and the reader is referred to 
them for details. In this thesis, we study a special class of interfaces which
are produced by non-uniform fields which vary sharply in the direction
perpendicular to the interface. The primary function of this field is to 
break the translational symmetry along the growth direction. Breaking this
symmetry implies a non zero energy cost for long wavelength fluctuations 
of the interface which primarily tends to stabilize the sort of planar
interfaces discussed in Section 1.3. Surprisingly, however, we discover, in
each case, the existence of residual fluctuations which produce novel 
phenomena.

In Chapter 2, we specialize our discussion of interfaces to interfaces in 
the Ising model. We introduce the model and discuss the equilibrium phases.
The importance of Ising
model in the sense that a variety of other statistical mechanical systems
can be described by it, is then pointed out. In particular, reference is made
to the mapping to lattice gas and binary alloy. Next we review existing
literature on interfaces in the ferromagnetic Ising model. A study of the
physics of single surfaces or interfaces begins with the characterization 
of the shape of the interface. In this regard, we study the structure of
the Ising interface. We derive the interfacial profile using the continuum
$\phi^4$ theory. We discuss the stability of a flat interface in the
two-dimensional Ising model. The role of fluctuations is pointed out in
detail. The time-dependent properties of an inclined interface
separating up and down spin regions in a two-dimensional nearest-neighbor 
Ising model evolving under Glauber dynamics in a non-zero but uniform field
is discussed. We show that the for a constant magnetic field, the equation of
motion for the interface reduces to the KPZ equation as expected. 
In the limit of large exchange coupling, the model reduces to the single-step 
model for ballistic growth and thence to the asymmetric exclusion process which 
describes a driven diffusive system of hard core particles on an one-
dimensional lattice. The drift velocity of the interface is found as a 
function of field, temperature and inclination, and interface correlation 
functions are related to sliding tag correlation functions in the particle 
system. The non-linear dependence of the current on density leads to 
kinematic waves which involves moving density fluctuations. 

In Chapter 3, we study the steady state structure and dynamics of a 
two-dimensional Ising interface placed in an {\em inhomogeneous} external field 
translated with velocity $v_{e}$. The non-uniform field has a profile with
a fixed shape which is designed to stabilize a flat interface.
For small velocities the interface is stuck
to the profile and is rippled with a periodicity which may be either
commensurate or incommensurate with the lattice parameter of the square
lattice. For a general orientation of the profile, the local slope
of the interface locks in to one of infinitely many rational directions
producing a ``devil's staircase'' structure. These ``lock-in'' or commensurate
structures disappear as $v_e$ increases through a kinetics driven commensurate
- incommensurate transition. For large $v_e$ the interface becomes detached
from the field profile and coarsens with KPZ exponents. The
complete phase~-diagram and the multifractal spectrum corresponding to these
structures are obtained numerically together with several analytic
results concerning the dynamics of the rippled phases. The fact that 
interfacial fluctuations are partially suppressed by the external field 
is manifested in the exact agreement between a mean field theory and 
simulation results. However small interfacial fluctuations produce a dynamical
phase diagram showing infinitely many dynamical phases and dynamic phase 
transitions. 

Next we turn our attention to more realistic systems viz., solid - liquid 
interfaces. In Chapter 4, we introduce and describe atomic systems and 
discuss the need to study such systems. Before we go on to study interfaces
in atomic systems we use this
chapter to introduce the models and computational techniques (e.g. Monte 
Carlo and molecular dynamics simulations) that are used
in this thesis. The 
relevance of these simulation techniques in the context of the thesis is
pointed out. Specific techniques to simulate hard systems are also
described. Bulk thermodynamic, structural and dynamic quantities that 
characterize different phases are described in some detail. We reproduce 
existing data in order to validate our 
computational methods to be used for subsequent calculations.

In Chapter 5, we show how one can create a solid - liquid interface using
a non - uniform external field. This is a direct analog of the Ising interface 
studied in Chapter 3. In contrast to the Ising interface, elastic 
deformations of the solid are allowed and play a prominent part in
determining the properties of the interface. In fact, we show that a thin
strip of solid placed within a bath of its own liquid relieves stress by
novel interfacial fluctuations which involve
addition or deletion of entire lattice layers of the crystal. Local analogues
of the bulk quantities defined in Chapter 4 are calculated along 
the length of the system in the direction perpendicular to the interface.
The ``layering'' transition is shown to be a generic feature of systems
interacting via any interatomic potential.

In Chapter 6, we show how these interfacial fluctuations influence mass,
momentum and energy transport properties across the interface. Tiny momentum 
impulses are seen to produce shock waves which travel through the liquid solid 
interface and causes the spallation of crystal layers into the liquid. We show 
how kinetic and energetic constraints prevents the spallation of partial layers
from the crystal, making this process suitable for creating nano-layer
and coatings or for making nano-wires. We also study heat transport through
the liquid solid interface and obtain the contact or Kapitza resistance of
the interface as a function of the depth of the potential well.

\chapter{Interfaces in the Ising model}

After an introduction to the structure and properties of interfaces in 
general, in this chapter we specialize our discussion to a study of interfaces
in the particularly simple context of the Ising model.The Ising model \cite{Ma} 
tries to 
imitate behaviour in which individual elements (e.g., atoms, animals, protein 
folds, biological membrane, social behavior, etc.) modify their behavior so as 
to conform to the behavior of other individuals in their vicinity. The Ising 
model has more recently been used to model phase separation in binary alloys 
\cite{Ma} and spin glasses \cite{spinglass}. In biology,
it can model neural networks \cite{neuralnet}, flocking birds \cite{birds}, 
or beating heart cells \cite{heartbeat}. It can
also be applied in sociology \cite{sociology}. 

In this chapter 
we start by giving an introduction of the Ising model and its importance
in explaining several phenomena. We discuss briefly
the mapping of the Ising model to binary alloy and lattice gas models. 
We then discuss the structure and stability of the Ising interface in
the low temperature limit when the interface may be described by a single
valued ``height'' function $h(x,t)$. Finally we discuss in detail the effect of
a homogeneous external field on the Ising interface with special emphasis on
the nature of fluctuations of the interface about the mean position 
$\overline h(t)$ of the interface.

\section{Ising model}
The Ising model was proposed in the doctoral thesis of Ernst Ising, 
a student of W. Lenz. Ising \cite{Ising} tried to explain certain empirically 
observed facts about ferromagnetic materials using a model proposed by 
Lenz \cite{Lenz}.
It was referred to in Heisenberg's \cite{Heisenberg} paper which used the exchange 
mechanism to describe ferromagnetism. The name became well-established with 
the publication of a paper by Peierls \cite{Peierls}, which gave a non-rigorous proof
that spontaneous magnetization must exist. A breakthrough occurred when it
was shown that a matrix formulation of the model allows the partition 
function to be related to the largest eigenvalue of the matrix (Kramers and 
Wannier \cite{KW}, Montroll \cite{Montroll1,Montroll2}, Kubo \cite{Kubo}. Kramers and Wannier 
calculated the Curie temperature using a two-dimensional Ising model, and a 
complete analytic solution was subsequently given by Onsager \cite{Onsager}.

Consider a lattice in $d$ dimensions of sites ${i}$ labelled 
$1,2,.......{\cal N}(\Omega)$, which we will take to be hypercubic, unless otherwise
stated. The degrees of freedom are classical spin variables, $S_i$, residing 
on the vertices of the lattice, which take only two values : up or down, or 
more usefully,
\begin{eqnarray}
S_i = \pm 1
\end{eqnarray}
The total number of states of the system is $2^{{\cal N}(\Omega)}$. The spins 
interact with an external field (in principle varying from site-to-site)
${H}_i$ and with each other through exchange interactions $J_{ij}, K_{ijk}, ..$
which couple two spins, three spins, ...... etc.

A general form of the Hamiltonian is
\begin{eqnarray}
- {\cal H}_{\Omega} = \sum_{i \in \Omega} {H}_{i}S_{i} + \sum_{ij}J_{ij}S_iS_j +
\sum_{ijk}K_{ijk}S_iS_jS_k + ...
\end{eqnarray}
The free energy is given by
\begin{eqnarray}
{\cal F}_{\Omega}({\cal T},{{H}_i},{J_{ij}}...) = -k_B{\cal T} \log Tr e^{-\beta {\cal H}_{\Omega}}
\end{eqnarray}
where the trace operation is a sum over all possible configurations of 
the spin $S_i$, and thermodynamic properties can be obtained by differentiating
the free energy. 

To discuss the phase transition of the Ising model, we consider the simple 
case of the nearest neighbour Hamiltonian
\begin{eqnarray}
- {\cal H}_{\Omega} = {H} \sum_{i=1}^{{\cal N}(\Omega)}S_{i} + J \sum_{<ij>}S_iS_j
\end{eqnarray}
where we have assumed that the external magnetic field ${H}$ is uniform in 
space, and that the only interaction between spins is that between 
{\em neighbouring} spins, with strength denoted by $J$. We consider the case 
$J > 0$. With a uniform external magnetic field, we can define the 
magnetization or magnetic moment per site, $M$ :
\begin{eqnarray}
M \equiv \frac{1}{{\cal N}(\Omega)}\sum_{i=1}^{{\cal N}(\Omega)}\langle S_i \rangle.
\end{eqnarray}
Now consider the case when ${H} = 0$. The spins then seek configurations 
that at constant ${\cal T}$, will minimize the Helmholtz free energy 
${\cal F}_{\Omega} = {\cal E - TS}$. At high ${\cal T}$, ${\cal F}_{\Omega}$ is clearly minimized by 
maximizing the entropy, {\cal S}. The maximally disordered state has the 
highest entropy, implying that the equilibrium state at high temperatures is 
the {\em paramagnetic} state with no average alignment of spins, i.e., no
magnetization. At low temperatures, the internal energy dominates over
${\cal TS}$, and the state that minimizes ${\cal F}_{\Omega}$ is one that minimizes ${\cal E}$.
States which minimize ${\cal E}$ have a non-vanishing magnetization, and the low
temperature equilibrium phase is the $ferromagnetic$ phase with nonzero
magnetization $M$. At some temperature ${\cal T}_c$, there is a phase transition
from the entropy dominated paramagnetic state to the energy dominated
ferromagnetic state. The magnetization $M$ is called the {\em order parameter}
of the ferromagnetic phase and takes two degenerate values 
$\pm M({\cal T}) \rightarrow \pm1$ as ${\cal T} \rightarrow 0$. The two degenerate states 
correspond to cases where the spins are either all up ($S_i = +1$) or all
down ($S_i = -1$). At ${\cal T} = 0$, all spins are perfectly aligned to either of 
the two possible values. At any nonzero temperature, domains of one phase
in the other are stabilized due to entropy. These domains cost positive 
interfacial energy and grow in response to external fields.

\subsection{Mapping to other models}
The usefulness of the Ising model stems to large extent, from its generality.
The Ising model maps onto a large number of simple statistical models for
complex phenomena in material science as well as biology, economics and
sociology. We show below three common mappings which are included as 
illustrative examples \cite{Ma}.
 
\bigskip
\noindent
{\em A. Ising antiferromagnetic model} : If we replace $J$ in Eqn. 2.4 by
$-J$ ($J > 0$) then energy is lowered for neighbouring spins with different
signs. At low temperatures neighbouring spins are antiparallel. This
becomes the antiferromagnetic model. In bipartite lattices where the lattice
naturally breaks up into two interpenetrating sub lattices (e.g., square
lattice in two dimensions and the body centered cubic lattice in three),
the Ising antiferromagnet is mathematically identical to the ferromagnetic
case. Indeed for, ${H} = 0$, if we write every 
alternate $S_i$ as $-S_i$, then ${\cal H}_{\Omega}$ is the same as before because 
the extra minus sign cancels that from $-J$. In other words, if ${H} = 0$, 
the sign of $J$ does not affect the thermodynamic potential. The order
parameter for the Ising antiferromagnet is the ``staggered'' magnetization,
$M_s$, which is the difference of the individual magnetizations of the
two sub lattices. The analogue of the external field, ${H}$, is the ``staggered 
field'', ${H}_s$, which has opposite signs for each sublattice. 

\bigskip
\noindent
{\em B. Model of binary alloy} : Let $S_i = 1$ represent an atom of type $A$ 
at point $i$ and $S_i = -1$ represent an atom of type $B$ at point $i$. Let 
$\epsilon_{AA}, \epsilon_{BB}, \epsilon_{AB}$ be the interaction energies 
between neighbouring atoms. A binary substitutional alloy which has either $A$
or $B$ types of atoms at each site may be described by the Ising model
\cite{tanusri}. The energy of a pair of
neighbouring atoms $i,j$ can be written as
\begin{eqnarray}
\nonumber
\epsilon_{ij} &=& -JS_iS_j - \frac{1}{\zeta}{H}(S_i + S_j) + {\cal K} \\ 
{\cal H}_{\Omega} &=& \frac{1}{2}\sum_{i,j}\epsilon_{ij}
\end{eqnarray}
where $\zeta$ is the number of nearest neighbours for each spin. 

Comparing the energies of a two atom cluster containing either $AA$, $BB$ or
$AB$ types and solving for $J$, ${H}$ and ${\cal K}$ we get :
\begin{eqnarray}
\nonumber
J &=& \frac{1}{2}\epsilon_{AB} - \frac{1}{4}(\epsilon_{AA} + \epsilon_{BB}) \\
\nonumber
{\cal K} &=& \frac{1}{2}\epsilon_{AB} + \frac{1}{4}(\epsilon_{AA} + \epsilon_{BB}) \\  
{H} &=& \frac{\zeta}{4}(\epsilon_{BB} - \epsilon_{AA})
\end{eqnarray}
Using Eqn. 2.7, Eqn. 2.4 now describes the binary alloy. 
If $(\epsilon_{AA} + \epsilon_{BB})/2 < \epsilon_{AB} (J > 0)$, at sufficiently
low temperatures the atoms $A$ will gather together while atoms $B$ will form 
another clsuter. Hence we have a model for a {\em de-mixing} transition. 
If $J < 0$ and the temperature is low, the atoms $A$
and $B$ will arrange alternately, leading to a {\em order-disorder} transition.
The paramagnetic phase at high temperature of course, represents a randomly
substituted alloy - the mixed phase.

\bigskip
\noindent
{\em C. Lattice gas model} : Let $S_i = 1$ represent a molecule at site 
$i$, and $S_i = -1$ represent no molecule at site $i$. We assume that the
same site cannot accommodate two or more molecules. Let the interaction 
energy of neighbouring molecules be $-\epsilon$. Then we may write the 
energy between two neighbouring lattice sites as 
\begin{eqnarray}
\nonumber
- J - \frac{2{H}}{\zeta} + {\cal K} &=& - \epsilon\,\,,\,\,\,\,\,S_i = S_j = 1 \\
\nonumber
- J + \frac{2{H}}{\zeta} + {\cal K} &=& 0\,\,\,\,\,,\,\,\,\,\,S_i = S_j = 1 \\
J + {\cal K} &=& 0\,\,\,\,\,,\,\,\,\,\,S_i = 1,\,\,S_j = - 1
\end{eqnarray}
Solving these, we get
\begin{eqnarray}
\nonumber
J = \frac{1}{4}\epsilon \\ 
{H} = \frac{\zeta}{4}\epsilon
\end{eqnarray}
Therefore, Eqn. 2.4 is also a model of gas molecules, where $- \epsilon$
represents the attraction between the gas molecules. The restriction of 
at most one molecule per site represents short-range repulsion. The high
density state, i.e. $\langle S_i \rangle > 0$ corresponds to the liquid 
phase while the low density state, i.e. $\langle S_i \rangle > 0$ 
corresponds to the gas phase. Although this model may seem rather contrived
since there is no underlying lattice in a real liquid-gas transition,
surprisingly, the critical properties of this model turn out to be identical
to real experimental gases due to universality \cite{Goldenfeld,Chaikin}.

\subsection{Structure of interfaces in the Ising model}
As mentioned above, at any non-zero temperature the Ising model contains 
domains populated by either one of the two ground states. At low enough
temperatures, these domains are separated by well defined interfaces
which fluctuate and move as one phase grows at the expense of the other.
In this section, we shall study Ising interfaces for zero external field,
where the two phases on either side of the interface are degenerate in
energy. We shall also implicitly assume here and elsewhere in this thesis
that the temperature is low enough compared to the coupling parameter $J$
so that the interface can be represented either as a flat boundary or at
least as a single valued curve which does not loop back on itself (i.e., 
contains no overhangs). To find the structure
and behavior of the Ising interface we start with the Ginzburg-Landau free
energy expansion
\begin{eqnarray}
{\cal F} = \int d{\bf r} [ - \frac{\Psi}{2} \phi({\bf r})^2 + 
\frac{c}{4}\phi({\bf r})^4 + \frac{B}{2}|\nabla\phi({\bf r})|^2 ] 
\end{eqnarray} 
where the coefficients $\Psi$, $c$ and $B$ are related to the microscopic
parameters and $\phi$ is the space dependent coarse grained magnetisation. 
In particular $\Psi$ vanishes at the critical point given by 
${\cal T}_c$. The function $\phi({\bf r})$ which minimizes ${\cal F}$ is determined by
the equation :
\begin{eqnarray}
\frac{\delta {\cal F}}{\delta \phi({\bf r})} = 
\frac{\partial f}{\partial \phi} - \frac{\partial}{\partial r_i}
\frac{\partial f}{\partial \phi_{r_i}} = 0
\end{eqnarray}
where $\phi_{r_i} = \partial \phi/\partial r_i$. The repeated index $i$
and ${\bf r}_i = (x,y,z)$ are summed over in Eqn. 2.11. The resulting equation
is 
\begin{eqnarray}
- \Psi \phi + c\phi^3 - B\nabla^2\phi = 0
\end{eqnarray}
with the boundary conditions that the system is uniform away from the 
interface. Far away from the interface, we expect the two phases to have 
their equilibrium values of the composition. Assuming that $\phi$ is small
we get,
\begin{eqnarray}
\phi_0 = \pm \sqrt{\frac{\Psi}{c}}
\end{eqnarray}

\begin{figure}[t]
\begin{center}
\includegraphics[width=8cm]{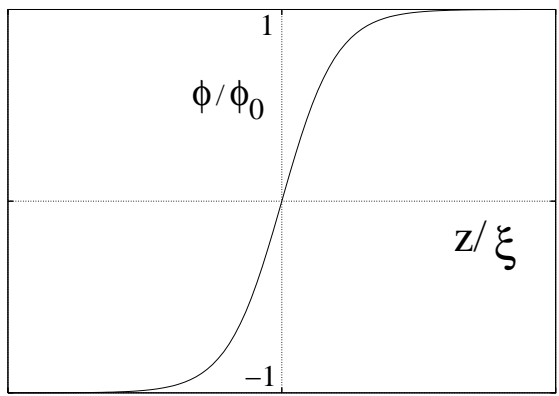}
\end{center}
\caption{A plot of the interfacial profile}
\label{fig2.1}
\end{figure}

Consider now a one-dimensional concentration variation, $\phi(z)$, with
boundary conditions $d\phi/dz = 0$ at $z = \pm\infty$, which ensure the
equilibrium phases at these limits and represents a flat interface. The solution
is then
\begin{eqnarray}
\phi(z) = \sqrt{\frac{\Psi}{c}}\tanh\frac{z}{\xi}
\end{eqnarray}
where the width of the interfacial region is given by 
$\xi = \sqrt{2B/\Psi}$, which is the bulk correlation length and diverges
as $\Psi \rightarrow 0$ or ${\cal T} \rightarrow {\cal T}_c$. Thus, for 
$z \rightarrow \pm\infty$, $\phi$ approaches its equilibrium, uniform
values of $\pm\sqrt{\Psi/c}$ as shown in the Fig.~\ref{fig2.1}. 
The width of
the interfacial region is proportional to $1/\sqrt\Psi$, so that
as the critical point is approached $(\Psi \rightarrow 0)$ the 
interfacial width diverges. 

\subsection{Stability of the Ising interface}
\begin{figure}[t]
\begin{center}
\includegraphics[width=12cm]{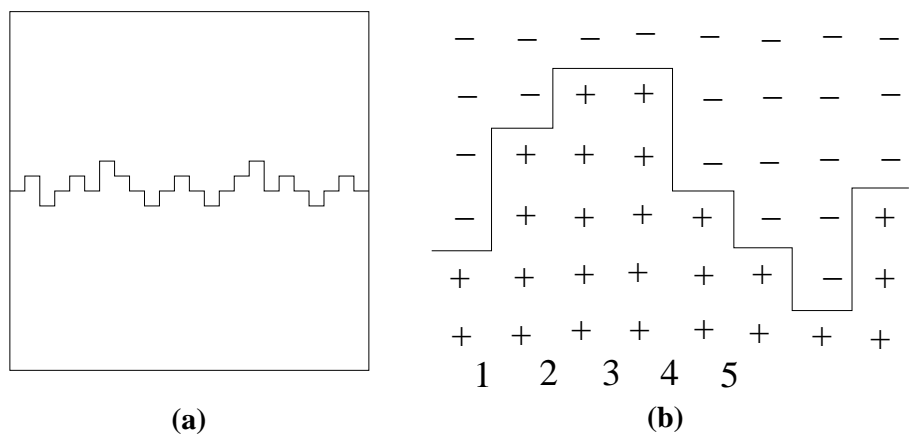}
\end{center}
\caption{(a) Boundary line for a square lattice. (b) An enlargement of a
portion of (a).}
\label{fig2.2}
\end{figure}

To discuss the stability of the flat Ising interface described above, let us 
consider a square lattice of Ising spins with a boundary line (interface) 
separating up and down spin regions as shown in the Fig.~\ref{fig2.2}.
The energy of the interface is its total length times $2J$. The total length is 
\begin{eqnarray}
L = L_0 + \sum_{k=1}^{L_0}|y_k|
\end{eqnarray}
Here, $y_k$ is the length of the $k-th$ boundary counting from the
left. In Fig.~\ref{fig2.2} 
$y_1 = 2, y_2 = 1, y_3 = 0, y_4 = -2, y_5 = -1,....$ etc. 
Therefore the thermodynamic potential ${\cal F}$ of the full interface in the 
limit of large $L_0$, is
\begin{eqnarray}
\nonumber
{\cal F} &=& - k_B{\cal T}\ln {\cal Z} \\
{\cal Z} &=& \sum_{y_1 = -\infty}^{\infty}\sum_{y_2 = -\infty}^{\infty}.....
\sum_{y_{L_0} = -\infty}^{\infty} e^{-2JL/k_B{\cal T}}
\end{eqnarray}
Now, in the k-th horizontal position, the height of the boundary is
\begin{eqnarray}
h_k = \sum_{j=1}^k y_j
\end{eqnarray}
Hence, the difference in height between two points on the interface is
\begin{eqnarray}
\Delta h \equiv h_{k+n} - h_{k} = \sum_{j=k+1}^{k+n} y_j
\end{eqnarray}
If $n << L_0$, each $y_j$ will be independent variables. Using the central
limit theorem we get the distribution of $\Delta h$ :
\begin{eqnarray}
\nonumber
f(\Delta h) &\simeq& \frac{1}{\sqrt{2\pi\sigma}}e^{-(\Delta h)^2/2\sigma^2} 
\\
\nonumber
w^2 &=& n\langle y^2 \rangle \\
&=& \frac{n}{2\sinh^2(J/k_B{\cal T})}
\end{eqnarray}
Hence the interface fluctuates violently in the same manner as a one 
dimensional random walker. The amplitude of fluctuation in the
middle is about $\sqrt{L_0}/\sinh(J/k_B{\cal T})$. Although $\sqrt{L_0}$ is much
smaller than $L_0$, it is still not a microscopic length scale. Therefore,
this interface is a wiggly and rough curve, not smooth and flat.
In this analysis we have not considered "overhangs" or isolated regions. 
However the conclusion that the interface is rough is valid even if we include
these effects.

\section{Driven Interfaces in the Ising model}
\begin{figure}[h]
\begin{center}
\includegraphics[width=6cm]{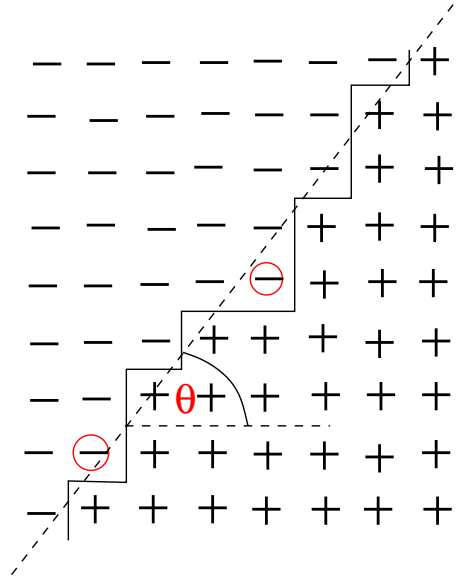}
\end{center}
\caption{An Ising interface (bold line) between up and down Ising spins.
Since $J$ is very large, only spins at the corners, like the circled ones 
(coloured red) can flip. The orientation $\theta$ of the interface is also
shown in the figure.
}
\label{fig2.3}
\end{figure}

When there is an external field, the interface in an Ising model moves in a 
way that increases the size of the domains which are aligned in the 
direction of the field. The general problem of such a driven interface even 
for a simple model system like the Ising model is quite complex. In this
section we show how, by using simplifying assumptions one can study such a 
driven interface in the two-dimensional Ising system. This is a reproduction
of earlier work on constant field driven interface by Majumdar and Barma
\cite{Majumdar2,Majumdar3,Barma2,Barma1}.
In the limit that the field $H$ and temperature ${\cal T}$ are both much 
smaller than the nearest neighbour exchange coupling $J$ (assumed isotropic),
one can neglect overhangs so that the interface can as before be represented
as a single valued function.
The interface moves in steady
state with a velocity $v_f$ which depends on the external field ${H}$ and on
the orientation $\theta$ of the interface, measured with respect to the
horizontal (Fig.~\ref{fig2.3}). 
As the Ising spins obey single-spin flip Glauber dynamics, and $H, {\cal T} << 
J$ holds, the interface moves by
``corner flips'' where a vertical bond and a horizontal bond at a corner
exchange places. 

We shall show how such driven interfaces may be described using the 
KPZ equation. We will derive this dynamical equation
first from the time dependent Ginzberg-Landau equation (TDGL) \cite{Chaikin}, 
and then once again after mapping the corner flip dynamics of the 
two-dimensional Ising model to the dynamics of a one-dimensional gas of hard 
core particles (the exclusion process) in the limit $H/J, {\cal T}/J 
\rightarrow 0$. We shall use the latter mapping to deduce some
important dynamical properties of this interface. 

\subsection{Derivation of KPZ from TDGL}
We consider, for the moment, that the inclination of the interface $\theta = 0$.
The interface is given by the function $h(x,t)$. The coarse grained,
space dependent magnetization $\phi$ is given by a function which is
uniform everywhere except near the interface, $h(x,t)$ such that 
$\phi = \phi(y - h(x,t))$. The function $\phi$ has a profile similar
to Fig. 2.1.
Model A dynamics \cite{Chaikin} for $\phi$ then implies,
\begin{eqnarray}
\frac{\partial \phi}{\partial t} &=& -\Gamma \frac{\delta {\cal H}_{T}}
{\delta \phi} + \zeta ({\bf r},t) 
\label{modelA}
\end{eqnarray}
where 
\begin{eqnarray}
{\cal H}_{T} &=& \int d{\bf r} [ a_{1}\phi^{2} + a_{2}\phi^{4} + 
a_{3}(\nabla \phi)^{2} - {H}\phi]
\end{eqnarray}
is the coarse-grained Hamiltonian (which, for the moment, ignores the 
lattice) of an Ising system in a external field ${H}$ and $\zeta$ 
is a Gaussian white noise with zero mean and  
\begin{eqnarray}
<\zeta ({\bf r},t)\zeta ({\bf r^{\prime}},t^{\prime})> &=& 2k_{B}T{\Gamma}
{\delta({\bf r - \bf r^{\prime}})}{\delta(t - t^{\prime})}
\end{eqnarray}

Using ${\cal H}_{T}$ in Eqn.\,\ref{modelA}, taking $y - h(x,t) = \nu$ and 
converting all derivatives to derivatives over the profile $h(x,t)$, we have, 
\begin{eqnarray}
-\phi^{\prime}(\nu) \frac{\partial h}{\partial t} = &-&\Gamma [2a_{1}\phi(\nu)
 + 4a_{2}\phi^{3}(\nu) - 2a_{3}\phi^{\prime \prime}(\nu) + 
2a_{3}\phi^{\prime}(\nu)\frac {\partial^{2} h}{\partial x^{2}} 
\nonumber\\
&-& 2a_{3}\phi^{\prime \prime}(\nu)
(\frac{\partial h}{\partial t})^{2} - {H}] + \zeta ({\bf r},t)
\end{eqnarray}
where primes denotes derivatives with respect to $\nu$.
We then choose a $\phi$ dependent mobility $\Gamma$ contributing to the lowest
order in $\phi$ consistent with symmetry viz.
$\Gamma = \Gamma_{0} + \Gamma_{1}(\nabla \phi)^{2}$.
Substituting for $\Gamma$ and integrating both sides of the equation with 
respect to $h$ between limits $(h - \chi/2)$ and $(h + \chi/2)$ i.e.
over the interfacial region, we finally get an equation of motion for the 
interface. 
\begin{equation}
\frac{\partial h}{\partial t} = \lambda_{1} \frac{\partial^{2} h}
{\partial x^{2}}-\lambda_{2} \Big(\frac{\partial h}{\partial x} \Big)^{2} 
-\lambda_{3} + \zeta(x,t)
\label{m-KPZ}
\end{equation}
where $\lambda_{1}$,$\lambda_{2}$ and $\lambda_{3}$ are parameters which 
involves the integral of $\phi(\nu), \phi^{\prime}(\nu)$ and 
$\phi^{\prime}(\nu)$ over the regions of the interface. 
Note that Eqn.\,\ref{m-KPZ} lacks Galilean invariance\footnote{Note that 
$f(Y,t)$ in Eq. (\ref{m-KPZ}) generates a space and time dependent,
  (annealed) random, Galilean boost. Random Galilean transformations often lead
  to multifractal steady states; see for eg. U. Frish, {\em Turbulence}
  (Cambridge University Press, 1995)} 
$h^{\prime} \rightarrow h + \epsilon x,\>\> x^{\prime} \rightarrow x - 
\lambda_{2}\epsilon t, \>\> t^{\prime} \rightarrow t$.
Eqn. 2.25 is the familiar KPZ equation and therefore the interface grows 
with KPZ exponents $\alpha = 1/2, \beta = 1/3$.

\subsection{Mapping to the Exclusion Process} 
In the exclusion process we consider a one-dimensional lattice of ${\cal N}_s$ sites, 
of which ${\cal N}_p = \rho {\cal N}_s$ are occupied by particles, and assume periodic 
boundary conditions \cite{Barma2,Majumdar2,Majumdar3}. In the simple exclusion process,
a particle chosen at random attempts to hop with probability $p$ to the right
and probability $q$ to the left, with $p+q = 1$.Because of the hard core 
constraint, the hop actually takes place if the sought site is vacant.${\cal N}_p$
such attempted hops constitute a single time step. 

In the steady state, every configuration of of the ${\cal N}_p$ particles is equally
likely {\cite{AEP1,AEP2}}.Let us label the particles $n = 1,2,.......,{\cal N}_p$  
sequentially at $t = 0$. The ordering is preserved by the dynamics of the 
exclusion process. The configuration of the system is specified by the set 
$\{h(n)\}$ where $h(n)$ denotes the location of the $n$th particle. The 
corresponding interface model, called the particle height (PH) model, is 
obtained by interpreting the tag label $n$ as a horizontal coordinate, and      $h(n)$ as a local height. Each configuration $\{h(n)\}$ then defines a one-
dimensional interface in the form of a staircase inclined to the horizontal 
with mean slope $\tan\theta = 1/\rho$. The interface coordinates satisfy 
$h(n+1) \ge h(n) + 1$, and the periodic boundary conditions translate into
$h(n+{\cal N}_p) = h(n) \pm {\cal N}_s$.The evolution rule is as follows: in each time step,
$h(n)$ tends to increase (or decrease) by 1 with probability $p$ (or $q$); it 
actually increases (or decreases) if and only if ${h(n+1) - h(n) > 1}$ 
(or ${h(n) - h(n-1) > 1}$).In the unbiased case ($p = q = 1/2$), the interface 
does not move with a net velocity, but fluctuates around its initial position.
But in the biased case ($p \ne q$), the interface moves vertically with the 
particle drift velocity $v_p$. The probabilities $p$ and $q$ can be related to 
the Ising model parameters on noting that ratio of the rates of up-down and 
down-up flips is $\exp(-2 \beta {H})$; thus 
$p = \exp(\beta {H})/(2 \cosh(\beta {H})$ and the bias 
$\Delta = p - q = \tanh(\beta {H})$.
\begin{equation}
v_p \equiv <(h(n,t)-h(n,0)> = (1 - \rho)(p - q)
\end{equation}
where $h(n,t)$ is the location of the $n$th particle at time $t$.

\subsection{Some Exact Results}

It is straight-forward to derive \cite{AEP1,AEP2} the expression for $v_p$ and the particle
current $j$. The particle velocity,
\begin{equation}
v_p \equiv <(h(n,t)-h(n,0)> = (1 - \rho)(p - q)
\end{equation}
which is the statement of the fact that in order for a particle to move one 
needs a bias {\em and} a hole to move into.
Although we have defined the exclusion process in terms of particle hopping,
it can equally well be viewed as the backward motion of holes(sites on which
there are no particles). The drift velocity for labelled holes is
$v_{ho} = -(p - q)\rho$ and the steady state current is 
\begin{eqnarray}
j = \rho v_p = (1 - \rho)v_{ho} = \rho(1 - \rho)(p - q)
\end{eqnarray}

In time $t$, the mean vertical shift of the interface is given by the average
number of particles which pass by a given hole, namely, $\rho(v_p -v_{ho})t$.
The magnitude of the normal velocity is then
$v_f = \rho (v_p -v_{ho}) cos \theta = j(\sin \theta + \cos \theta)$
which may be rewritten as
\begin{equation}
v_f = \frac {\tanh {\beta {H}}}{(\sec {\theta} + {\rm cosec} {\theta})}.
\end{equation} 
\begin{figure}[h]
\begin{center}
\includegraphics[angle=270,width=7.0cm]{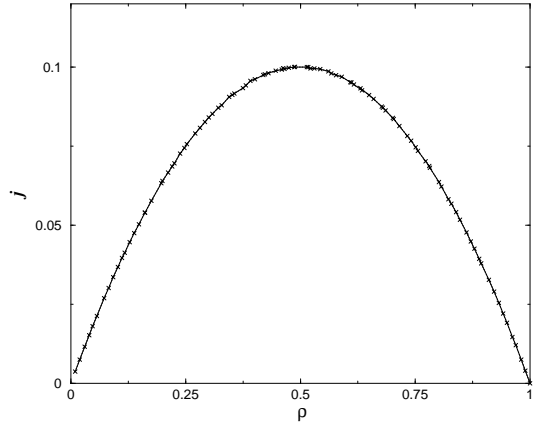}
\end{center}
\caption{Plot of j versus density}
\label{Figure}
\end{figure}

The space dependent coarse-grained density $\rho(n)$ and current 
${\it J(n)}$ are related through the continuity equation \cite{Barma1}
\begin{eqnarray}
\frac {\partial \rho}{\partial t} + \frac {\partial J(n)}{\partial n} &=& 0.
\end{eqnarray} 

Now the current ${\it J(n)}$ can be written as \cite{beijeren},
\begin{eqnarray}  
J(n) &=& -D\frac {\partial \rho}{\partial n} + j(\rho) + \eta,
\end{eqnarray} 
where the first term is the usual diffusion of particles with diffusion
constant $D$, $\eta$ is a Gaussian white noise and $j(\rho)$ is the current
due to the local density $\rho \equiv \rho(n)$. We may replace $j(\rho)$
by its macroscopic value (Eqn. 2.27) and expand it in power series in the 
density deviation $\Lambda = \rho - \overline\rho$ around the average
density $\overline\rho = {\cal N}_p/{\cal N}$.  
\begin{eqnarray}
j(\rho) &=& \sum j_m \Lambda^m ,
\end{eqnarray}  
with $m!j_m = \partial^m J/\partial \rho^m$. The density 
fluctuation field $\Lambda(n,t)$ is related to the height variable $h(n,t)$ which  measures the transverse displacement of the interface:
\begin{eqnarray}
h(n,t) &=& \int_{n_0}^n \Lambda(n',t)dn'. 
\end{eqnarray}  
Differentiating Eqn. 2.32 and using Eqn. 2.29 and Eqn. 2.30 one can show that
{\it h} satisfies the equation by KPZ 
\cite{KPZ}. 
The equation of motion of the interface is
\begin{eqnarray}
\frac{\partial h}{\partial t}  &=& D\frac{\partial^{2}h}{\partial n^{2}} +         \sum_m j_m[\frac{\partial h}{\partial n}]^m + \eta (n,t)
\end{eqnarray}

We consider the height fluctuation correlation function $S(n,t)$ defined by
\begin{eqnarray}
S^2(n,t) &=& <(h(n,t) - h(n,0) - v_pt)^2>  
\end{eqnarray} 
The large-distance large-time behaviour of S(n,t) depends strongly on the 
first few coefficients $j_m$ in Eqn. 2.33. The constant $j_0$ is equal to the 
macroscopic current (Eqn. 2.27) and determines the mean rate of growth of the 
interface. It can be eliminated by the boost $h \to h + j_0t$.
\begin{figure}[h]
\begin{center}
\includegraphics[width=7.0cm]{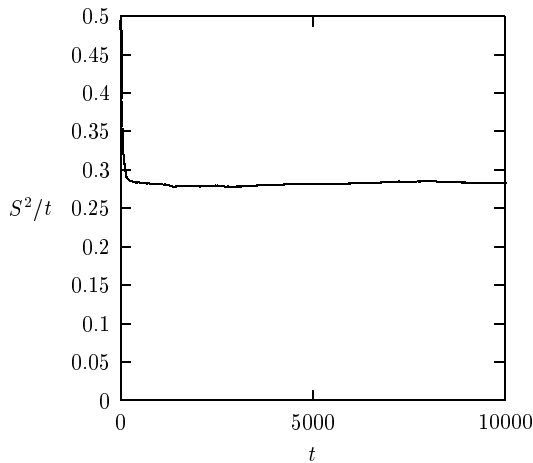}
\end{center}
\caption{The second moment of the height distribution $S^2/t$ as a function
of time $t$.}
\label{Figure}
\end{figure}

The first-order gradient term $j_1$ is equal to $U$, the speed of local
density variations or kinematic waves \cite{Lighthill}. Kinematic waves,
which transport density fluctuations in the lattice system, correspond to the
movement of transverse fluctuations in the interface problem and owes its
existence to the conservation of particle number.
The term $j_1$ can be eliminated from  Eqn. 2.34 by making the Galilean
shift $x \to x + j_1t$.This has the effect of moving to a coordinate system 
in which the wave is stationary.
If the $j_0$ term is eliminated but $j_1$ is not, it can be shown that $S(0,t)$
grows as $t^{1/2}$.Since the sliding of the wave is the dominant physical 
effect so $S(0,t)$ is approximated by the equal time correlation 
function $S(n=Ut,0)$. Since equal-time height fluctuations are determined by 
fluctuations of particle number,these grow as $n^{1/2}$, and it follows that
\begin{eqnarray} 
S(0,t)  \sim  (Ut)^{1/2}.
\end{eqnarray}

We have performed Monte Carlo simulations \cite{Binder1,Binder2} with random sequential updates 
of the particles to reproduce these exact results. This constituted the 
benchmarking and validation for our codes which were subsequently used to
study the system discussed in Chapter 3. In Fig. 2.4 we show a plot of the
measured current as a function of the density together with the exact result.
${\cal N}_s = 10000$ was used in this run. In Fig. 2.5 we have shown our results
for the second moment of the height distribution function.  Our Monte Carlo
results clearly depicted the $t^{1/2}$ behaviour. We used ${\cal N}_s = 80000$ for
this calculation. 


\chapter{Ising Interfaces in spatially varying external field : Novel 
fluctuations}

In the last chapter we introduced the Ising model and the Ising interface.
We studied the structure of this interface in two-dimensions and discovered
that at any non zero temperature the interface is never flat but fluctuates 
rather strongly. The random fluctuations of the interface diverge as the
system size as $L^{1/2}$ and with time as $t^{1/3}$.

In this chapter, we discover how to control these interfacial fluctuations
and create an essentially flat interface. The ability to grow flat solid 
surfaces \cite{magnet1,magnet2} is often of major technological concern, 
for example, in the fabrication of magnetic materials 
for recording devices where surface roughness \cite{degrade} causes a sharp
deterioration of magnetic properties. We show that non-uniform external
fields\footnote{There are several practical examples where non-uniform
fields drive interfaces. Some of them include zone purification of Si where the 
controlled motion of a temperature field profile is used to preferentially
segregate impurities \cite{Haasen}, magnetization of a 
bar of iron with a permanent magnet, phase transitions induced by a travelling
heat (welding) or pressure (metamorphosis of rocks) fronts etc.} which have
a sharp gradient at the position of the interface, suppresses height
fluctuations to a large extent. This is a direct consequence of the fact 
that such spatially varying fields break the translational symmetry of the
interface in the direction of growth so that long wavelength fluctuations
now do cost non zero interfacial energy.
Surprisingly however we show that this suppression of long wavelength 
fluctuations generate new {\em short wavelength} structures and patterns.
Driving the ineterface by using a moving field profile causes interesting
dynamical transitions between patterns which are either commensurate or
incommensurate with the underlying lattice structure over which the Ising
model is defined.

Commensurate-in~-commensurate (C-I) transitions \cite{Chaikin} have been 
extensively studied over almost half a century following early experiments on 
noble gases adsorbed on a crystalline substrate \cite{Kr} eg. Kr on graphite. 
Depending on coverage and temperature, adsorbates may show high 
density periodic structures the reciprocal lattice vectors (RLVs) 
of which are either a rational (commensurate) or irrational 
(in~-commensurate) multiple  of a substrate RLV. By changing external 
parameters (eg. temperature) one may induce phase-transitions between 
these structures. Recently, the upsurge of interest in the fabrication
of nano-devices have meant a renewed interest in this field following a
large number of experimental observations on ``self-assembled'' domain patterns
(stripes or droplets) on epitaxially grown thin films for eg. Ag films on
Ru(0001) or Cu-Pb films on Cu(111) \cite{films} etc. The whole area of 
surface structure modifications and surface patterns has tremendous 
technological implications for example in opto-electronics, recording
industry, coatings and paints, etc.

Almost universally, C-I transitions may be understood 
using some version of the simple Frenkel~-Kontorova \cite{Talopov} model, 
which models them as arising from a 
competition between the elastic energy associated with the distortion of 
the adsorbate lattice and substrate~-adsorbate interactions. A 
complicated phase diagram involving an infinity of phases corresponding to 
various possible commensuration ratios (rational fractions) is obtained as 
a function of the two energy scales. In-between two commensurate structures 
one obtains regions where the periodicity of the adsorbate lattice is 
in~-commensurate.

The structures obtained in our interfacial problem, on the other hand, undergo
C-I transitions which are entirely dynamical in origin and can be controlled
by adjusting the velocity of the interface. Short periodic structures are
stabilized at lower velocities. As the velocity of the interface approaches
a limiting velocity $v_{\infty}$, the patterns disappear and the interface 
begins to fluctuate over all length scales. Beyond $v_{\infty}$, KPZ behaviour
takes over. 

\section{The Model}

\begin{figure}[h]
\begin{center}
\includegraphics[width=7cm]{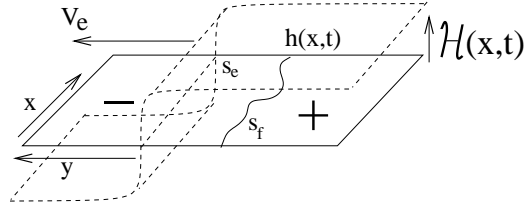}
\end{center}
\caption{
An Ising interface $h(x,t)$ (bold curved line) between regions of positive 
(marked $+$) and
negative (marked $-$) magnetization in an external, inhomogeneous field 
with a profile which is as shown(dashed line). The positions of the edge of the
field profile and that of the interface are labelled $S_e$ and $S_f$ 
respectively.
}
\label{ising}
\end{figure}

We show in Fig.~\ref{ising} a one-dimensional interface $h(x,t)$ between phases with 
magnetization, $\phi(x,y,t) > 0$ and $\phi(x,y,t) < 0$, in a two-dimensional 
Ising model on a square lattice\footnote{$\phi$ represents any scalar order 
parameter with $h(y,t)$ a field conjugate to $\phi$, eg. number density and 
chemical potential for a growing gas~-solid surface.} obeying single-spin flip 
Glauber dynamics \cite{Men} in the limit $H/J , {\cal T}/J \rightarrow 0$. 
An external non-uniform field   is applied
such that $H = {\it H_{max}}$ in the +ve  
and $-{\it H_{max}}$ in the -ve $\phi$ regions separated by
a sharp {\it edge}. The {\it edge} of the field (i.e. where the 
field changes sign) lies at $S_e$. The {\it front} or interface, $h(x,t)$,  
(no overhangs !) separates up and down spin phases. The interface 
is shown as a bold curved line with the average position  $S_f$.
To move the interface we move the edge with velocity ${\it v_e}$; 
in response the front moves with velocity ${\it v_f}$. Parts of the front 
which leads (lags) the edge of the 
field experience a backward (forward) force pulling it towards the edge.  
The driving force therefore varies in both space and time and depends on the 
relative position of the front compared to that of the edge of the dragging 
field. In the low temperature limit the 
interface moves solely by random corner flips \cite{Barabasi} , 
the fluctuations  
necessary for nucleating islands of the minority phase in any region being
absent. We study the behaviour of the front velocity $v_f$ and the 
structure of the interface as a function of $v_e$ and orientation.
\begin{figure}[h]
\begin{center}
\includegraphics[width=7cm]{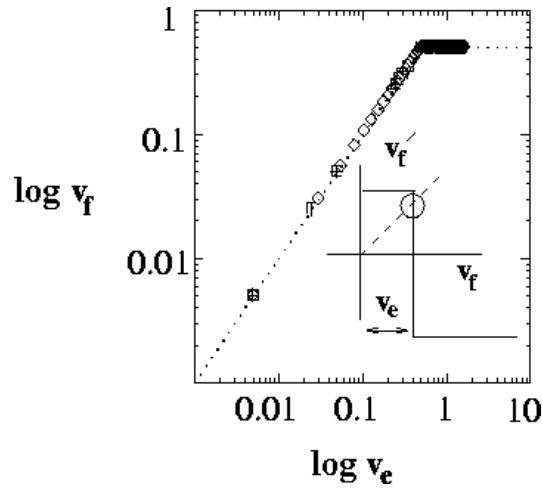}
\end{center}
\caption{
The interface velocity $v_f$ as a function of the velocity of the
dragging edge $v_e$ for ${\cal N}_s = 100(\Box), 1000(\Diamond), 10000(+)$
and $\rho = 0.5$. All the data ($\Box,\Diamond,+$) collapse on the
mean field solution (dashed line). Inset shows the graphical solution
(circled) of the self-consistency equation for $v_f$; dashed line
represents $v_f = v_f$.
}
\label{modvel}
\end{figure}

Naively, one would expect fluctuations of the interfacial coordinate $h(x,t)$
to be completely suppressed in the presence of a field profile. This 
expectation, is only partially true. While, as we show below, a mean field 
theory gives the exact behaviour of the front velocity $v_f$ as a function of 
$v_e$; small interfacial fluctuations produce a dynamical phase 
diagram showing infinitely many dynamical phases. 
For $v_e < v_{\infty}$ 
the interface is stuck to the profile $v_f = v_e$. The stuck phase
has a rich structure showing microscopic, ``lock-in'', commensurate
ripples. These disappear at high velocities through a dynamical  
C-I transition.
For large $v_e$ the interface detaches from the profile and moves at 
velocity $v_{\infty}$ corresponding to an uniform field of magnitude 
${H} = {H}_{max}$ independent of $v_e$ and coarsens with KPZ exponents.

\section{Continuum Description}
In the continuum description of this model, we need to substitute 
$H = H(h,t) = H_{max}f(h,t)$ with $f(h,t) = \tanh ((h - v_{e}t)/\chi)$ 
where $\chi$ is the width of the profile (see Fig~\ref{ising}) in the
equation for the Hamiltonian in Eqn. 2.22. Carrying out an analogous procedure
for obtaining the equation of motion for the interface we obtain a generalized
KPZ equation for the dynamics of the interface in a spatially varying field.
The result analogous to Eqn. \ref{m-KPZ} is
\begin{eqnarray}
\frac{\partial h}{\partial t} = \lambda_{1} \frac{\partial^{2} h}
{\partial x^{2}}-\lambda_{2} \Big(\frac{\partial h}{\partial x} \Big)^{2} 
f(h,t)-\lambda_{3} f(h,t)+ \zeta(x,t)
\label{mod-KPZ}
\end{eqnarray}  
A mean field calculation amounts to taking $h \equiv h(t)$ i.e. neglecting 
spatial fluctuations of the interface and noise. 
For large times ($t \rightarrow \infty$), $h \rightarrow v_{f}t$, where 
$v_{f}$ is obtained by solving the self-consistency equation;
\begin{eqnarray}
v_{f} &=& \lim_{t\to \infty} -\lambda_{3}\tanh \Big(\frac{(v_{f} - v_{e})t}{\chi} \Big) \nonumber \\
      &=& -\lambda_{3}\,{\rm sign}(v_{f} - v_{e})
\label{mft-v}
\end{eqnarray}

For small $v_e$ the only solution to Eqn. \ref{mft-v} is 
$v_f = v_e$ and for  $v_{e} > v_{\infty}$, where 
$v_{\infty} = \lambda_{3}$ we get 
$v_{f} = \lambda_{3} = v_{\infty}$. We thus have a sharp transition\, 
(Fig.~\ref{modvel}) from a region where the interface is stuck to the edge 
to one where it moves with a constant velocity.
How is this result altered by including spatial fluctuations of $h$ ?
We answer this question by mapping the interface problem to an 
asymmetric exclusion process \cite{Barabasi,AEP1,AEP2} and studying
the dynamics both analytically and numerically using computer simulations.

\section{Beyond Mean Field Theory}

The mapping to the exclusion process follows
 \cite{Barma2,Majumdar1,Majumdar3,AEP1,AEP2} as in the previous chapter by distributing
${\cal N}_p$ particles among ${\cal N}_s$ sites of a 1-d lattice. 
The particles are labelled $i = 1,2,.......,{\cal N}_p$
sequentially at $t = 0$. Any configuration of the system is specified 
by the set of integers 
$\{h_i\}$ where $h_i$ denotes the location of the $i$th particle. In the
interface picture $i$ maps onto a  horizontal
coordinate ($x$ in Fig.~\ref{ising}), and $h_i$ as the local height $h(x)$. 
Each configuration $\{h_i\}$ defines a one-dimensional interface inclined
to the horizontal with mean slope $\tan \theta_f = 1/\rho$ where 
$\rho = {\cal N}_{p}/{\cal N}_{s}$. The $h_i$ satisfy the hard core constraint 
$h_{i+1} \geq h_i + 1$. The local slope near particle $i$ is given 
by $h_{i+1} - h_i$ and is equal to the inverse {\em local} density 
$\rho_i$ measured in a region around the $i^{\rm th}$ particle. 
Alternatively, one associates a vertical bond with 
a particle and a horizontal bond with a hole \cite{Majumdar1,Majumdar3}, 
in which case, again, we 
obtain an interface with a slope $\tan \theta_f^\prime = \rho/(1-\rho)$. The
two mappings are distinct but equivalent. 
Periodic boundary conditions amount to setting
$h_{i+{\cal N}_p} = h_i \pm {\cal N}_s$.
Motion of the interface, by corner flips
corresponds to the hopping of particles. In each time step
(${\cal N}_{p}$ attempted hops with particles chosen randomly and 
sequentially \cite{AEP1,AEP2}),
$h_i$ tends to increase (or decrease) by 1 with probability $p$ (or $q$); it
actually increases (or decreases) if and only if ${h_{i+1} - h_i > 1}$. 
The dynamics involving random sequential updates is known to introduce the 
least amount of correlations among $h_i$ which enables one to 
derive exact analytic expressions for dynamical quantities using simple 
mean field arguments \cite{AEP1,AEP2} (see chapter 2). 
The right and left jump probabilities $p$ and $q$ ($p+q = 1$) 
themselves depend on the relative position of the interface 
$h_i$ and the edge of the field profile $i/\rho + v_e t$. Note that this 
relative position is defined in a moving reference frame
with velocity $v_f(t)$, the instantaneous average particle velocity defined
as the total number of particles moving right per time 
step. We use a bias 
$\Delta_i(t) = p-q = \Delta\,{\rm sign}(h_i- i/\rho - v_e t)$ 
with $\Delta = 1$ unless otherwise stated. Our model is thus a generalization
of the one-dimensional exclusion process with space dependent, dynamic 
jump probabilities \cite{Barma2,Majumdar2}. 
In addition to the front velocity $v_f$, we also examine the behaviour of 
the average position,
\begin{eqnarray}
<h(t)> = {\cal N}_p^{-1}\sum_{i=1,{\cal N}_p}h_i(t)
\end{eqnarray}
\noindent
and the width of the interface: 
\begin{eqnarray}
w^{2}(t) &=& {\cal N}_p^{-1}\sum_{i=1,{\cal N}_p}<(h_i(t) - <h_i(t)>)^2>
\end{eqnarray}
as a function of time and system size ${\cal N}_s$. Here, $<h_i(t)> = i/\rho + v_et$.
Angular brackets denotes an average over the realizations of the 
random noise.  Note that the usual particle hole symmetry for an exclusion 
process \cite{Barabasi,AEP1,AEP2}
is violated since exchanging particles and holes changes the relative position
of the interface compared to the edge. This violation is, of course, 
completely equivalent to the breaking of translation symmetry of the 
interface in the growth direction as mentioned before (chapter 1). 

We perform numerical simulations of the above 
model for ${\cal N}_s$ upto $10^4$ to obtain $v_f$ for the steady 
state interface as a function of $v_e$ as shown in Fig.~\ref{modvel}.  
A sharp dynamical transition from an initially stuck interface with 
$v_f = v_e$ to a free, detached interface with 
$v_f = v_\infty = \Delta (1-\rho)$ is clearly evident as predicted 
by mean field theory. The detached interface coarsens with KPZ 
exponents \cite{ising-physica}.
Note that, even though the mean field solution for $v_f(v_e)$ neglects 
the fluctuations present in our simulation, it is exact. The detailed 
nature of the stuck phase 
($v_f = v_e$ and $w$ bounded) is, on the other hand,
considerably more complicated than the mean field assumption $h(x,t)=h(t)$.
Below we analyse the nature of the stuck phase starting from the ground
state configurations at $v_e = 0$.

\subsection{The Ground State and the Devil's Staircase}

The ground state of the interface in the presence of a stationary 
($v_e = 0$) field profile is obtained by minimizing 
$E = \sum_i (h_i- i/\rho - c)^2$ with respect to the set $\{h_i\}$ 
and $c$. This maybe shown from Eqn \ref{mod-KPZ}
by neglecting terms containing spatial derivatives of $h$; the resulting 
equation of motion, for small deviations of $h$ from the edge may be derived 
from the effective Hamiltonian $E$. 
The form of $E$ leads 
to an infinite range, non-local, repulsive, interaction between particles 
in addition to hard core repulsion and the minimization is subject to the 
constraint that  
$h_i$ be an integer. For our system, the result for the energy may be obtained
exactly for density $\rho = m/n$, an arbitrary rational fraction.
For even $m$ we have the following expression.

\begin{eqnarray}
E &=& \frac{1}{m}\left[ (\frac{1}{m}-c)^2 + (\frac{1}{m}+c)^2 + 
(\frac{2}{m}-c)^2 + (\frac{2}{m}+c)^2 \right.
\nonumber\\
&+&....+ \left. (\frac{\frac{m}{2}}{m}-c)^2 + (\frac{\frac{m}{2}}{m}+c)^2 + 
(\frac{1}{2}-c)^2 + c^2\right]
\nonumber\\
&=& \frac{1}{6}(\frac{1}{2}-\frac{1}{m})(1-\frac{1}{m}) + \frac{1}{m}(mc^2
+ \frac{1}{4} -c)
\end{eqnarray} 

Minimising with respect to $c$, we get $c = \frac{1}{2m}$. Substituting for
$c$ we get the expression for minimized energy as

\begin{eqnarray}
E &=& \frac{1}{6}(\frac{1}{2}-\frac{1}{m})(1-\frac{1}{m})+
\frac{1}{4m}-\frac{1}{4m^2}
\end{eqnarray}

\noindent
Similarly for odd $m$ we have the following expression for energy,

\begin{eqnarray}
E &=& \frac{1}{m}\left[ (\frac{1}{m}-c)^2 + (\frac{1}{m}+c)^2 + 
(\frac{2}{m}-c)^2 + (\frac{2}{m}+c)^2 \right.
\nonumber\\
&+&............+ \left. (\frac{\frac{m}{2}}{m}-c)^2 + 
(\frac{\frac{m}{2}}{m}+c)^2 + c^2\right]
\nonumber\\
&=& \frac{1}{12}(1 - \frac{1}{m^2}) + c^2
\end{eqnarray} 

\noindent
which when minimised with respect to $c$ gives $c = 0$. Hence, the minimised
energy for odd $m$ is given by

\begin{eqnarray}
E &=& \frac{1}{12}(1 - \frac{1}{m^2})
\end{eqnarray}

\begin{figure}[h]
\begin{center}
\includegraphics[width=12cm]{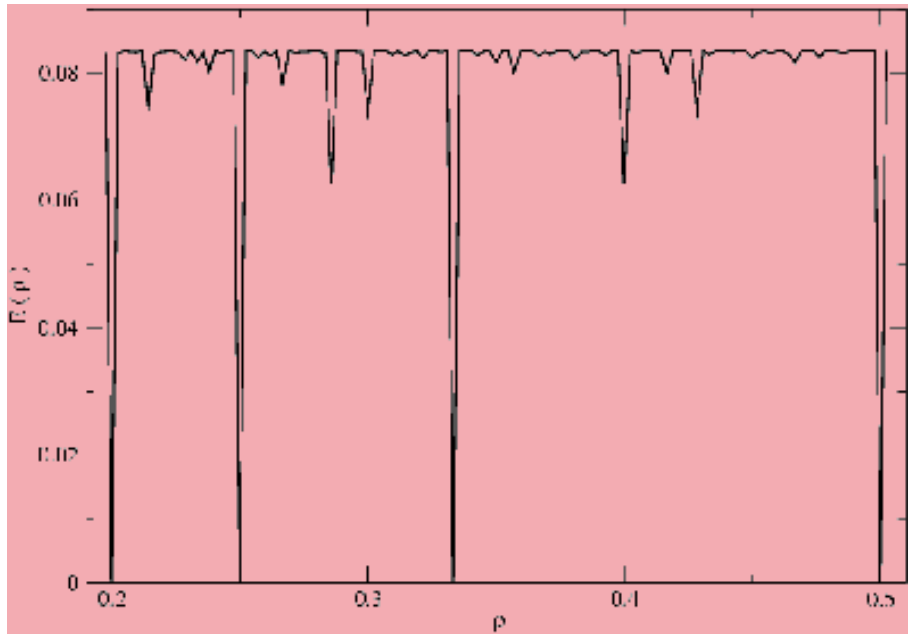}
\end{center}
\caption{
Plot of $E(\rho)$ for a ${\cal N}_s = 420$ site system.
}
\label{ener}
\end{figure}

\begin{figure}[h]
\begin{center}
\includegraphics[width=10cm]{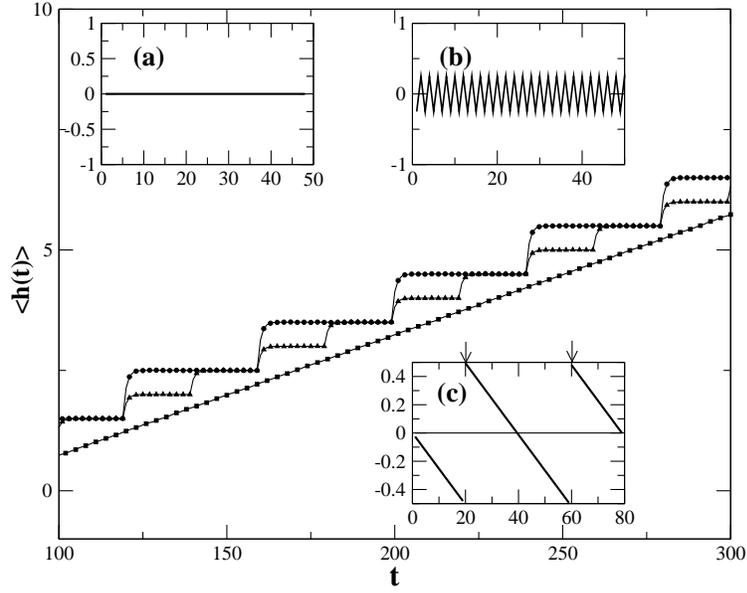}
\end{center}
\caption{ Variation of $<h(t)>$ with $t$ for 
$v_{e} = 0.025$ and $p = 1.0$. Lines denote analytic
results while points denote Monte Carlo data for $\rho = 1/5$ (uppermost curve),
$2/5$ and an incommensurate $\rho$ near $1/3$.
Inset (a)-(c) shows the corresponding ground state 
interfaces ($h_i-i/\rho$). The arrows in (c) mark the positions of two 
discommensurations.
}
\label{hops}
\end{figure}

\noindent
The resulting ground state profiles 
are shown in Fig.~\ref{hops}. The lower bound for $E(\rho)$ is zero 
which is the energy for all $\rho = 1/n$.
For irrational $\rho$ the energy is given by $\lim_{m\to \infty} E(m/n) = 1/12$
for both even and odd $m$ which constitutes an upper bound (Fig.~\ref{ener}).
For an arbitrary $0\,<\,\rho\,<\,1$ the system ($\{h_i\}$) therefore prefers 
to distort, conforming within local regions, to the nearest low-lying 
rational slope $1/\tilde{\rho}$ interspersed with ``discommensurations'' 
of density $\rho_d = |\rho - \tilde{\rho}|$ and sign +ve (-ve)
if these regions are shifted towards (away) from each other by $1$.  
A plot of $\tilde{\rho}(\rho)$ shows a ``Devil's staircase'' 
structure \cite{FK,Bak2}. We observe this in our simulations by analysing
the instantaneous distribution of the local density of the particle-hole
system to obtain weights for various rational fractions. A time average
of these weights then give us the most probable density $\tilde{\rho}$ - 
distinct from the average $\rho$ which is constrained to be the inverse
slope of the interface.

\subsection{Dynamics of interfacial patterns for small $v_e$}
What are the low lying excitations of the ground state structures ?
For low velocities and density where correlation effects due to the 
hard core constraint are negligible, the dynamics of the interface may also be 
obtained exactly. Under these circumstances the ${\cal N}_p$ 
particle probability distribution for the $h_i$'s, $P(h_1,h_2,\cdots,h_{{\cal N}_p})$
factorizes into single particle terms $P(h_i)$. Knowing the time development
of $P(h_i)$ and the ground state structure the motion of 
the interface at subsequent times may be trivially computed as a sum of 
single particle motions. 
A single particle (with say index $i$) moves with
the bias $\Delta_i(v_e\,t)$ which, in general, may change sign at 
$h < i/\rho+v_e\,t < h + 1$. Then $P(h_i)$ satisfies the following set of
master equations,
\begin{eqnarray}
\dot P(h_i) & = & -P(h_i) + P(h_i + 1) \,\,\,\,\,\,\,\,\,\,\,\,\,\,\,\,\,\,\,\,\, {\rm for\,\,\, h_i\, >\, h+1} \nonumber\\
\dot P(h_i) & = & P(h_i - 1) - P(h_i) + P(h_i + 1)\,\,\,\,\,\,\,{\rm for\,\,\, {h_i}\, =\, h,h+1} \nonumber\\
\dot P(h_i) & = & -P(h_i) + P(h_i - 1)\,\,\,\,\,\,\,\,\,\,\,\,\,\,\,\,\,\,\,\,\, {\rm for\,\,\, {h_i}\, <\, h}.
\label{master}
\end{eqnarray}

\noindent

Note that the average position of the particle
is given simply by $<h_i(t)> = \sum_{h_i=-\infty}^{\infty} h_i\,P(h_i)$ and 
the spread by $w^{2}(t) = \sum_{h_i=-\infty}^{\infty} 
(h_i - <h_i(t)>)^2\,P(h_i)$.
Solving the appropriate set of master 
equations we obtain,
for $v_e << 1$ the rather obvious steady state solution 
$P(h_i) = 1/2(\delta_{h_i,h}+\delta_{h_i,h+1})$ and the particle 
oscillates between $h$ and $h+1$. Subsequently, when $i/\rho+v_e\,t \geq h+1$,
the particle jumps to the next position and $P(h_i)$ relaxes exponentially
with a time constant $\tau = 1$ to it's new value with $h \to h+1$.
For $\rho = 1/n$ the entire interface moves 
as a single particle and the average position advances in steps with a 
periodicity of $1/v_e$ (see Fig.~\ref{hops})

In general, for rational $\rho = m/n$, the motion
of the interface is composed of the independent motions of $m$ particles
each separated by a time lag of $\tau_{L} = 1/m\,v_e$.
The result of the analytic calculation for small $v_e$ and $\rho$ has been 
compared to those from simulations in Fig.~\ref{hops}  for 
$\rho = 1/5$  and $2/5 $. 
For a general irrational $\rho < 1/2$, $m \to \infty$ consequently, 
$\tau_{L} \to 0$. The 
$h_i$'s are distributed uniformly around the mean implying $w^2 = 1/3$
independent of system size and time.
For $\rho > 1/2$ the width $w^2 = (1-\rho)/3\rho$
since the number of mobile particles 
decreases by a factor of $(1-\rho)/\rho$. 

In chapter 2, we encountered kinematic waves which were travelling density 
modulations in the exclusion process which, in the interface picture, 
represent height fluctuations which travel along the interface with 
velocity $U = \partial J/\partial \rho$. In this case too, the forward 
motion of the interface is accompanied by the motion of discommensurations
along the interface. The velocity of these kinematic waves in this system,
however, is constrained to be equal to $v_e$.

As the velocity $v_e$ is increased, the time lag, $\tau_L$, decreases and the 
patterns begin to get distorted $\tau_L$ becomes comparable with the MC
time step. The instantaneous values of $\tilde{\rho}$ begins to make 
excursions to other nearby fractions and eventually becomes free. Since 
$\tau_L$ is smaller for large $m$, higher order fractions becomes unstable
earlier. 
\begin{figure}[h]
\begin{center}
\includegraphics[width=6.8cm]{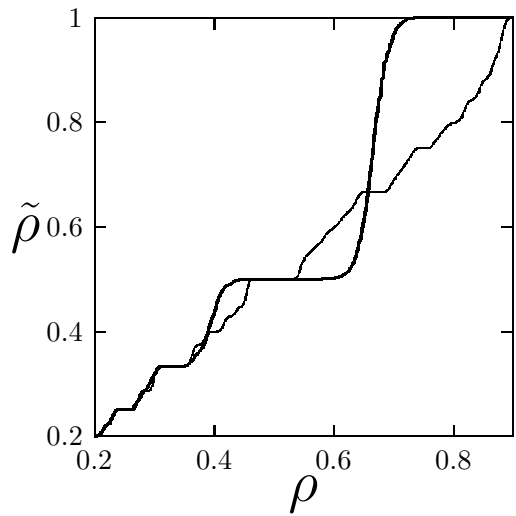}
\end{center}
\caption{
Devil's staircase structure for two different velocities. Steps disappear
as velocity of the edge increases. 
}
\label{veldev}
\end{figure}
In Fig.~\ref{veldev}, thus effect is clearly visible; the Devil's staircase
for the $\tilde{\rho}(\overline{\rho})$ curve for the higher velocity has
fewer steps. Steps corresponding to $\tilde{\rho} = m/n$ disappear 
(i.e. $\tilde{\rho} \rightarrow \rho$) sequentially in order of decreasing
$m$ and the interface losses the ripple patterns. 

\subsection{Dynamical transitions and the dynamical phase diagram}
\begin{figure}[t]
\begin{center}
\includegraphics[width=18cm,height=13cm]{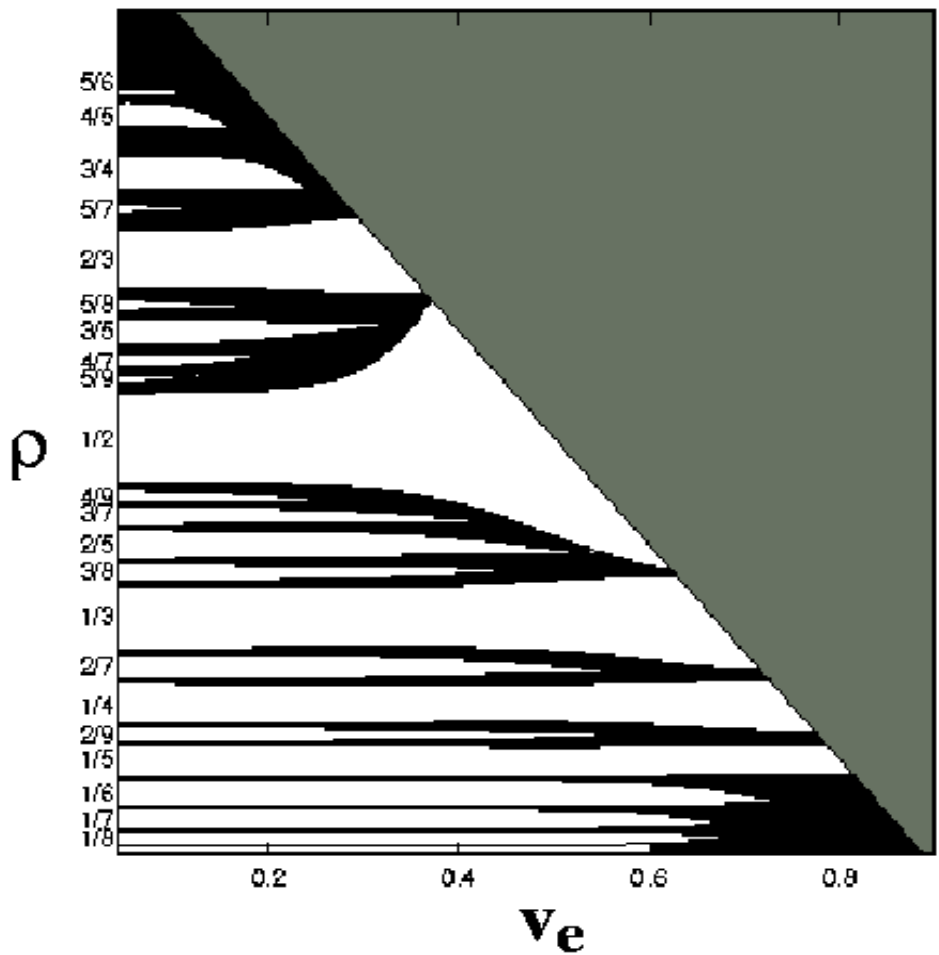}
\end{center}
\caption
{The dynamical phase diagram in $v_e$ and $\rho$ plane. The numbers on the 
$\rho$ axis mark the fractions $\tilde{\rho}$, which determines
the orientation of the lock-in phase. The three regions white, black and grey 
correspond to the rippled, the disordered and the detached 
phases respectively.} 
\label{pdia}
\end{figure}

The beginning and end points of each step in the 
$\tilde{\rho}(\overline{\rho})$ curve (Fig.~\ref{veldev}) mark the stability limits
for each of the patterns which may be regarded as distinct dynamical phases.
The locus of these points as a function of $v_e$ trace out the dynamical 
phase diagram which is shown in Fig.~\ref{pdia}. 
As expected, fractions with 
higher values of $m$ disappear as $v_e$ increases making way for simpler
structures. We thus have true dynamical transitions from more complicated
to simpler patterns as $v_e$ approaches $v_{\infty}$. Beyond $v_{\infty}$,
the interface detaches and we get back the KPZ interface.

\begin{figure}[h]
\begin{center}
\includegraphics[width=6.5cm]{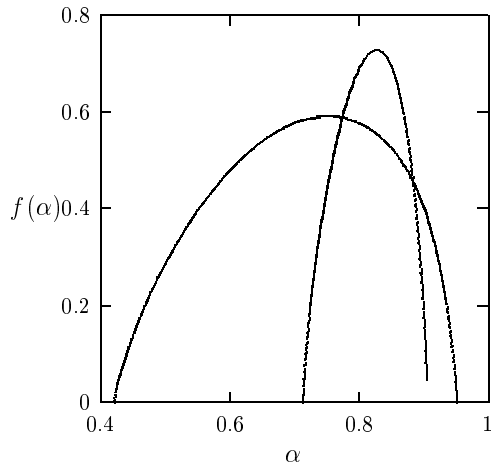}
\end{center}
\caption[]
{The multifractal spectrum $f(\alpha)$ for two different velocities 
$v_e/(1-\rho) = 0.1$ and $ 0.5$.. Note that with increasing velocity, 
$\alpha_{min} \to 0$ (see text).} 
\label{falpha}
\end{figure}

The sequential smoothening of the Devil's staircase may be naturally
described using the language of multifractals. For every $v_e$, we first
obtain the generalized dimension $D_k$. For this purpose we obtain the
scales $l_i$ which are the differences in $\rho$ between subsequent
rational fractions in the Farey construction (e.g. $\rho = 0, 1, 1/2, 1/3,
2/3,$ etc). For every scale, the corresponding measure $\pi_i$ is given
by the jump in $\tilde{\rho}$ which depends on $v_e$ and the structure
of the Devil's staircase. The generalized dimension $D_k$ is obtained
for any given value of $k$ by solving 
\begin{equation}
\sum_{i}(\pi_{i}^{k}/l_{i}^{(k - 1)D_{k}}) = 1.
\end{equation}
Once $D_k$ is obtained as a function of $k$, $f(\alpha)$, the multifractal
spectrum of singularities, is obtained by a Legendre transform of 
$(k - 1)D_k$ (see Fig.~\ref{falpha}).
\begin{eqnarray}
\alpha(k) = \frac{d}{dk}[(k - 1)D_{k}]
\nonumber \\
f(\alpha) = k \frac{d}{dk}[(k - 1)D_{k}] - (k - 1)D_{k}
\end{eqnarray}
In the vicinity of the primary steps corresponding to $\rho = 1/n$, 
$\tilde\rho \sim (\rho - \rho_{max})^{\xi}$ where $\rho_{max}$ is the
largest density at which the step is stable. This universal critical 
exponent \cite{modelock} $\xi = D_{\infty}$ has been determined to be 
$0.71 \pm .001$ from our data at $v_e / (1-\rho) = 0.1$. The exponent
$\xi$ determines how strongly $\tilde{\rho}$ diverges away from its
value at a step for small changes in $\rho$. Since the density $\rho$
is related to the average orientation of the interface which is in turn
determined by the orientation of the external field profile, $\xi$
determines the stability of the ripple pattern to small changes in 
the external field. As $v_e$ increases, $\xi \rightarrow 0$ implying
that the $1/n$ steps become extremely stable at the expense of those
corresponding to higher order rational fractions. 

The nature of the C-I transitions discussed here is essentially new and
fundamentally different from that described within the Frenkel-Kontorowa (F-K)
model. Indeed the control parameter for our phase diagram is $v_e$ - a 
dynamical parameter. Also, the non local $E(\rho)$ and the non linear 
constraint $h_i = integer$ makes it impossible to devise a natural mapping
of this problem to an effective F-K model. We therefore use a different
approach as discussed below.

\section{Langevin Dynamics}

While the stability of mismatch domains \cite{Bak2} is 
decided, mainly, by competition between mechanical, long-ranged 
(elastic) and short-ranged (atomistic) interactions \cite{FK}, dynamical 
ripples vanish with increasing $v_e$ through increased fluctuations. 
We argue that it is sufficient to project the 
entire configuration space ${h_i}$ of the stuck interface onto the single 
variable $\rho$.  
As is obvious from the energy versus rho diagram (Fig.~\ref{ener})
$E(\rho)$ has a structure similar to the free energy surface of a 
1-d ``trap'' model \cite{trap} often used to describe glassy dynamics.
\begin{figure}[t]
\begin{center}
\includegraphics[width=12cm]{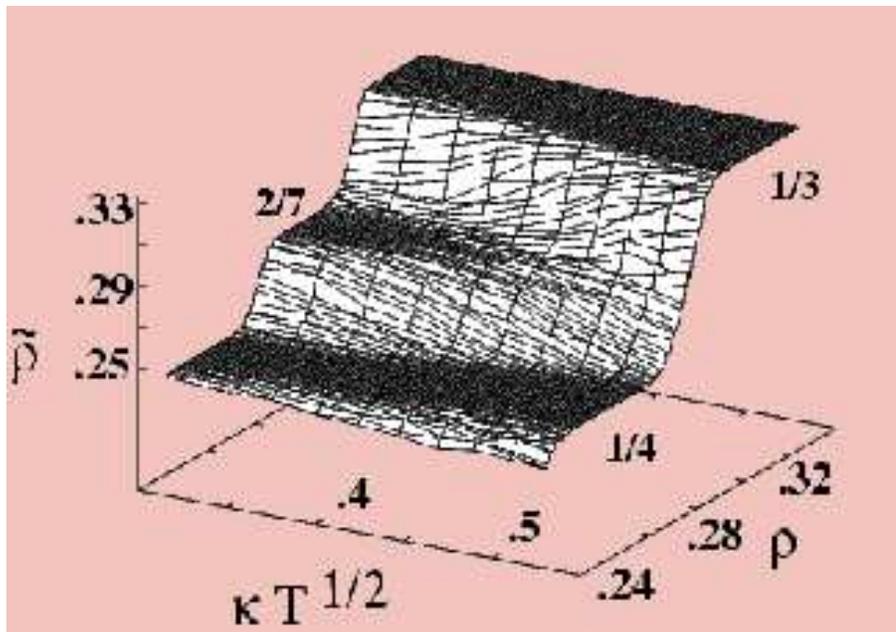}
\end{center}
\caption{
A surface plot of $\tilde{\rho}$ vs $\rho$ and $\sqrt{T} \kappa$ showing 
steps for the fractions $1/4,2/7$ and $1/3$. Note that the 
step corresponding to $2/7$ 
vanishes at $\sqrt{T} \kappa > 0.5$.
}
\label{lang}
\end{figure} 
The distinction, of course, is the fact that the energies of the traps in 
this case are highly correlated. We now 
attempt to describe the dynamics of the stuck interface as the Langevin 
dynamics viz.
\begin{eqnarray}
\dot\rho^\prime & = & -\frac{d\,F}{d\,\rho^\prime} + \eta_{\rho^\prime} \nonumber \\
<\eta_{\rho^\prime}(t)\eta_{\rho^\prime}(t^\prime)> & = & 2\,T\,\delta(t-t^\prime)
\label{slang}
\end{eqnarray}  
of a single particle with coordinate $\rho^\prime$ diffusing on a 
energy surface given by, 
\begin{equation}
F(\rho^\prime) = E(\rho^\prime) + \kappa\,(\rho^\prime - \rho)^2.
\label{free}
\end{equation}  

The particle is kicked by a Gaussian white noise ($\eta_{\rho}$) of 
strength $T$. The second term, containing the modulus $\kappa$, 
ensures that for infinitely large times $\rho^\prime$ always relaxes to the 
true global minimum $\rho^\prime = \rho$.
At intermediate times, however, 
the system may get trapped indefinitely in some nearby low-lying minimum 
with $\rho^\prime = \tilde{\rho}$ if the noise strength $T$ is not large 
enough. When the time taken for jumping between minima exceeds the  
residence time in any minimum because of enhanced noise, we have 
a fluctuation induced C-I transition (Fig.~\ref{lang}) from that density $\rho$.
To show this we obtain the limiting value of $\rho^\prime$ averaged over 
realisations of the noise as a function of $\rho$ and $T$ from a 
numerical solution of Eqn. \ref{slang}. While it is 
difficult to make a direct quantitative connection between $T$ and $v_e$ 
symmetry considerations would require $T \propto v_e^2$.
In Fig.~\ref{lang} we have 
plotted $\tilde{\rho}(\rho,T)$ for $1/4 < \rho < 1/3$ showing prominent 
steps for $\tilde{\rho}$ corresponding to the rational fractions 
$1/4, 1/3$ and $2/7$. The steps for $2/7$ 
vanish at the square root of the noise amplitudes as $1/2$ 
which is as expected.

\section{Behavior at the transition point}

We want to determine scaling form for $w(t)$ at the transition 
point viz. the growth exponent $\beta$, the 
roughness exponent $\alpha$ and the dynamic exponent $z$. In the detached phase 
we know from renormalization group analysis that the 
exponents are in the KPZ universality class
\cite{Barabasi,KPZ} viz.
 $\beta = 1/3$, $\alpha = 1/2$  and $z = \alpha/\beta 
= 3/2$. To determine these exponents at the transition point we make use 
of {\sl Family-Vicsek} scaling relation \cite{Barabasi}

\begin{eqnarray}
w(L,t) \sim {\cal N}_s^{\alpha}f(t/{\cal N}_s^{z})
\end{eqnarray}

\begin{figure}[h]
\begin{center}
\includegraphics[width=9cm]{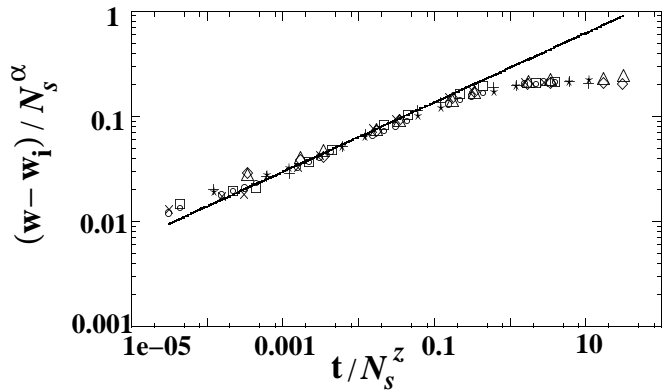}
\end{center}
\caption{Monte Carlo data $(w-w_i)/{\cal N}_s^{\alpha}$ vs $t/{\cal N}_s^{z}$ 
for $p = 1.0$, $p = 0.7$ and with ${\cal N}_s = 1000$ [{$\times,\bigcirc$}] ,
$800$ [{$\Box,\circ$}] , $400$ [{$+,\star$}] , $200$ [${\Diamond,\triangle}$]. 
All the curves collapse to a single universal function showing KPZ scaling.
}
\label{fig6}
\end{figure}

\noindent
Fig.~\ref{fig6} shows the variation of $t/{\cal N}_s^{z}$ with $w(L,t)/{\cal N}_s^{\alpha}$ for 
different $p$ and different system sizes ${\cal N}_s$. The curves collapse onto 
one curve once an intrinsic width $w_{i}$, arising from finite-size and 
crossover effects \cite{Barabasi}, is subtracted out. 
The exponents were found to be KPZ.

To understand why this happens we go back to our modified KPZ equation 
Eqn. 3.1 and make the transformation $h = h^{\prime} + v_{f}t$. We get

\begin{eqnarray}
\frac{\partial h^{\prime}}{\partial t} + v_{f} &=& \lambda_{1} 
\frac{\partial^{2} h^{\prime}}{\partial h^{2}} - 
\lambda_{2}\Big(\frac{\partial h^{\prime}}
{\partial h}\Big)^{2} \tanh\Big(\frac{(v_{f} - v_{e})t + h^{\prime}}{\chi}\Big) 
\nonumber\\
&&- \lambda_{3}\tanh \Big(\frac{(v_{f} - v_{e})t + h^{\prime}}{\chi}\Big)
+ \zeta^{\prime}(h^{\prime},t)
\end{eqnarray}

\noindent
Now substituting the mean field result for $v_f$ (Eqn.~\ref{mft-v}), making use 
of the fact that at the transition point $v_{f} = \lambda_{3}$ and simplifying
one can show that the above equation reduces to the familiar KPZ equation in 
$h^{\prime}$.

\section{Overview, caveats and conclusion}
\begin{figure}[h]
\begin{center}
\includegraphics[width=6.0cm]{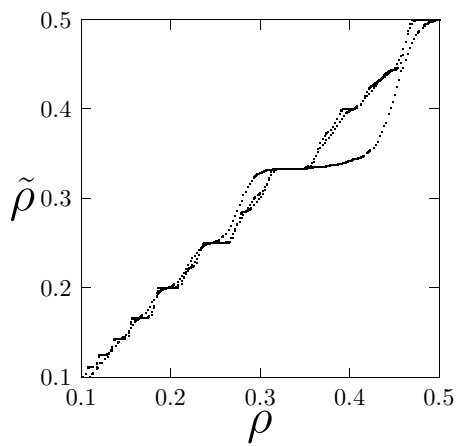}
\end{center}
\caption{
Devil's staircase structure for $v_e = 0.05$ and $1/\chi = 0.01, 1, 5$.  
}
\label{tanh}
\end{figure}
In this chapter we have shown how Ising interfaces in spatially varying,
dynamical fields can give rise to interesting and new phenomena like 
dynamical C-I transitions. We hope that our study may lead to experimental
work on real interfaces in carefully controlled and shaped external fields
which may be used to carve static or dynamical patterns with technological
applications. There are several problems which need to be sorted out before
these techniques begin to have any practical application.

Firstly, the sharpness of the field profile directly influences the 
``resolution'' of the patterns, Even at small velocities, as the width of 
the profile $\chi$ increases, the steps in the Devil's staircase
corresponding to particular patterns disappear (Fig.~\ref{tanh}) - simpler patterns
being more stable as expected. In order to obtain the more interesting
patterns therefore one needs the field profile to be atomistically sharp.
Secondly, the most useful interfaces are two dimensional and therefore
our studies need to be extended to higher dimensions and for other
underlying crystal structures. We believe that there are exciting 
possibilities to be explored in this direction. Lastly, Ising interfaces
miss two very important characteristics of real interfaces. (1) Elastic
distortions and the effects of interfacial stress and (2) the possibility
of particle transfers between the two phases. It is this last possibility
that we explore in the subsequent chapters where we look at a 
two dimensional liquid solid interface stabilized by a spatially varying 
chemical potential field.



\chapter{Atomic Systems}

\section{Introduction}
In the last two chapters we have concentrated on interfaces in the
two dimensional Ising model. Ising interfaces are particularly simple
because they represent spatial variation of only a single scalar order
parameter - the magnetization. Real interfaces in condensed matter
systems are much more complex. In a liquid - solid interface for example,
in principle, one has to consider the variation of an infinity of order
parameters corresponding to all the Fourier components of the non-uniform
solid density $\rho(r)$. We shall study liquid - solid interface in
Chapters 5 and 6. This will involve simulations of systems composed of
particles, or atoms, with interatomic potentials. Before we go on to describe 
our work on atomic systems, we use this chapter to introduce the model 
interatomic potentials we consider. We then briefly describe the 
computational techniques employed to compute a host of useful thermodynamic,
structural and dynamical quantities after defining each of them
appropriately. Wherever possible, we have compared our results with
those available in the literature. This constitutes a check on our methods
and a verification of the codes that we use later to perform simulations
of the liquid - solid interface.

\section{Models for atomistic systems}
Consider a system containing ${\cal N}$ atoms. We may divide the potential 
energy into terms depending on the coordinates of individual atoms, pairs, 
triplets, etc. as :
\begin{eqnarray}
{\cal V} = \sum_{i}u_{1}({\bf r}_{i}) + \sum_{i}\sum_{j>i}u_{2}
({\bf r}_{i},{\bf r}_{j}) + \sum_{i}\sum_{j>i}\sum_{k>j>i}u_{2}
({\bf r}_{i},{\bf r}_{j},{\bf r}_{k}) + ......
\label{atomic}
\end{eqnarray}
$\sum_{i}\sum_{j>i}$ notation indicates a summation over all distinct 
pairs $i$ and $j$ without counting any pair twice; similarly for triplets, etc.
The first term 
$\sum_{i}u_{1}(\bf r_{i})$ represents the effect of an external field on
the system. The remaining terms represent particle interactions. The second
term, $u_2$, the pair potential, depends on the magnitude of the pair 
separation $r_{ij} = |{\bf r}_{i} - {\bf r}_{j}|$, so it may be written 
$u_2(r_{ij})$. The triplet $u_3$ term is ignored in our thesis
The pairwise approximation gives a remarkably good description of most 
atomic systems because the average three-body effects can be usually included 
by defining an "effective" pair potential. To do this, we rewrite Eqn. 
(\ref{atomic}) in the form
\begin{eqnarray}
{\cal V} \approx \sum_{i}u_{1}({\bf r}_{i}) + \sum_{i}\sum_{j>i}u_2^{eff}
({\bf r}_{i},{\bf r}_{j})
\end{eqnarray}
For simplicity, we use the notation $u(r_{ij})$ or $u(r)$ for the pair
potential.

\subsubsection{Effective pair potentials}
In our simulations to model atomistic systems we have chosen different pair
potentials. We discuss them briefly with special emphasis on two-dimensional
results. 
\begin{itemize}
\item The hard core potential is defined by a pair interaction between two 
classical particles with a non-overlap condition. It is given as
\begin{eqnarray}
u^{HS}(r) &=& \infty \,\,\,(r < \sigma) \nonumber \\
&=& 0 \,\,\,\,\,\,(\sigma \le r)
\end{eqnarray}
where $\sigma$ is the diameter of the spheres and $r$ is the distance between
the two centers of the spheres. A peculiarity of the hard core potential is
that it sets a length scale, $\sigma$, but it does not set any energy scale.
A configuration of two overlapping spheres cost an infinite energy.
Since $u^{HS}$ can only take two values $0$ and $\infty$, therefore, for the 
Boltzmann factor, we get, $e^{-\beta u^{HS}} = e^{-u^{HS}}$, 
$\beta = 1/k_B{\cal T}$. Thus, the temperature scales out trivially. Or, in 
other words, hard 
objects are {\em athermal}; all their structural and thermodynamical 
properties do not depend on temperature so that the density $\rho = N/V$
(or the packing fraction $\eta$, which is the ratio of the volume of the 
$N$ spheres and the total accessible volume, $V$) is the only relevant
thermodynamical variable. Due to this temperature independence, the hard
sphere model is the simplest non-trivial model for an interaction. A lot of
results exist for the hard sphere model (see \cite{fun} and references therein)
which makes it useful as a reference 
system for systems with more complicated interactions and particle shapes. 
Moreover, the equilibrium thermodynamic properties of the hard sphere model 
can actually be probed in nature by examining suspensions of spherical 
sterically-stabilized colloidal particles \cite{fun}.

In the two-dimensional case of hard disks the close-packed area fraction
is $\eta_c \equiv \pi\rho_c\sigma^2/4 = \pi/2\sqrt{3} = 0.907...$ 
corresponding to a perfect triangular lattice with long-ranged translational
order. The existence of only two phases, fluid and solid, and a freezing
transition from one to the other as the packing fraction is changed has
been confirmed by extensive simulations of this model system \cite{2dmelting,
snb}. Hard disks
form a key system in our understanding of both equilibrium and dynamical 
aspects of a model liquid-solid interface. 

\item The soft sphere potential is defined as 
\begin{eqnarray}
u^{SS}(r) = \epsilon\left(\frac{\sigma}{r}\right)^{\nu} = ar^{-\nu}
\end{eqnarray}
where $\nu$ is a parameter, often chosen to be an integer. $\sigma$ and
$\epsilon$ have the dimensions of length and energy. The soft-sphere
potential becomes progressively "harder" as $\nu$ is increased. Soft sphere
potentials contain no attractive part. This retains some of the simplicity
of hard spheres in the sense that there is again one thermodynamic variable.
To understand this, note that we can eliminate the explicit temperature
dependence by defining $\sigma* = (\beta\epsilon)^{1/\nu}\sigma$ and
rewriting Eqn. 4.4 as $\beta u^{SS}(r) = ({\sigma*}/r)^{\nu}$. The 
dimensionless excess (over the ideal gas) thermodynamic properties then depend 
only on the single dimensionless state variable
\begin{eqnarray}
\rho* \equiv \rho{\sigma*}^2 = \rho\sigma^2(\beta\epsilon)^{2/\nu}
\end{eqnarray}
Here again we have only two phases but unlike hard systems, soft sphere 
potential is analytic everywhere. In our simulation we use $\nu = 12$. For the 
two dimensional system of soft disks (with $\nu = 12$), a lot of simulation 
results exist, including the phase diagram
and an empirical equation of state \cite{bgw}.

\item The Lennard-Jones 12-6 potential :
\begin{eqnarray}
u^{LJ}(r) = 4\epsilon\left[\left(\frac{\sigma}{r}\right)^{12} - 
\left(\frac{\sigma}{r}\right)^{6}\right]
\end{eqnarray}
This potential has a long-range attractive tail of the form $- 1/r^6$, a 
negative well of depth $\epsilon$, and a steeply rising repulsive wall
at distances less than $r \sim \sigma$. The Lennard-Jones potential is
extensively used to model atomistic systems and indeed provides a fair
description of the interaction between pairs of rare-gas atoms \cite{Hansen}.  
\end{itemize}

\section{Methods for simulations}
We shall now introduce the various simulation methods that are generally 
used in simulating atomic systems and in particular those that we have
used in this thesis to understand the behavior of liquid - solid interfaces.
We begin by briefly discussing the generic Monte Carlo and molecular dynamics
methods. Special emphasis has been given to explain the simulation of hard 
systems which require some care during simulations to handle the 
associated non analyticities. We include this section in this thesis mainly for
completeness. There exist, of course, excellent textbooks \cite{daan,allen,Binder1,Binder2}
which explains these techniques in great detail.

\subsection{The Monte Carlo method}

Consider the ${\cal N}$ particle system. The classical expression for the 
thermal average of an observable ${\cal A}$ is given by
\begin{eqnarray}
\langle{\cal A}\rangle = \frac{\int{d{\bf p}^{N}d{\bf r}^{N}
{\cal A({\bf r}^{N},{\bf p}^{N})} \exp[-\beta {\cal H}({\bf r}^{N},
{\bf p}^{N})]}}{\int{d{\bf p}^{N}d{\bf r}^{N}
\exp[-\beta {\cal H}({\bf r}^{N},{\bf p}^{N})]}}
\end{eqnarray}
where ${\bf r}^{N}$ stands for the coordinates of all N particles, and 
${\bf p}^{N}$ for the corresponding momenta and $\beta = 1/k_{B}T$. 
${\cal H}({\bf r}^{N},{\bf p}^{N})$ is the Hamiltonian of the system.
As ${\cal H}$ is a quadratic function of the momenta, the integration over 
momenta can be carried out analytically. The integral over particle
coordinates is carried out numerically by the Monte Carlo method or, more
precisely, the Monte Carlo importance-sampling algorithm introduced in 1953
by Metropolis {\em et. al.}.

\subsubsection{A basic Monte Carlo algorithm}

In the approach introduced by Metropolis {\em et. al.} the following 
scheme is proposed : 
\begin{itemize}
\item Select a particle at random, and calculate its energy 
${\cal U}({\bf r}^N)$.
\item Give the particle a random displacement; $r^{\prime} = r + \Delta$, 
and calculate its new energy ${\cal U}({{\bf r}^\prime}^N)$
\item Accept the move from ${{\bf r}}^N$ to ${{\bf r}^\prime}^N$ with 
probability
\begin{eqnarray}
acc(o \rightarrow n) = 
min(1,\exp\{-\beta[{\cal U}(n) - {\cal U}(o)]\})
\end{eqnarray}
In order to decide whether to accept or reject the trial move, we generate 
a random number, denoted by  Ranf, from a uniform distribution in the 
interval [0,1]. The probability that Ranf is less than 
$acc(o \rightarrow n)$ is equal to $acc(o \rightarrow n)$. Therefore, we accept
the trial move if ${\rm Ranf} < acc(o \rightarrow n)$ and reject it otherwise.
This rule guarantees that the probability to accept a trial move from 
o to n is indeed equal to $acc(o \rightarrow n)$.
\end{itemize}

Using the Monte Carlo method it is particularly easy to to simulate a system
of particles interacting via hard potentials. The same Metropolis procedure 
is used, except that, in this case, the overlap of two spheres results in an 
infinite positive energy change and 
$\exp\{-\beta[{\cal U}(n) - {\cal U}(o)]\} = 0$.
Thus we immediately reject all trial moves involving an overlap since 
$\exp\{-\beta[{\cal U}(n) - {\cal U}(o)]\} < {\rm Ranf}$. Equally all moves 
that do not involve overlap are immediately accepted. As in soft potentials, 
in the case a move is rejected, the old configuration is recounted in the 
average.

\subsection{The Molecular Dynamics method}

As described in the last section, in a  Monte Carlo simulation we compute the 
average behavior of a system in a purely static sense : {\em ensemble} average.
In most experiments, however, we perform a series of measurements during
a certain time interval and then determine the average of these 
measurements. The basic idea behind molecular dynamics simulations is 
precisely that we can study the average behavior of system simply by
computing the natural time evolution of that system numerically. After that
we average the quantity of interest over a sufficiently long time. The time 
averaged value of some observable ${\cal A}$ is therefore given by
\begin{eqnarray}
\overline{A(r)} = \lim_{t\rightarrow\infty}\frac{1}{t}\int_0^t 
dt^{\prime}A(r;t^{\prime})
\end{eqnarray}
``Ergodic hypothesis'' states that, if we wish to compute
the average of a function of the coordinates and momenta of a many-particle
system, we can {\em either} compute the quantity by time averaging (the ``MD''
approach) {\em or} by ensemble averaging (the ``MC'' approach). The 
molecular dynamics approach is also useful if we want to look at systems
which are driven far from equilibrium where it is hard to define an
``ensemble'' average. We shall use the MD approach for our studies of the 
dynamics of liquids and the liquid solid interface in Chapter 6.

\subsubsection{A basic molecular dynamics algorithm}
\begin{itemize}
\item We read in the parameters that specify the conditions of the run
(e.g., initial temperature, number of particles, density, time step).
\item We initialize the system (i.e., we select initial positions and 
velocities).
\item We compute the forces on all particles.
\item We integrate Newton's equations of motion. The step and the previous
one make up the core of the simulation. They are repeated until we have
computed the time evolution of the system for the desired length of time.
To integrate the equations of motion we use the ``velocity Verlet'' 
algorithm which is implemented in two stages. Firstly, the
new positions at time $t+\Delta t$ are calculated using the equation, 
\begin{eqnarray}
r(t+\Delta t) = r(t) + v(t)\Delta t + \frac{f(t)}{2m}\Delta t^2
\end{eqnarray}
and the velocities at mid step are computed using 
\begin{eqnarray}
v(t+\frac{1}{2}\Delta t) = v(t) + \frac{1}{2m}f(t)\Delta t
\end{eqnarray}
The forces and accelerations at time $t+\Delta t$ are then computed, and 
the velocity move completed
\begin{eqnarray}
v(t+\Delta t) = v(t+\frac{1}{2}\Delta t) + \frac{1}{2m}f(t+\Delta t)\Delta t
\end{eqnarray}
At this point, the kinetic energy at time $t+\Delta t$ is available. The
potential energy at this time will have been evaluated in the force loop.
Note that if we use the kinetic energy per particle as a measure of the
instantaneous temperature, then we would find that, in the canonical
ensemble, this temperature fluctuates. To avoid this, we use either
simple temperature rescaling or proper heat baths \cite{daan}.
\item After completion of the central loop, we compute and print the 
averages of measured quantities, and stop.
\end{itemize}

If we have a system of particles interacting via hard potentials then 
we have to take a different approach. Whenever two such particles come
close enough to reach a point of discontinuity in the potential, then
there is a ``collision''. Therefore, in such a system, the primary aim
is to calculate the time, collision partners, and all impact parameters, 
for every collision occurring in the system. Unlike soft potentials, 
hard potential programs evolve on a collision-by-collision basis.
The general scheme is as follows :
\begin{itemize}
\item locate next collision
\item move all particles forward until collision occurs
\item implement collision dynamics for the colliding pair
\item calculate any properties of interest, ready for averaging, before 
returning to the first step.
\end{itemize}
Although the above scheme is the basis for simulating hard core systems,
the introduction of an external potential renders the simulation technique
difficult to implement. This is because in the presence of an external potential
the particles will experience a force ${\bf f_{ij}}$ and the collision equation 
that we need to solve is given by
\begin{eqnarray}
|{\bf r}_{ij}(t + t_{ij})| = |{\bf r}_{ij} + {\bf v}_{ij}t_{ij} + 
(1/2){\bf f}_{ij}t_{ij}^2| = \sigma
\end{eqnarray}
where ${\bf r}_{ij} = {\bf r}_i - {\bf r}_j$ and ${\bf v}_{ij} = {\bf v}_i - 
{\bf v}_j$ and $\sigma$ is the hard core diameter.
Therefore we now have a quartic equation in $t_{ij}$ and it is highly non-
trivial and time consuming to find the solution. Therefore, to simulate a hard 
disk system in the presence of an external potential, we use multiple time step 
molecular dynamics that we use for soft potentials. However, in the 
position and velocity update scheme, we introduce the additional restriction
that if two particles overlap, then they are retained in their old positions
but their velocities are interchanged. This of course introduces an error in
the simulation procedure but we choose a sufficiently small time scale 
and benchmark our simulation against standard results for the velocity of
sound in hard disk liquid. In our simulations, the unit of time, $t$, is 
$\tau = \sqrt{m \sigma^2/k_B T}$, where $m (= 1)$ is the mass of the hard 
disks. We find that a time step of $\Delta t = 10^{-4}$
conserves the total energy to within $1$ in $10^3$ (at worst) - $10^6$.

\section{Bulk variables}
From computer simulations, we get information at the microscopic level - 
atomic and molecular positions, velocities, etc. With the help of statistical
mechanics we convert this very detailed information into macroscopic
terms such as pressure, internal energy, etc. In this chapter, we focus on 
the determination of bulk quantities which are averaged over the
entire system and are meaningful only for homogeneous phases. The analogous
space dependent local quantities at interfaces are described and discussed
in the next chapter. 
\subsection{Thermodynamic variables}
\begin{figure}[h]
\begin{center}
\includegraphics[width=9.0cm]{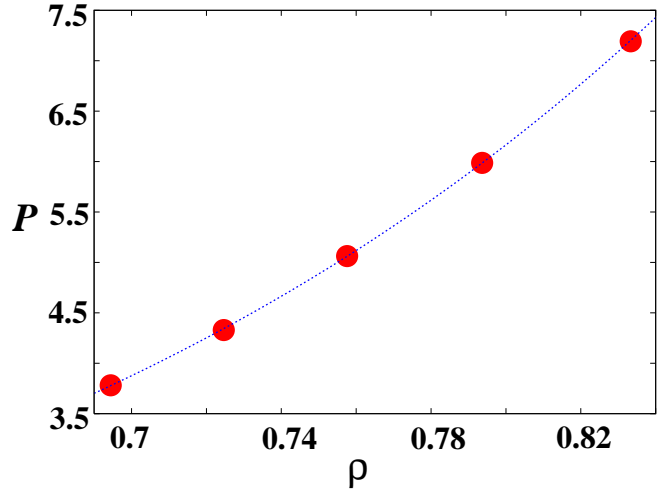}
\end{center}
\caption{The equation of state of the soft disk liquid. Symbols are our
results, line from \cite{bgw}} 
\label{sdeos}
\end{figure}

\begin{figure}[h]
\begin{center}
\includegraphics[width=9.0cm]{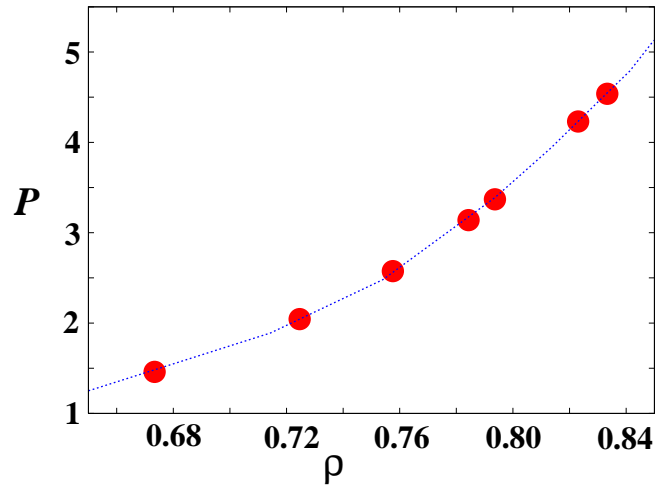}
\end{center}
\caption{The equation of state of the Lennard-Jones disk liquid. Symbols are 
our results, line from \cite{toxvaerd}} 
\label{ljeos}
\end{figure}

Bulk thermodynamic variables such as density, pressure, stress etc. can
be easily calculated. For a system of ${\cal N}$ particles in a volume 
${\cal V}$, in a homogeneous phase the average density is given simply as 
$\rho = {\cal{N/V}}$. 
The pressure may be calculated from the equation ($\langle \cdot \rangle$ 
denotes the ensemble average)
\begin{eqnarray}
{\cal{PV}} = {\cal N}k_B\cal{T} + \langle\cal{W}\rangle
\end{eqnarray}
where ${\cal W}$ is the `internal virial' given as
\begin{eqnarray}
{\cal W} = \frac{1}{2}\sum_{i = 1}^{\cal N} {\bf r}_i\cdot {\bf f}_i
\end{eqnarray}
in two-dimensions.
${\bf f}_i$ is the force on some particle $i$. For pairwise interactions,
$\sum_i{\bf r}_i\cdot {\bf f}_i = \sum_i\sum_{j>i}{\bf r}_{ij}\cdot {\bf f}_i$
where $r_{ij} = r_i - r_j$. We can thus calculate the pressure for various
densities. In Figs.~\ref{sdeos} and~\ref{ljeos} we plot the equation of 
states for the soft core and Lennard-Jones liquid in two-dimensions. In both 
cases,
we perform Monte-Carlo simulations in the constant ${\cal NAT}$ ensemble with 
${\cal N} = 1200$ and ${\cal T} = 1$. The potentials are truncated at
$r_c = 2.5$. In our simulations we have used reduced units ($\sigma = \epsilon
= m = 1$). The time step of $\delta t = 10^{-3}$ conserves the energy.
To compare our simulation results for the soft disk liquid we use the empirical
equation of state \cite{bgw} given as
\begin{eqnarray}
\nonumber
\frac{\beta \Delta {\cal P}}{\rho} = 1.77306\rho* + 2.36241{\rho*}^2
+ 1.798198\rho*^3 - 5.648177\rho*^4 \\  
+ 78.65712\rho*^5 - 197.57241\rho*^6 + 212.37417\rho*^7 - 
79.57456\rho*^8
\end{eqnarray}
For the hard disk system the forces are impulsive and one cannot directly
use Eqn. 4.15. We however make use of the pair 
distribution function to calculate the pressure. The pair distribution
function is defined in the next section. Thus, the equation of state for the 
hard disk 
system is \cite{hoover} 
\begin{eqnarray}
Z \equiv \frac{\beta {\cal P}}{\rho} = 1 + 2\eta g(\sigma)
\end{eqnarray}
where $g(\sigma)$ is the radial distribution function at $r = \sigma$, the 
hard disk diameter. We again perform MC simulation in the canonical ensemble
with the same number of particles and at the same temperature. We average over 
$10^4$ configurations to obtain the radial distribution functions for various 
$\eta$'s and from them derive the corresponding $g(\sigma)$'s.
\begin{figure}[h]
\begin{center}
\includegraphics[width=9.0cm]{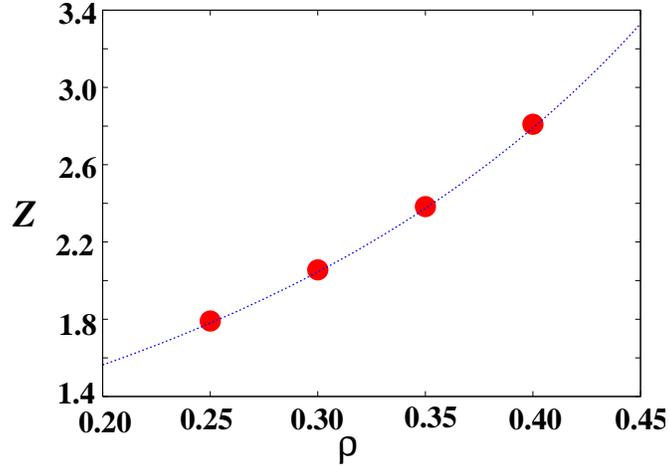}
\end{center}
\caption{The equation of state of the hard disk liquid. Symbols are our
results, line from \cite{hdsound}} 
\label{}
\end{figure}
We compare our results with the semi-empirical equation of state given 
as \cite{santos},
\begin{eqnarray}
Z = \left(1 - 2\eta + \frac{2\eta_{c} - 1}{\eta_{c}^2} 
\eta^2\right)^{-1}
\label{hdsan}
\end{eqnarray}

\subsection{Structural quantities}
One important function that characterizes the structure of a homogeneous 
phase is the pair distribution function $g_2({\bf r}_i,{\bf r}_j)$, or
$g_2(r_{ij})$ or simply $g(r)$. It is defined as the probability of
finding a pair of atoms a distance $r$ apart, relative to the probability
expected for a completely random distribution at the same density. A form
for $g(r)$ useful for computer simulation purpose is given as
\begin{eqnarray}
g(r) = \rho^{-2}\langle\sum_i\sum_{j\ne i}\delta({\bf r}_i)\delta({\bf r}_j
- {\bf r})\rangle = \frac{{\cal V}}{{\cal N}^2}\langle\sum_i\sum_{j\ne i}
\delta({\bf r}_j - {\bf r})\rangle
\end{eqnarray}
In simulations, the delta function is replaced by a function which is 
non-zero in a small range of separations, and a histogram is compiled of
all pair separations falling within each such range. 
The pair distribution function is extremely important because we may also
express thermodynamic quantities like the energy and the pressure in terms
of this function.

One other quantity that is very useful in characterizing the structure of 
a phase is the structure factor $S(k)$ where $k$ is the wavevector. In a
simulation with periodic boundaries, $k$ is restricted by the periodicity
of the system, i.e. with the simulation box. The structure factor describes 
Fourier components of density fluctuations as
\begin{eqnarray}
S(k) = {\cal N}^{-1}\langle \rho(k)\rho(-k) \rangle
\end{eqnarray}
where $\rho(k) = \sum_{i=1}^{\cal N}\exp(i{\bf k}\cdot{\bf r}_i)$.
The structure factor is essentially the Fourier transform of the pair 
distribution function and is directly measurable in X-ray or neutron
diffraction experiments \cite{Chaikin}.

Apart from these two quantities we may also define the order parameters that
distinguish two phases. Here we discuss two such quantities \cite{snb} 
that would be 
useful to characterize the fluid and solid phases in our analysis of the
liquid-solid interface of a two-dimensional system in Chapter 5. 

\noindent
(i) {\bf Bond orientational order parameter} : In a periodic crystal, there
is only a discrete set of directions defined by vectors between nearest 
neighbour particles, which occupy sites on a lattice. These directions are 
the same throughout the lattice and define a long-range orientational order
often called the {\em bond orientational} order or the {\em bond-angle} order.
The orientational order of a two-dimensional hard disk system can be 
described by the (global) bond orientational order parameter 
$\langle \psi_6 \rangle$ where as usual $\langle \cdot \rangle$ denotes 
ensemble average. The local value of $\psi_6$ for a particle $i$ 
located at ${\bf r}_i = (x,y)$ is given by
\begin{eqnarray}
\psi_{6,i} = \frac{1}{N_i}\sum_j \exp(6i\theta_{ij})
\end{eqnarray}
where the sum on $j$ is over the $N_i$ neighbours of this particle, and 
$\theta_{ij}$ is the angle between the particles $i$ and $j$ and an arbitrary
but fixed reference axis.
The (global) bond orientational order parameter is then defined as 
\begin{eqnarray}
\langle \psi_{6,i} \rangle = \langle |\frac{1}{N}\sum_{i=1}^N \psi_{6,i}| 
\rangle
\end{eqnarray}
where $N$ is the particle number of the system. For a perfect triangular
structure $\psi_6 = 1$. Thus a solid or a hexatic phase will have 
$\psi_6 \ne 0$. However for the disordered liquid phase $\psi_6$ will average
to zero for $N \rightarrow \infty$.

\noindent
(ii) {\bf Solid order parameter} : In an external field however the bond
orientational order may be non-zero even in the fluid phase. This is because,
we can now have a ``modulated liquid'' in which the local hexagons consisting
of the six nearest neighbours of a particle are automatically oriented by the
external field for example, near the liquid solid interface. Thus $\psi_6$ is 
non-zero both in the (modulated) liquid and the crystalline phase and  
cannot be used to study the freezing transition or liquid solid 
interface. The order parameters corresponding to the solid phase 
are the Fourier components of the (nonuniform) density-density correlation 
$\langle \rho({\bf r}_i)\rho({\bf r}_j) \rangle$ calculated at the
reciprocal lattice points $\{{\bf G}\}$. This (infinite) set of numbers are all
zero (for ${\bf G} \ne 0$) in a uniform liquid phase and nonzero in a solid.
We restrict ourselves to the star consisting of the six smallest reciprocal
lattice vectors of the two dimensional triangular lattice. In modulated 
liquid phase, the Fourier components corresponding to two out of these six
vectors, eg., those in the direction perpendicular to the interface,
${\bf G}_1$, are nonzero. The other four 
components of this set which are equivalent by symmetry (${\bf G}_2$)
are zero in the (modulated) liquid and
nonzero in the solid (if there is true long ranged order). Thus, we use the
following order parameter :
\begin{eqnarray}
\langle \psi_{{\bf G}_k} \rangle = \langle |\sum_{i,j=1}^N \exp(-i {\bf G}_k 
\cdot {\bf r}_{ij})|\rangle
\end{eqnarray}
where ${\bf r}_{ij} = {\bf r}_i - {\bf r}_j$. Note that though 
the order parameter $\langle \psi_{G_2} \rangle$ decays to zero with 
increasing system size in the two-dimensional solid - quasi long ranged order -
this decay, being weak, does not hinder us from distinguishing, in a finite
system, a modulated liquid from the solid phase with positional order. 

\subsection{Dynamic quantities}
The important dynamical quantities for our purposes are those which 
characterize the transfer of mass, momentum and energy across a liquid solid
interface. In Chapter 6, we shall study the speed of sound, the sound
absorption coefficient and the thermal conductivity of an inhomogeneous 
system consisting of solid and liquid regions. In this section, we shall
show how non-equilibrium molecular dynamics simulations may be used to obtain
the sound velocity and the absorption coefficient for a homogeneous hard
disk liquid. Consider a rectangular simulation box of length $L_x$ and width
$L_y$. Sound propagation is studied by producing a momentum ($v_y = V_0$)
impulse over a thin rectangular region in $y$ spanning the
width $L_x$ of the cell. This generates a weak, acoustic, shock \cite{lanlif2}. 
The shock rapidly evolves into a Gaussian pulse which propagates
non-linearly with a speed $c_{pulse} \ge c_0(\rho)$, the speed of sound
in the liquid which is a function of the density $\rho$ of the hard disk 
liquid. 
\begin{figure}[h]
\begin{center}
\includegraphics[width=6.0cm]{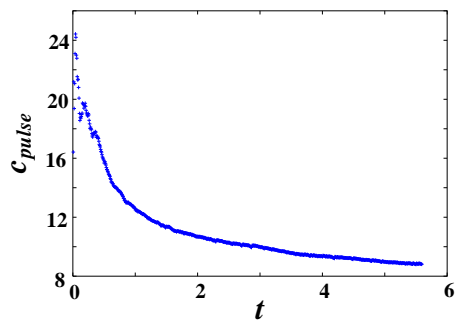}
\end{center}
\caption{Speed of propagation of a momentum pulse in the hard disk liquid
as a function of time} 
\label{pulse}
\end{figure}
The deceleration of the pulse follows a pattern which is strongly reminiscent
of the propagation of acoustic shock in water after the collapse of a
sonoluminiscent bubble. If we fit the time dependent pulse velocity(Fig.
~\ref{pulse}) to a form $c_{pulse}(t) = c_0(\rho) + c_1/t^{\nu}$, we obtain 
the limiting velocity $c_0(\rho)$ with $\nu$ close to $0.5$. We now compare the
speed of sound $c_0(\rho)$  with the equation for the speed of
sound in a hard disk gas given by \cite{hdsound}
\begin{eqnarray}
c_0 = \sqrt{\frac{2k_BT}{m}\frac{\partial}{\partial \eta}[\eta Z(\eta)]}
\end{eqnarray}
where for $Z(\eta)$ we use the familiar equation of state Eqn. 4.18. We find
that the speed of sound measured in this way compares remarkably well with 
the speed of sound in the hard disk liquid (Fig.~\ref{sos}). The calculated 
$c_0(\rho)$ is of course independent of the initial strength $V_0$ of the shock.
\begin{figure}[h]
\begin{center}
\includegraphics[width=6.0cm]{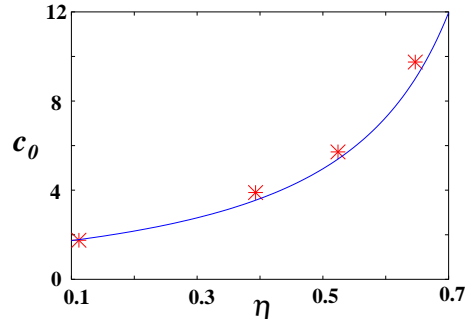}
\end{center}
\caption{Extrapolated limiting pulse speed $c_0$ as a function of the
packing fraction, $\eta$. The line is a plot of the speed of sound obtained 
from the equation of state Eqn. 4.24} 
\label{sos}
\end{figure}
Viscous dissipation spreads the pulse into a smooth Gaussian with a width
$\Delta^2$ which, at late times, increases linearly with distance $s$, and 
the sound absorption \cite{lanlif2} coefficient $\alpha = \Delta^2/4 c_0^2 s$ 
is also independent of $V_0$ (Fig.~\ref{del}).
\begin{figure}[h]
\begin{center}
\includegraphics[width=6.0cm]{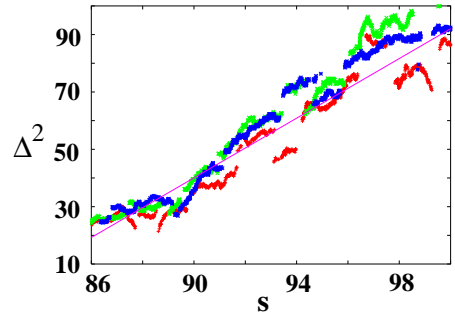}
\end{center}
\caption{The absorption coefficient $\Delta^2$ showing the expected linear
dependence with distance of propagation $s$. The different coloured 
symbols represent different values of the initial momentum shock $V_0$
red : ...; green : ...., etc.} 
\label{del}
\end{figure}
The $q^{th}$ Fourier component of the velocity therefore 
decays with $s$ as $v_q(s) = v_q(0) \exp(-\alpha c_0^2 q^2 s)$.
In order to obtain a 
measurable Gaussian we average all quantities over $\sim 200$ initial 
velocity configurations chosen from a Maxwellian 
distribution with temperature $k_B T = 1$.

\chapter{Liquid Solid interfaces : Equilibrium properties}

\section{Introduction and background}
In this chapter we shall use the ideas and methods discussed in the
last chapter to study interfaces between a liquid and a solid in two
dimensions.  We shall show that such interfaces may be constructed
using external chemical potential fields similar to the simple Ising
interface studied in Chapters 2 and 3. Small wavenumber fluctuations,
known as crystallization waves, which are dominant in field free
liquid solid interfaces are, of course suppressed for reasons similar
to the Ising case viz. the explicit breaking of translational
invariance of the system perpendicular to the interface. Our
interfaces therefore remain planar for the parameters used in our
study. However, as we shall see shortly, the planar liquid solid
interface is not inert. Particles are transferred across the interface
in new and interesting ways. We believe that some of our predictions
may be directly checked for liquid solid interfaces in atomic, as well
as, colloidal systems where the chemical potential field may be
provided either by a laser trap or by adsorbing colloidal particles on
a patterned substrate. In the next chapter we shall discover how these
interfacial fluctuations influence transport properties of the
interface contributing to the resistance offered to the transfer of
momentum impulses and of heat.

A liquid is homogeneous and isotropic. The local density $\rho({\bf
r})$ is independent of the spatial coordinate ${\bf r}$ and has
Fourier components which are nonzero only for the $q=0$ mode. This
translational invariance is partially broken in the solid which has
Fourier components $\rho_{\bf G}$ which are nonzero for all
wavevectors ${\bf G}$ which are members of the reciprocal lattice
vector set of the solid. The solid can therefore be described
completely using the countably infinite number of Fourier components
$\rho_{\bf G}$. Across a liquid solid interface all the order
parameters are expected to go from zero to their value in a solid
(close to unity) within a region of width $\xi$, the interfacial
thickness. While, the structure of real liquid-solid interfaces is
difficult to determine experimentally in molecular detail, the
interface is expected to be sharp with the order parameters building
up within a few atomic layers. This is simply a reflection of the fact
that the freezing transition is sharply first order in three
dimensions \cite{RY} with a finite correlation
length. Also most liquid solid interfaces are not planar. The
finiteness of heat and mass transport coefficients at the interface
generates instabilities, first analyzed by Mullins and Sekerka
\cite{mullins,godreche}, which grow into non-compact
dendritic structures typically observed in most real situations. There
has, however, been a number of theoretical and computer simulation
studies starting from the pioneering work of Haymet and Oxtoby
\cite{haymet1,haymet2}.  Most
of these studies refer to a planar interface.

In the next section we shall introduce the local analogs of the
thermodynamic and structural quantities introduced in the earlier
chapter. All of these quantities will have different values on either
side of a liquid solid interface and will be used to define the
interface. Next we shall show how non uniform chemical potential
fields may be used to generate liquid solid interfaces and shall try
to make contact with experiments by calculating approximately the
parameters required for constructing a suitable laser trap. We shall
plot local thermodynamic and orderparameters to characterize the
interface.  Our main result of this chapter, namely the existence of
complete layer transfers will then be presented for three different 
systems viz. the hard disk, soft disk and Lennard-Jones system. We shall 
attempt to understand our result from thermodynamic arguments, for
the simplest of the three systems, namely hard disks.

\section{The liquid solid interface}
\begin{figure}[h]
\begin{center}
\includegraphics[width=7.0cm]{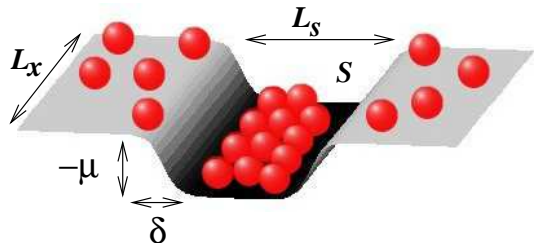}
\end{center}
\caption{A cartoon diagram of the system} 
\label{tanhp}
\end{figure}

In this section, we explore the possibility of creating a patterned sequence 
of confined solid and liquid regions using an external, space-dependent, 
chemical potential field $\phi({\bf r})$. 
Consider a two dimensional system (see Fig ~\ref{tanhp}) of $N$ atoms
of average density (packing fraction) $\eta = \pi N/4 A$ within
a rectangular cell of size $A = L_x \times L_y$ where the central region 
${\cal S}$ of area $A_s = L_x \times L_s$ is occupied by $N_s$
atoms arranged as a crystalline solid of density $\eta_s > \eta$, 
while the rest of the cell is filled with liquid of density
$\eta_l < \eta$. 
The difference in density is produced by an external 
field $\phi({\bf r}) = -\mu$ for ${\bf r} \in {\cal S}$; increasing sharply 
but smoothly 
to zero elsewhere with a hyperbolic tangent profile of width $\delta_{\phi}$. 
How may $\phi({\bf r})$ be realized in practice? In model solids like 
colloids \cite{hamley}, one may use a surface template to create a static 
pattern \cite{AMOLF}. In real systems, as well as colloids \cite{baumgartl}, 
one may be able to use laser traps \cite{phillips} or non~-uniform electric 
or magnetic fields. Usual laser traps for alkali metals or rare gas atoms are 
in the range of $10mK$ for which a power of about $100mW$ is required.
 
For specificity and simplicity we first choose the atoms to interact with a   
{\em hard disk} potential \cite{jaster} although we show towards the end of this 
chapter that qualitative results for more realistic potentials, e.g.,
soft sphere or Lennard-Jones are the same.
We have chosen $\delta_{\phi} = \sigma/4$, where $\sigma$ is the hard disk
diameter and sets the scale of length. 
The energy scale for this system is set by $k_B {\cal T}$ where $k_B$ is the 
Boltzmann constant and ${\cal T}$ the temperature. In our simulation we
set $\sigma = k_B{\cal T} = 1$. Most of the qualitative phenomena 
discussed subsequently are independent of the exact nature of the interactions.

The full configuration dependent Hamiltonian is 
${\cal H} = \sum_{ij} V_{ij} + \sum_{i}\phi({\bf r}_i)$. We have carried 
out extensive MC simulations with usual Metropolis moves \cite{daan}, 
periodic boundary conditions in both directions and in the constant 
number, area and temperature ensemble to obtain the equilibrium behaviour of 
this system for different $\mu$ at fixed $\eta$. 
$N=1200$ particles occupy an area $A = 22.78 \times 59.18$ with the 
solid occupying the central third of the cell of size $L_s = 19.73$
The initial configuration is chosen to be a liquid with 
$\eta = .699$; close to but slightly lower than the 
freezing \cite{jaster} density 
$\eta_f = .706$. On equilibration, ${\cal S}$ contains a solid 
 with the closest~-packed planes  parallel to the solid~-liquid interfaces 
which lie, at all times,
along the lines where $\phi(y) \to 0$. The equilibration time 
is large and many ($\sim 10^7$) Monte Carlo steps (MCS) are discarded
before results shown in Figs.~\ref{dens} and ~\ref{system} are obtained. 

\section{Local averages}
Since in this chapter we study a liquid solid interface, thermodynamic 
and structural quantities which are averaged over the entire system makes
no sense. We have to, therefore, construct local analogs of these quantities 
as discussed in Chapter 4. To this purpose, we divide the simulation box
into thin rectangular strips with the longer dimension parallel to the 
interface. Each of these strips are indexed by an integer $i$ which runs from
$1$ to $M$, the total number of strips. We shall see that the external
potential induces a solid which has a layered structure. Therefore, it is
important that while breaking the system up into strips of equal width,
the edge of the strips in the solid region always falls between two layers.
The index $i$ multiplied by the width of the strip gives the distance in the 
direction perpendicular to the interface.

The thermodynamic quantities may always be defined as intensive variables 
containing sums over all the particles of the system divided by the total 
number of particles. In this case the sums run only over the particles in
a strip. Thus for example,
\begin{eqnarray}
\rho_i = \frac{Number\,\,of\,\,particles\,\,in\,\,strip\,\,i}{Strip\,\,width}
\end{eqnarray}
where $\rho_i$ is the local density in strip i, and similarly for other
quantities.

Structural quantities are typically sums over pairs of particles. Here,
we take one of the particles within the strip while its companion lies
either in the same strip or in neighbouring strips.  
The orientational order parameter is calculated in
each strip and is averaged over equilibrium configurations. To calculate 
the local solid order however, we choose area within the solid region 
and count the number of particles in this area. We then find out the solid 
order parameter for these particles as a function of the length perpendicular
to the interface. We use the same trick for the liquid region and finally
merge the two at the interface to get the variation of the solid order
parameter for the entire system in the direction perpendicular to the 
interface. Just as for the orientational order parameter, averaging is done
over equilibrium configurations. 

In all these definitions, we have of course assumed that there is no systematic
variation of any of the quantities in the direction parallel to the interface,
in accordance with the symmetry of the problem.

\section{Order parameter profiles and structure factor}
\begin{figure}[t]
\begin{center}
\includegraphics[width=7.0cm]{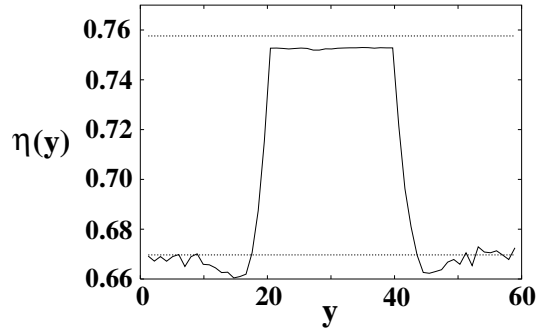}
\end{center}
\caption{The density
profile $\eta(y)$ coarse grained over strips of width $\sigma$
(averages taken over $10^3$ MC configurations each separated by $10^3$
MCS), varies from $ \eta_{\ell}$ to $\eta_s$ as we move into ${\cal
S}$. The lines show the predictions from a simple thermodynamic theory
presented in Section 5.5.1} 
\label{dens}
\end{figure}

\begin{figure}[h]
\begin{center}
\includegraphics[width=10.0cm]{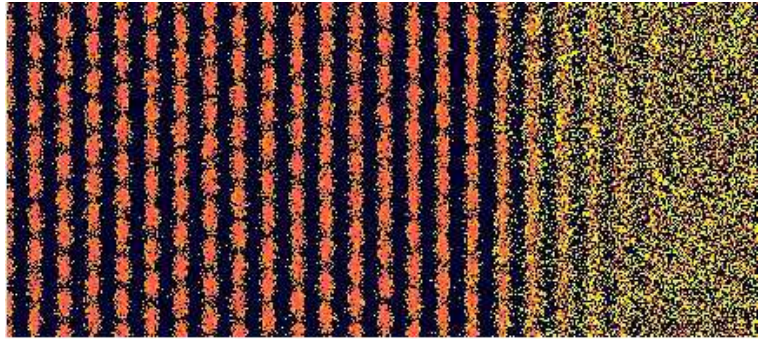}
\end{center}
\caption{Superposition of particle configurations from the MC run showing a 
solid like order (red : high $\eta$) gradually vanishing into the fluid 
(yellow : low $\eta$) across a well defined solid-fluid interface.} 
\label{system}
\end{figure}

In this section we present all results for the local thermodynamic and
structural quantities calculated as above.
The density $\eta(y)$ coarse grained over strips of width $\sim \sigma$ 
varies from its value $\eta_l$ in the liquid 
to $\eta_s$ as we move into (and away from) ${\cal S}$ (Fig. ~\ref{dens}). 
Averages taken over $10^3$ MC configurations each separated by $10^3$ MC steps.
The trap depth $\mu = 6$, supports an equilibrium solid of density 
$\eta_s = .753$ in contact with a fluid of density $\eta_{\ell} = .672$. 
The horizontal lines are predictions of a simple free-volume based theory 
(discussed in later sections) for $\eta_s$ and $\eta_{\ell}$.
A superposition of atomic positions 
shows a static, flat, liquid solid interface with the 
solid like order gradually vanishing into the liquid (Fig. ~\ref{system}).
We have thus created a thin nano-sized crystal which is about $21$ atomic
layers wide (for a trap depth, $\mu = 6$) and is flanked on either side by 
liquid separated by two liquid solid interfaces.

The bond orientational order parameter $\langle \psi_6(y) \rangle$ coarse 
grained over strips 
of width $\sigma$ (averages taken over $10^3$ MC configurations each separated
by $10^3$ MCS) shows a sharp rise from zero to a value close to one, as we 
move into the region $\cal S$.
\begin{figure}[b]
\begin{center}
\includegraphics[width=7.0cm]{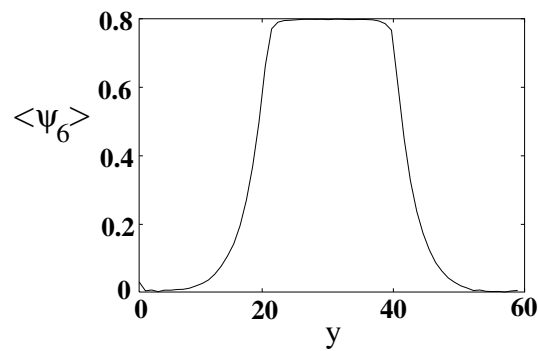}
\end{center}
\caption{Bond orientational order parameter across the liquid solid 
interface for a $21$ layered solid surrounded by liquid on both sides.} 
\label{}
\end{figure}
This indicates that the particles in this region are arranged approximately
hexagonally. However, this does not necessarily justify the phase to be 
a solid. Therefore, the need to calculate the solid order parameter. The
solid order parameter $\langle \psi_{\bf G} \rangle$ in the direction 
${\bf G}_2$ and the
other equivalent to it by symmetry is nonzero in the region $\cal S$, 
(Fig.~\ref{g1g2})indicating without doubt the nucleation of a solid phase. 
Note that $\langle \psi_6 \rangle$
shows a larger interfacial region than that obtained from 
$\langle \psi_{{\bf G}_2} \rangle$. This is because the liquid near the liquid
solid interface is highly orientationally ordered.
\begin{figure}[h]
\begin{center}
\includegraphics[width=12.0cm]{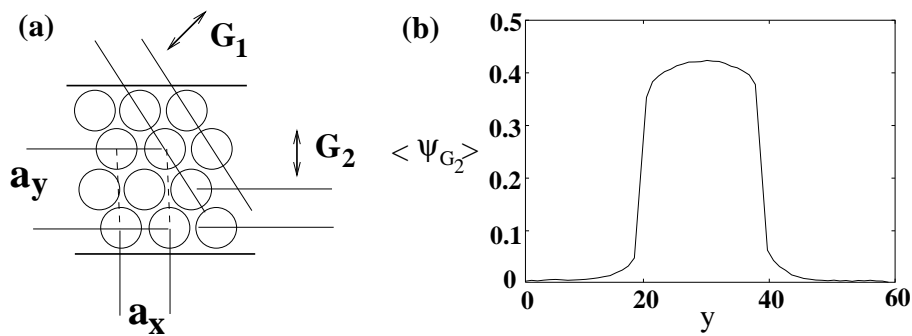}
\end{center}
\caption{(a)The reciprocal lattice vectors ${\bf G}_1$ and ${\bf G}_2$, and
the rectangular unit cell. (b) Solid order parameter corresponding to 
${\bf G}_2$, is nonzero in the region ${\cal S}$, indicating a solid} 
\label{g1g2}
\end{figure}

\begin{figure}[h]
\hskip -1.7cm
\includegraphics[width=21.0cm]{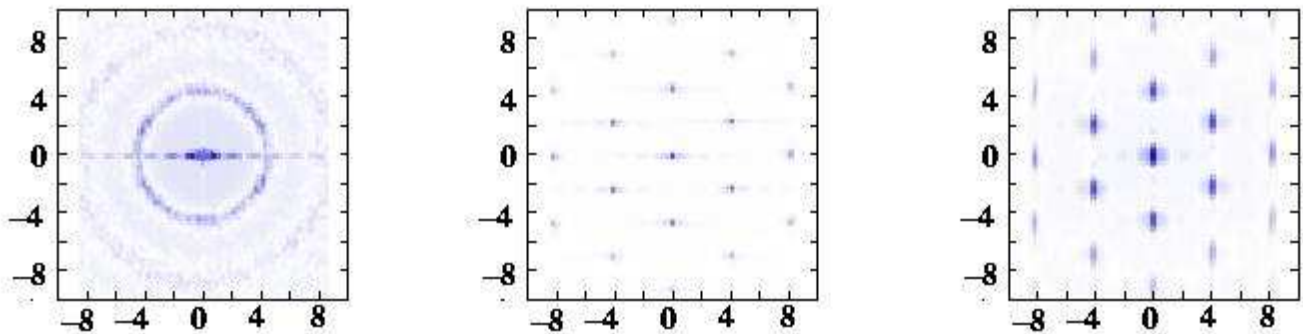}
\caption{Two-dimensional structure factor for the liquid, solid and 
interfacial regions respectively.} 
\label{sq}
\end{figure}

The two-dimensional structure factor shows sharp Bragg peaks for the solid 
region and isotropic pattern for the liquid (Fig.~\ref{sq}). 
Although the structure factor for the solid shows well defined peaks
with six-fold symmetry of the triangular lattice, the quasi-onedimensional
nature of our system implies that true solid~-like order may not be
present. However, we do not pursue this question in detail in this thesis.
In the interfacial region the peaks are seen to be diffuse indicating once 
more that the interfacial region is highly orientationally ordered.
 
\section{The layering transition}
\begin{figure}[!htbp]
\begin{center}
\includegraphics[width=10.0cm]{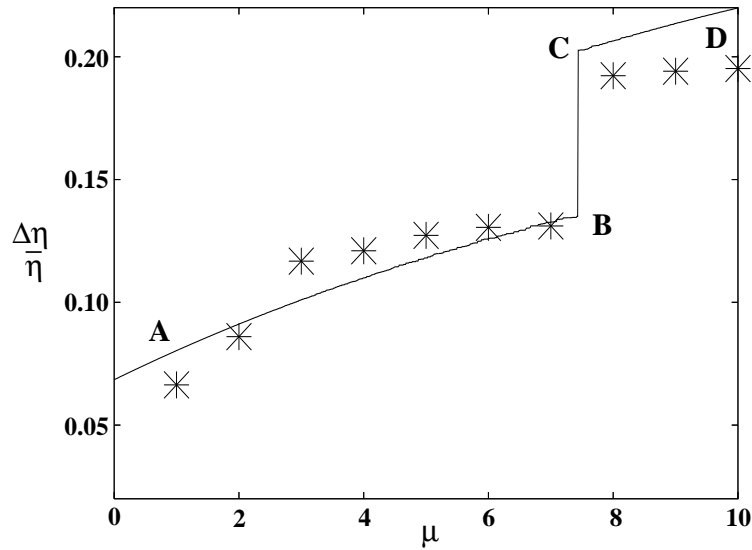}
\end{center}
\caption{ 
Plot of the equilibrium fractional density change $\Delta\eta/\eta$ as a 
function of $\mu$ (points (MC data), thick solid line (approximate theory)), 
showing discontinuous jump at $\mu \approx 8$.}
\label{step}
\end{figure}

\begin{figure}[h]
\begin{center}
\includegraphics[width=7.0cm]{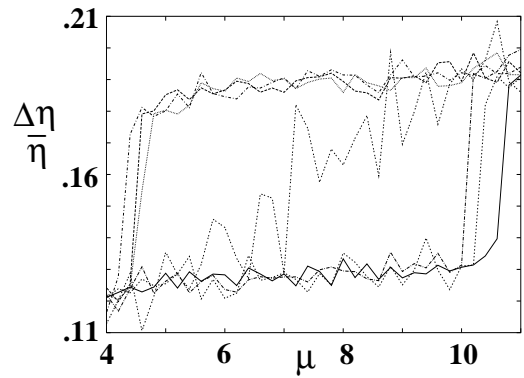}
\end{center}
\caption{Cycle-averaged hysteresis loop as $\mu$ is cycled at the rate of $.2$ 
per $10^6$ MCS. The central jagged line is the result of the initial cycle
when a single dislocation pair was present in the central solid region.} 
\label{hyst}
\end{figure}

We now calculate the difference in densities between the solid and liquid
regions $\Delta \eta = (\eta_s - \eta_l)$ as a function of the strength of
the external field $\mu$. 
While $\Delta\eta/\eta$ increases with increasing $\mu$ as expected, the 
smooth increase is punctuated by a sharp jump (Fig.~\ref{step}). An examination
of the
particle configuration shows that the jump occurs when an extra close~-packed 
layer enters ${\cal S}$ increasing the number of solid layers by one.
For the parameters in our simulation, the jump occurs at $\mu \approx 8$ with
the number of layers increasing from $21$ to $22$.
The value of $\mu$ at the jump is a strong function of $L_s$.
The solid structure is seen to be a defect free triangular lattice  
with a small rectangular distortion $\varepsilon_d(\eta_s,L_s)\,$\cite{debc}. 
We have examined the variation of $\Delta\eta(\mu)/\eta$ by cycling $\mu$ 
adiabatically around the region of the jump. This yields a prominent 
hysteresis loop as shown in Fig.~\ref{hyst} which indicates that 
`surface' steps (dislocation pairs) nucleated in the course of adding 
(or subtracting) a solid layer, have a vanishingly short lifetime.
Consistent with this we find that the jump in
$\Delta\eta(\mu)/\eta$ vanishes when the system is minimised at each $\mu$ 
with a constraint that the solid contains a single dislocation pair 
(Fig.~\ref{hyst}).
Interestingly, a dislocation pair forced initially into the bulk, rises 
to the solid-fluid interface due to a gain in strain energy \cite{lanlif1}, 
where they form surface indentations flanked by kink-antikink pairs. This costs
energy due to the confining potential, as a result, the kink-antikink pair gets
quickly annealed by incorporating particles from the adjacent fluid.
The jump in  $\Delta\eta(\mu)/\eta$ is also seen to decrease with increasing 
$\delta_{\phi}$.

\subsection{The thermodynamic theory}
The qualitative features of these results may be obtained by a simple 
thermodynamic theory (Fig.~\ref{step}) with harmonic distortions of the solid, 
ignoring contributions from spatial variations of the density.
thermodynamic theory. We first write down the free energy densities of
the bulk fluid and solid phases.
\subsubsection{Free energy of the solid} 
For a solid which is only a few atomic layers wide, it is impossible to
separate surface and bulk contributions to the free energy. We therefore
assume that the dominant effect of confinement of the solid is the 
introduction of an uniform strain $\varepsilon_d$ to be calculated from 
a reference triangular lattice with the same number of atomic layers \cite{debc}.
Therefore, we have two terms in the free energy of the bulk solid.

\noindent
(i){\bf Free volume part} : In a given solid lattice, the particles have 
a bulk mean particle distance $a_0$ which is the distance between nearest
neighbours of the lattice. We consider the particles to be confined in
independent Wigner-Seitz (or Voronoi) cells of the solid. The cell has a 
volume $V_b = g_ba_0^D$ where $g_b$ is a geometrical prefactor that depends
on the lattice type and on dimensionality $D$. Each center-of-mass coordinate
of the hard disks can move within a free volume of 
\begin{eqnarray}
V_{bf} = g_b(a_0 - \sigma)^2
\end{eqnarray}
without touching the neighbouring disks. Hence, one obtains a lower bound
for the bulk partition function, $Q \ge ({\cal V}_{bf}/\Lambda^2)^{\cal N}$ which provides
an upper bound to the bulk free energy density (free volume contribution) 
\begin{eqnarray}
f_{\Delta} \le -{\cal N}k_B{\cal T}\ln\left(\frac{g_b(a_0 - \sigma)^2}{\Lambda^2}\right)
\end{eqnarray}
The upper bound becomes asymptotically exact for close packing. 
In our calculation for the two-dimensional hard disk solid we calculate
$g_b = 1/(0.5231\sigma)^2$ and $a_0 = \sigma \sqrt{\pi/(2\sqrt3\eta_{s})}$.

\noindent
(ii){\bf Elastic contribution} : In order that the solid channel accommodates
$n_l$ layers of a homogeneous triangular lattice with lattice parameter
$a_0$ of hard disks of diameter $\sigma$, we need
\begin{eqnarray}
L_s = \frac{\sqrt3}{2}(n_l - 1)a_0 + \sigma
\label{els}
\end{eqnarray}
Defining 
\begin{eqnarray}
\chi(\eta_s,L_s) = 1 + \frac{2(L_s - \sigma)}{\sqrt3 a_0}
\end{eqnarray}
Eqn.~\ref{els} then implies $\chi = integer = n_l$ and violation of 
Eqn.~\ref{els} implies a rectangular strain away from the reference triangular 
lattice of $n_l$ layers. The lattice parameters of a centered rectangular 
lattice (CR) unit cell are $a_x$ and $a_y$ (Fig.~\ref{g1g2}). In general, for 
a CR lattice with given $L_s$ we have $a_y = 2(L_s - \sigma)/(n_l - 1)$ and, 
$a_x = 2/\rho a_y$. The normal strain $\varepsilon_d = \varepsilon_{xx} -
\varepsilon_{yy}$ is then,
\begin{eqnarray}
\varepsilon_d = \frac{n_l - 1}{\chi - 1} - \frac{\chi - 1}{n_l - 1},
\end{eqnarray}
where the number of layers $n_l$ is the nearest integer to $\chi$ so that
$\varepsilon_d$ has a discontinuity at half-integral values of $\chi$.
For large $L_s$, this discontinuity and $\varepsilon_d$ itself vanishes
as $1/L_s$ for all $\eta_s$. We therefore have an elastic contribution
of  $(1/2)K_{\Delta}\varepsilon_d^2$ in the free energy of the solid.

Incorporating these two factors we may write down the free energy density
of the solid phase as 
\begin{eqnarray}
f_s = f_{\Delta} + \frac{1}{2} K_{\Delta}\varepsilon_d^2,
\label{}
\end{eqnarray}
where $K_{\Delta}$ is the Young's modulus of an undistorted triangular lattice. 

\subsubsection{Free energy of the liquid}
The free energy density of the liquid bulk phase may be simply written as
\begin{eqnarray}
f_l = \int_0^{\eta_l} d\eta_l^{\prime}\frac{P/\rho_l - 1}{\eta_l^{\prime}} + f_{id}
\end{eqnarray}
where the ideal gas Helmholtz free energy per particle $f_{id} = \ln(\rho) - 1$
and we use the semiemperical equation of state for the hard disk liquid 
\cite{santos} given by Eqn.~\ref{hdsan}

\subsubsection{Free energy of the system}
We now write down the total free energy density of the system (fluid + solid 
regions) using the free energy density expressions for solid and liquid bulk 
phases as
\begin{eqnarray}
f = x \left[ f_s(\eta_s,L_s) - 4 \eta_s \mu/\pi\right] + 
(1-x) f_l(  \eta_{\ell})
\label{}
\end{eqnarray}
We then minimise this free energy density with the constraint that the average
density is fixed, 
$\eta = x \eta_s + (1-x)   \eta_{\ell}$, where $x$ is the area fraction 
occupied by ${\cal S}$. The result of this calculation is 
shown in Fig.~\ref{step} where it is seen to reproduce the jump 
in $\Delta\eta(\mu)/\eta$. 

\begin{figure}[h]
\begin{center}
\includegraphics[width=7.0cm]{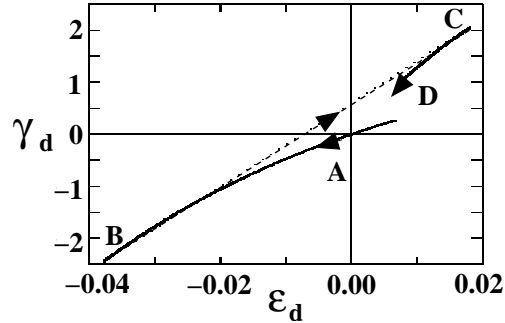}
\end{center}
\caption{ 
A plot of the tensile stress $\gamma_d$ against strain $\varepsilon_d$. The 
arrows show the behaviour of these quantities as $\mu$ is increased from the 
points marked A -- D.
}
\label{stsn}
\end{figure}

Why does the solid incorporate layers of atoms from the liquid ? This 
question may be answered elegantly if one calculates the uniaxial stress
in the solid region as a function of the depth of the strain, $\varepsilon_d$.
The stress may in fact be obtained in a straight forward fashion from
the expression of the free energy. Differentiating the free energy of the 
solid w.r.t. $\varepsilon_d$ we obtain
\begin{eqnarray}
\gamma_d = \frac{\partial f_s}{\partial \varepsilon_d}
\end{eqnarray}
When $\gamma_d$ is plotted versus the uniaxial strain $\varepsilon_d$, we 
observe that the solid is not stress free for any arbitrary value of a 
combination of $\mu$ and $L_s$. In fact, for our parameters, initially the $21$
layered solid is under tension. We follow the variation of the tensile stress
with the strain as $\mu$ is increased from the points $A-D$ in the 
Fig.~\ref{stsn}.  
The state of stress in the solid jumps discontinuously from tensile to 
compressive from B\,$\rightarrow$\,C due to 
an increase in the number of solid layers by one accomplished by 
incorporating particles from the fluid. This transition is reversible
and the system relaxes from a state of compression to tension by ejecting 
this layer as $\mu$ is decreased. 
As $\mu$ is increased, the tension increases 
till it reaches about $-2.45$ when the corresponding strain is about $-0.038$.
At this point a layer enters the solid region and the stress and strain 
switches from tensile to compressive. Further increase in $\mu$ will now 
decrease the stress and drive the solid to a state of zero stress. 
Thus, the layering transition from $21$ to $22$ layers as observed by us is a 
mechanism for relieving stress. 

Solids subject to large uniaxial deformations, relieve stress either by the 
generation and mobility of dislocations \cite{Haasen} and/or by the 
nucleation and growth of cracks \cite{Haasen,marder}. What is the nature of 
stress relaxation when conditions are arranged such that these conventional 
mechanisms are suppressed ? Nano-indentation experiments \cite{uzi,sutton1} 
show that 
if small system size prevents the generation of dislocations \cite{debc}, 
solids respond to tensile forces by shedding atoms from the surface layer.
What we have shown here using extensive computer 
simulations and theory, is that a small crystal trapped within a potential 
well \cite{metcalf} and in contact with its own fluid, a situation easily 
realised using optical traps, responds to large compressive stresses
by a novel mechanism --- the transfer of complete lattice layers across the 
fluid solid interface. In Chapter 6, we study the importance of this 
``layering'' transition by looking at momentum and heat transport across the
liquid solid interface and the role this transition plays on transport 
properties.

\subsection{Layering in other potentials}
If the layering transition observed in the hard disk system is actually a new 
nechanism for relieving stress in a thin crystal, it should be independent
of the details of the potential. Our main results trivially extend to particles
interacting with any form of repulsive potential, or even when the interactions
are augmented by a short range attraction, provided we choose $\mu$ deeper than
the depth of the attractive potential.
\begin{figure}[t]
\begin{center}
\includegraphics[width=7.0cm]{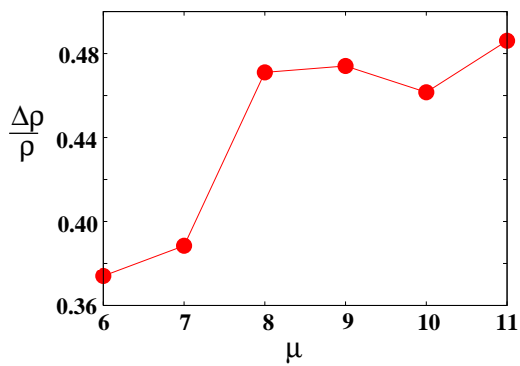}
\end{center}
\caption{ 
Plot of the equilibrium fractional density change $\Delta\rho/\rho$ as a 
function of $\mu$ (points (MC data) for soft core; line denotes approximate fit}
\label{step-soft}
\end{figure}
\begin{figure}[h]
\begin{center}
\includegraphics[width=7.0cm]{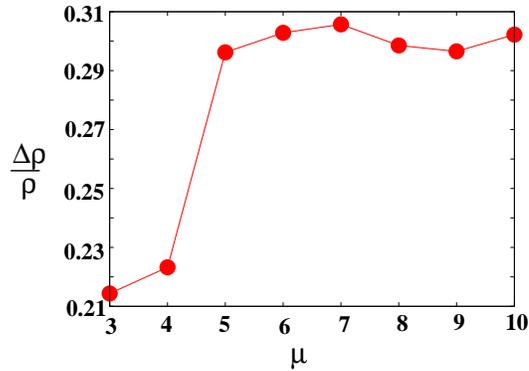}
\end{center}
\caption{ 
Plot of the equilibrium fractional density change $\Delta\rho/\rho$ as a 
function of $\mu$ (points (MC data) for Lennard-Jones system; line denotes 
approximate fit}
\label{step-lj}
\end{figure}
In this section we show explicitly that the layering transition is present
in the soft core and Lennard-Jones system. Once again we perform MC simulations
in the constant ${\cal NAT}$ ensemble with periodic boundary conditions and
with the external chemical potential $\mu$. The relevant parameters 
corresponding to these potentials, namely $\epsilon$ and $\sigma$ in 
Eqn. 4.4 and Eqn. 4.5, set the energy scale and the length scale respectively.
In our simulation $\epsilon = \sigma = 1$ and ${\cal N} = 1200$ particles
occupy an area ${\cal A} = 24\times 60$ with the solid occupying the central
third of the cell of size $L_s = 20$. The average density of the system is
therefore $\rho = 0.833$ to be compared with the freezing density of 
$\rho \sim 1.0$ and $\rho \sim 1.0$ for the soft core and Lennard-Jones systems at this
temperature.For the soft core potential, the $\mu$ value at the jump in 
density (Fig.~\ref{step-soft}) is even quantitatively comparable to the 
corresponding hard core system. In the LJ system the jump happens earlier as 
expected due to the cohesive energy, $e_{coh}$ of the LJ atoms 
(Fig.~\ref{step-lj}). We estimate, $\mu^{\prime} = \mu_{HD} - e_{coh}$, where
$\mu_{HD}$ is the jump potential for the hard disk. Thus 
$e_{coh} \approx -3.5$. The shift
of the layering transition may be used to obtain the cohesive energy of the
trapped solid. 

\chapter{Liquid solid interfaces : Dynamical properties}

\section{Introduction}
In the last chapter, we showed how one might produce a liquid solid
interface using a non-uniform external potential. We characterized
this interface using a variety of thermodynamic and structural
quantities, which were measured as a function of the perpendicular
distance from the interface. Finally, we showed that as a function of
the depth of the potential well, the trapped solid undergoes, what we
called, "layering" transitions which involved the addition (or
removal) of an entire layer of solid from or into the surrounding
liquid through the interface. The layering transition was accompanied
by a sharp jump in the density of the solid. We showed that this
layering transition is a novel mechanism by which a stressed
nano-solid constrained by an external potential can respond
plastically to large stresses {\em without} nucleating
dislocations. Finally, we established that this phenomenon is general
and is independent of the particular interatomic potential used.

In this final chapter of the thesis we shall explore how mass,
momentum and energy is transported across the liquid solid interface
and especially the role of the layering transition on the transport
coefficients. We shall show in this chapter that (1) fluctuations
associated with this transitions are of a special kind always
involving the transfer of complete layers of solid (2) these
fluctuations offer resistance to the transfer of momentum and energy
through the interface (3) the resistance is maximum when the energy
matches that required to raise a complete lattice layer from within
the potential well into the surrounding liquid. In the next section we
shall begin by studying the stability of surface kinks at the liquid
solid interface in the hard disk system. We shall then study the
response of the interface to weak acoustic shocks. In the last section
we shall use non-equilibrium molecular dynamics to study heat
transport through the liquid solid interface in soft disks and obtain
the contact or Kapitza resistance of the interface as a function of
the depth of the potential well.

In our investigations, to be reported in this chapter we shall
exclusively use molecular dynamics simulations since we shall be
interested in dynamical quantities.

\section{Kinetics of layering}

The large hysteresis loops associated with the layering transition
obtained in Chapter 5 makes it clear that the kinetics of this
transition is slow. To study the lifetime of the kink-antikink pairs
(surface step), we resort to a MD simulation.


Starting with an equilibrium configuration for the hard disk system,
taken from our Monte Carlo runs as
discussed in the last chapter (Section 5.2) at $\mu = 9.6$
corresponding to a $22$-layer solid, we create a unit surface step of
length $l$ by displacing a few interfacial atoms from the solid region
into the liquid and `quench' across the transition to $\mu=4.8$, where
a $21$-layer solid is stable.  The rest of the parameters are kept
identical to those given in chapter 5.  We observe that the
fluctuation thus created rapidly relaxes back and the surface step
vanishes as the atoms are pulled back into the solid. We illustrate this 
by plotting the number of hard disks within the solid region as a function 
of the MD time steps (Fig. 6.1). 
\begin{figure}[h]
\begin{center}
\includegraphics[width=8.6cm]{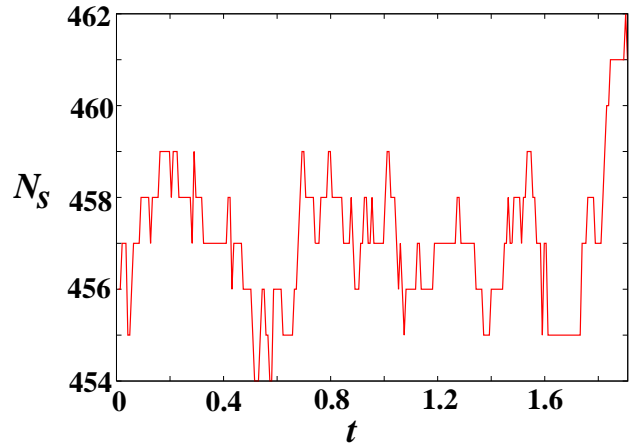}
\end{center}
\caption{Plot of the number of particles in the solid region (${\cal N}_s$)
as a function of time $t$ clearly shows that the displaced particles are pushed
back into the region.
}
\end{figure}
As soon as a step is
created, the line of atoms in the portion of the solid thus exposed
bend to fill up the gap created between the atomic layers and the edge
of the potential trap. This generates considerable local elastic
stress. Also, the liquid layer lying immediately adjacent to the solid
has a lot of orientational and solid-like order. For short times it
responds elastically to the presence of an increased local density of
atoms. The combined elastic response therefore pushes the displaced
atoms back into the solid region thereby annihilating the
step. Indeed, a free energy audit involving a bulk free energy gain
$\Delta F \sim 1/L_s$, going from a 22 to 21 layered solid, and an elastic
energy cost $\sim \log(l)$ for creating a step of size $l$, reveals that a
surface step is stable only if $l\geq l^* \sim 1/L_s$. For small
$L_s$, the critical size $l^*$ may therefore exceed $L_x$, the total
length of the interface. Of course, if the step spans the entire
length of the interface, there is no bending of the atomic lines and
there is no elastic energy cost. This explains the slow kinetics since
the system has to wait till a rare random fluctuation, which displaces
all the atoms in a solid layer across the interface coherently, is
required for the layer transition to occur. Although we have
explicitly demonstrated this for the hard disk system, we believe that
similar considerations should be appropriate for the soft disk and
Lennard Jones systems too.

The slow kinetics of the layering transition may have an impact on the
transfer of momentum across the liquid solid interface in the form of
regular sound waves or acoustic shocks. The large effective
compressibility of the solid at the layering transition as evidenced
by the jump in the density as the chemical potential is increased by
an infinitesimal amount (Fig.~\ref{step}) should reduce the velocity of
sound considerably. The propagation and scattering of sound in an
inhomogeneous region with coexisting phases has been studied
extensively \cite{lanlif2,zener,isakovich,onuki} in the past. The
transfer of mass between coexisting phases at inter-phase boundaries
is known to slow down and dissipate sound waves traveling through the
system. Our system has an artificially created inhomogeneity, which
should have a similar effect on its acoustic properties.

Further, the mechanism of stress relaxation of a thin ($L_s$ small)
solid via the transfer of an entire layer of atoms may be exploited
for a variety of practical applications. Provided we can eject this
layer of atoms deep into the adjoining fluid and enhance its lifetime
we may be able to use the ejected layer of atoms to create monolayer
atomic films or coatings. Highly stressed mono-atomic layers tend to
disintegrate or curl up \cite{novoselov} as they separate off from the
parent crystal. It may be possible to bypass this eventuality, if the
time scale of separation is made much smaller than the lifetime of the
layer. Can acoustic spallation \cite{zeldovich} be used to cleave
atomic layers from a metastable, stressed nanocrystal?

In the next section, therefore, we study the response of the liquid
solid interface in our system of hard disks to acoustic shocks with a
view to studying the effect of the layering transition on acoustic
shock propagation and dissipation.

\section{Effect of shock wave}
Imagine, therefore, sending 
in a sharp laser (or ultrasonic) pulse, producing
a momentum impulse ($v_y(t=0) = V_0$) over a thin region in $y$ spanning the
length $L_x$ of the simulation cell, which results in a weak acoustic 
shock \cite{zeldovich} (corresponding to a laser power $\approx 10^2$ mW and a 
pulse duration $1 {\rm ps}$ for a typical atomic system). 
The initial momentum pulse travels through the solid and 
emerges at the far end (Fig.~\ref{pulse}) as a broadened Gaussian 
whose width, $\Delta$, is a measure of absorption of the 
acoustic energy of the pulse due to combined dissipation in the liquid, 
the solid and at the interfaces \cite{lanlif2,zener,cahill}. 
For large enough pulse strengths $V_0$, this is accompanied by
\clearpage 
\begin{figure}[h]
\begin{center}
\includegraphics[width=12.0cm,height=18.0cm]{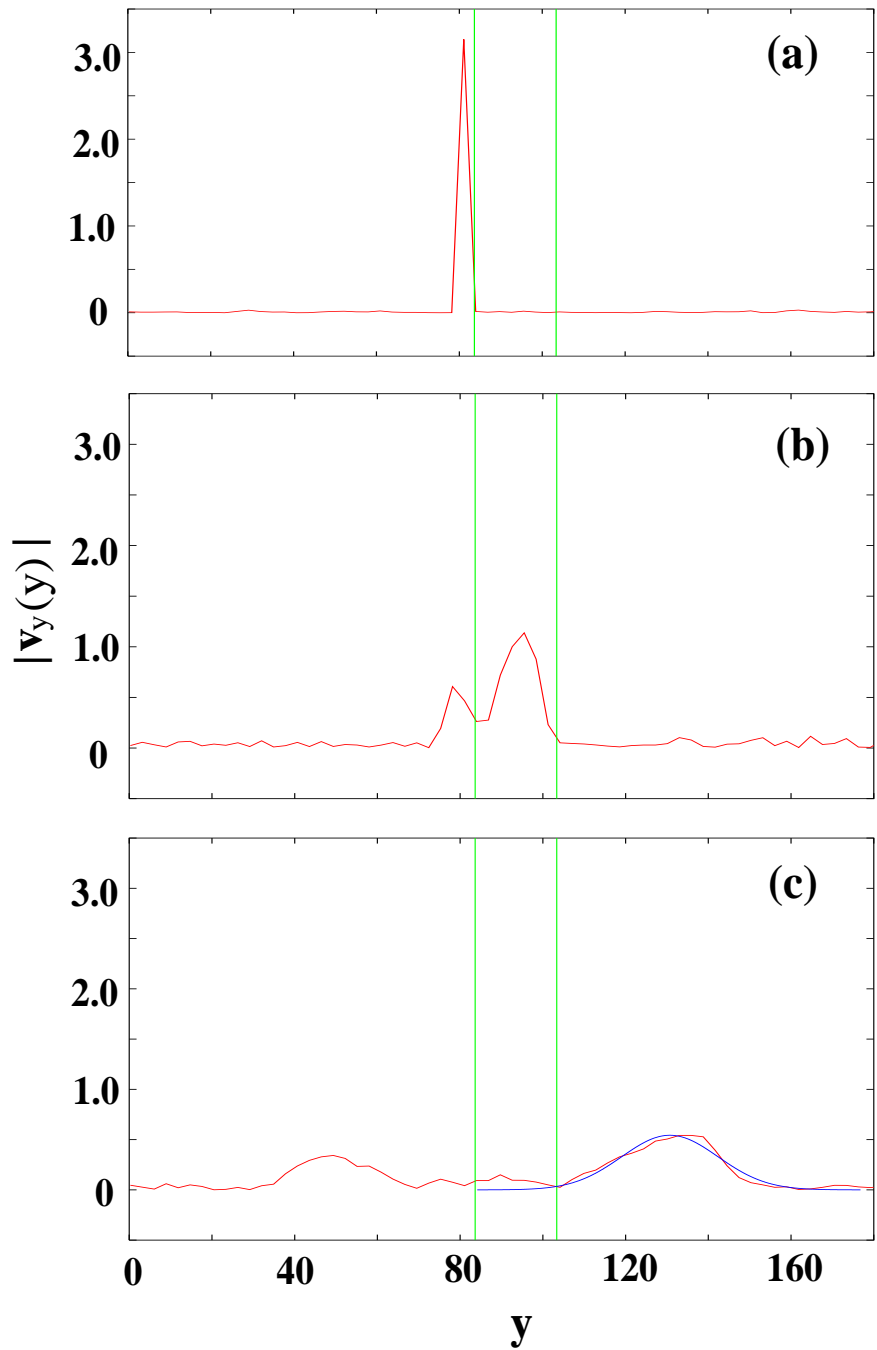}
\end{center}
\caption{
(a)-(c) Plot of the absolute value of the momentum $|v_y(y)|$ for molecular
dynamics times $t = .0007$ (a), $.2828$ (b) and $2.8284$ (c). The 
green lines show the position of the solid~-fluid interfaces. 
The fit to a Gaussian (blue line) is also shown in (c).  
Curves such as in (a)-(c) are obtained by averaging 
over $100-300$ separate runs using different realizations of the initial
momentum distribution.}
\label{pulse}
\end{figure}
\clearpage 
\noindent
a {\it coherent ejection of the (single) outer layer of atoms into the fluid}.  
Note that in our MD simulations, to reduce interference from the reflected 
pulse through periodic boundary conditions, we increase the fluid regions on 
either side, so that for the MD calculations we have a cell of size 
$22.78\times186.98$ comprising $3600$ particles. In Fig.~\ref{pulse} the 
initial momentum pulse with strength $V_0 = 6.$ is given within a narrow strip 
of size $\sim \sigma$, just to the left of the solid region and the curves are
fitted to a Gaussian (and the width $\Delta^2$ extracted) when the maximum of 
the pulse reaches a fixed distance of $44.1$ from the source. A reflected pulse can also be seen. 

\begin{figure}[b]
\begin{center}
\includegraphics[width=7.5 cm]{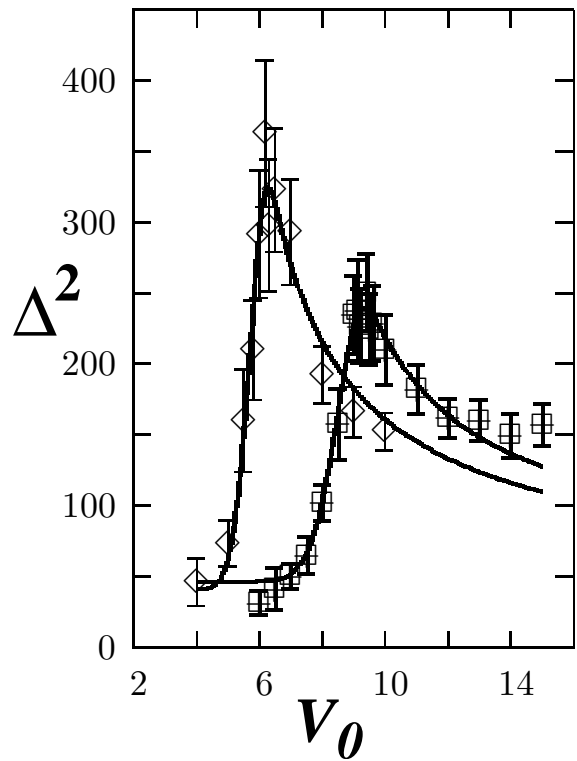}
\end{center}
\caption{Plot of the squared width $\Delta^2$ of the momentum 
pulse after it emerges from the solid as a function 
of $V_0$ for $\mu = 4.8\,(\,\Diamond)$ and $9.6\,(\,\Box)$.  The 
solid line is a guide to the eye.  
The peak in $\Delta^2(V_0)$ so produced is more prominent for 
the metastable $22$ layered solid $\mu = 4.8 $ than for the 
stable ($\mu = 9.6$) system showing a more coherent momentum 
transfer in the former case. 
}
\label{reson}
\end{figure}
\begin{figure}
\vskip 1cm
\includegraphics[width=17.0cm]{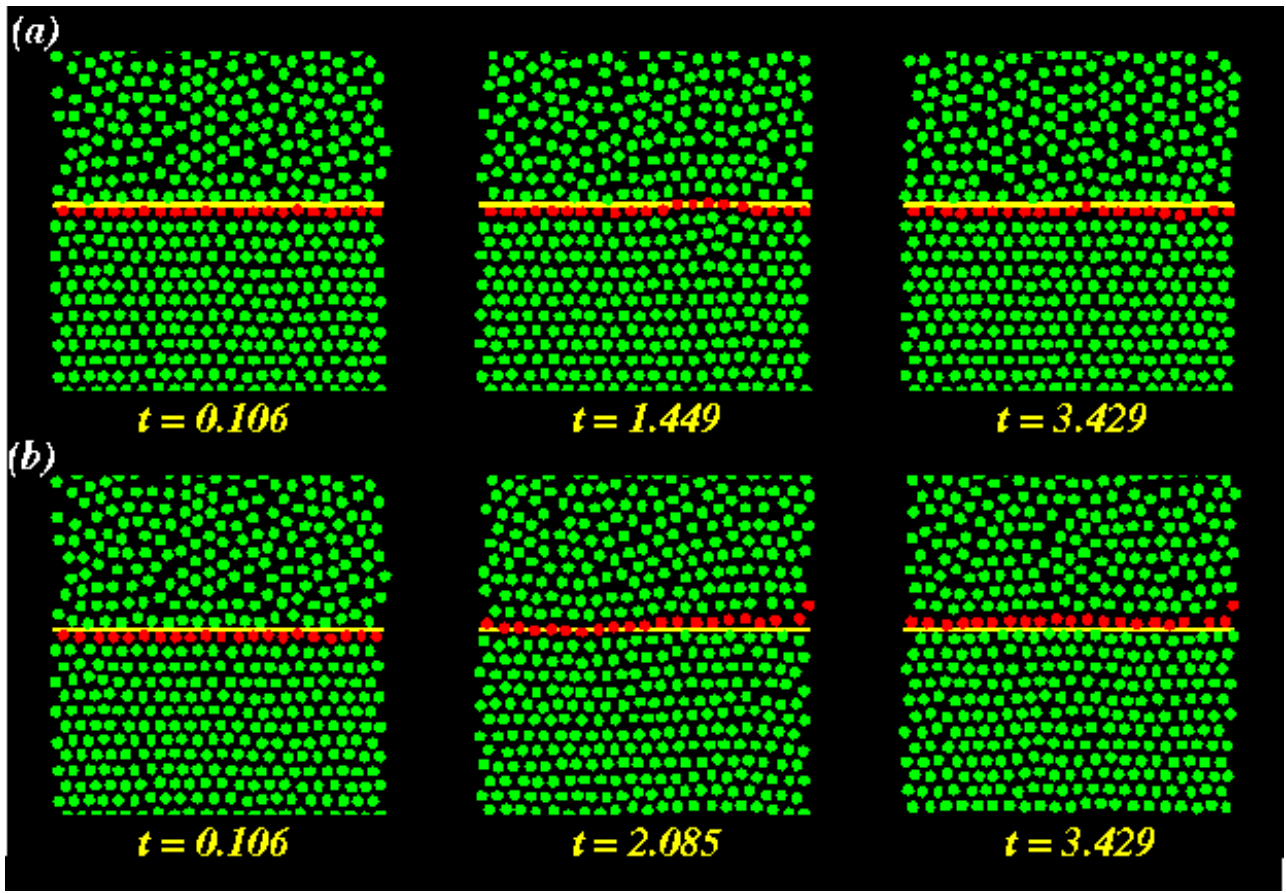}
\vskip 1cm
\caption{(a)
Configuration snapshot from a portion of our MD cell showing 
hard disk atoms (green circles) at the solid (bottom)- liquid 
interface (yellow line) as a weak momentum pulse ($V_0 = 2.$) emerges 
into the liquid, at three different times, for $\mu = 4.8$. The pulse 
initially ejects a few atoms from the solid but are subsequently pulled 
back due to the large elastic strain cost in bending. 
of the interfacial crystalline layer (red circles) of a metastable 
(b) The same for a stronger momentum pulse, $V_0 = 6$. This time the pulse
strength is sufficient to eject the layer.
}
\label{movie}
\end{figure}

When a shock wave, which propagates through a conventional solid
emerges from the free surface, the compressed material expands to zero
pressure \cite{zeldovich}. The unloading (rarefaction) wave travels backwards into the
material with the speed of sound. The response of the solid now
depends on the specific nature of the shock front. For a shock wave
with an approximately Gaussian profile as in our case, significant
negative pressures can develop at the interface where the shock
emerges due to the interaction of the forward and the reflected waves
and a portion of the solid may split off by a process known as
"spallation". Spallation in bulk solids like steel needs acoustic
pressures in excess of $10^5$ N/cm$^2$ \cite{zeldovich} usually
available only during impulsive loading conditions; the ejected layer
is a ``chunk'' of the surface. In contrast, the pressures generated by
the shock wave in our system causing coherent {\it nanospallation} involves 
surface stresses of the order $k_B T/\sigma^2 \approx  10^{-5}$ N/cm$^2$, and 
are therefore much lower. This difference comes
about because unlike a bulk system, a strained nanocrystal on the
verge of a transition from a metastable $n+1$ to a $n$ layered state
readily absorbs kinetic energy from the pulse.
The fact that surface indentations are unstable (Fig.~\ref{movie}) unless of 
a size comparable to the length of 
the crystal, $L_x$, ensures that a full atomic layer is evicted almost 
always, leading to coherent absorption of the pulse energy.  
The coherence of this absorption 
mechanism is markedly evident in a plot of $\Delta^2$ against $V_0$
which shows a sharp peak (Fig.~\ref{reson}). Among the two systems studied by 
us, {\it viz.}, a metastable ($\mu = 4.8$) and a stable ($\mu = 9.6$)  $22$ 
layered solid, the former shows a sharper resonance. 
Note that the absorption of momentum 
is largest when the available kinetic energy of the pulse exactly 
matches the potential energy required to eject a layer. To elucidate this 
fact further, we plot the configurations of the metastable system 
($\mu = 4.8$) as the pulse travels through the system, for two different
pulse strengths, $V_0 = 2$ and $V_0 = 6$ (Fig.~\ref{movie}). The weaker
momentum pulse ($V_0 = 2$) initially ejects a few atoms of the interfacial
crystalline layer of the metastable $22$ layered
solid. However, the resulting large non-uniform elastic strain 
evidenced by the bending of lattice layers causes these atoms 
to be subsequently pulled back into the solid. This effect is the same
as that seen in Section 6.2. Only a stronger pulse,
$V_0 = 6$, capable of ejecting a complete lattice layer succeeds in reducing 
the number of solid layers by one leading to overall lower elastic energy.

The eviction of the atomic layer is 
therefore assisted by the strain induced interlayer transition and 
metastability of the $22$ layered solid discussed above. Spallation
is also facilitated if the atomic interactions are anisotropic so that
attraction within layers is stronger than between layers (eg. graphite
and layered oxides \cite{novoselov}), for our model, purely repulsive, hard disk
solid, an effective, intralayer attractive potential of mean force is 
induced by the external potential \cite{debc}.
 
\begin{figure}
\begin{center}
\includegraphics[width=9.0cm,height=11cm]{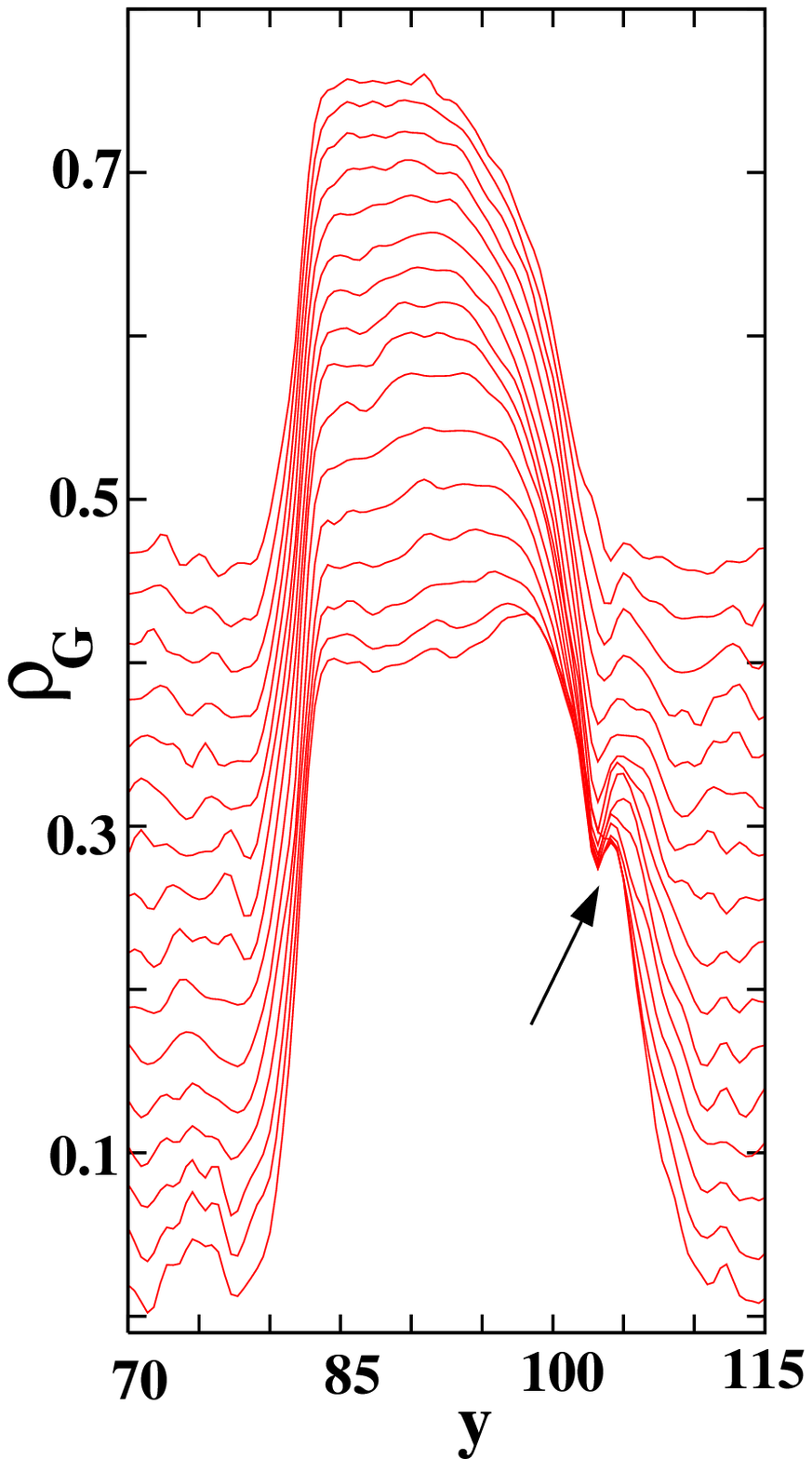}
\end{center}
\caption{A plot of the time development $\rho_{\bf G}(y,t)$  
The solid ejects a layer (shown by an arrow) which subsequently dissolves in 
the fluid. The curves from bottom to top correspond to time slices 
at intervals of $\Delta t = .07$ starting from 
$t = 1.06$ (bottom). We have shifted each curve upward by $.03 t/\Delta t$ for 
clarity. }
\label{life}
\end{figure}

The spallated solid layer emerges from the solid surface into the fluid, and 
travels a distance close to the mean free path; whereupon it disintegrates due 
to viscous dissipation (Fig.~\ref{life}). 
To obtain the life time of the spallated layer we calculate the time 
development of the Fourier component 
of the local density correlation $\rho_{\bf G}(y,t)$  
obtained by averaging, at each time slice $t$, the sum 
$\sum_{j=1,N} \exp(-i\,{\bf G\,\cdot}\,({\bf r_j - r_i}))$ over all particle 
positions ${\bf r_i}$ within a strip of width $\sim \sigma$ centered 
about $y$ and spanning the system in $x$. The wavenumber 
${\bf G} = (2 \pi/d) {\bf \hat{n}}$ where
$d = .92$ is the distance between crystal lines in the 
direction ${\bf \hat{n}}$ normal to the fluid-solid interface. The 
solid (central region with $\rho_{\bf G}(y,t) \ne 0$) ejects a layer (shown by 
an arrow in Fig.~\ref{life}) which subsequently dissolves in the fluid.
The lifetime of the layer is around 
2-3 time units ($\tau$) which translates to a few ps for typical atomic
systems. The lifetime increases with decreasing viscosity of the surrounding 
fluid. Using the Enskog approximation \cite{chapman} to the hard disk 
viscosity, we can calculate the bulk viscosity for a hard disk fluid to be
\cite{chapman,wood}
\begin{eqnarray}
\zeta_E = \frac{16}{\pi}\zeta_{00}\eta^2g(\sigma)
\end{eqnarray}
where $\zeta_{00}$ is a constant and $g(\sigma)$ is the pair-correlation
function at contact which can be calculated using Eqn. (4.17) and Eqn. (4.18).
For a system of hard disks with $m = \sigma = \beta = 1$, we can calculate
\cite{wood} $\zeta_{00} = 1/2\sqrt\pi$. Thus, $\zeta_E \propto \eta^2$ and 
we estimate that by lowering the fluid density one may increase 
the lifetime of the layer considerably. The lifetime enhancement is 
even greater if the fluid in contact is a low density gas (when the 
interparticle potential has an attractive part \cite{LJ-visc}). 


\section{Effective liquid theory}

\begin{figure}[t]
\begin{center}
\includegraphics[width=7.0cm]{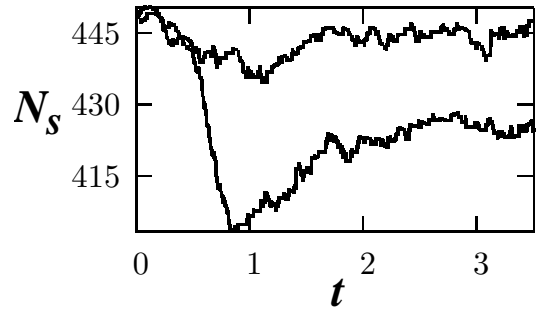}
\end{center}
\caption{A plot of the total number of particles $N_s$ within the solid region 
($\mu = 4.8$) as a function of time for $V_0 = 1.$ (top) and $6.$.
Note oscillations in $N_s$; only the stronger 
pulse changes the number of solid layers from $22$ to $21$. } 
\label{nst}
\end{figure}

The absorption line~-shape may be understood within a phenomenological 
``effective liquid'' approximation. The extra 
absorption producing the prominent peak in Fig.~\ref{reson} is due to the 
loss of a whole layer from the solid into the liquid (Fig.~\ref{movie}). For 
small $V_0$, the confined solid responds by centre of mass fluctuations 
($q\to 0$ phonons) shown by oscillations of $N_s$ with time (Fig.~\ref{nst}).
\begin{figure}[b]
\begin{center}
\includegraphics[width=7.0cm]{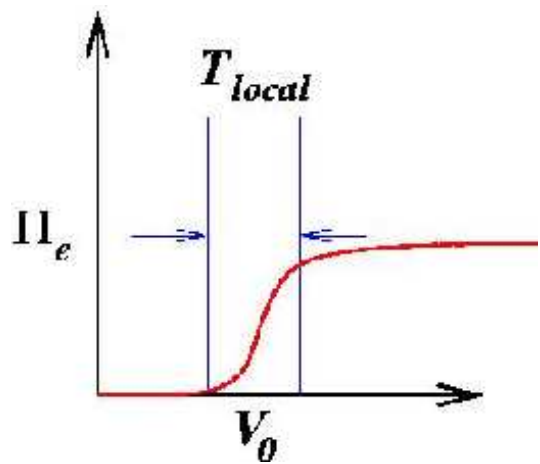}
\end{center}
\caption{A cartoon diagram showing the momentum transfer as assumed in our
phenomenological theory} 
\label{mom-theory}
\end{figure}
Scattering from this and other sources \cite{cahill,lanlif2,zener,isakovich,
onuki} constitute a background which we ignore, as a first approximation, for 
simplicity. Within our approximation, the momentum loss 
at the interface is modelled as 
regular dissipation within a liquid strip of (fictitious) width $\xi$ 
(see Fig.~\ref{mom-theory}). The
expected momentum transfer at the interface $\Pi_e = \Pi_0 \times$ the 
probability that the momentum $\Pi_0$ required to eject the layer, exists.
If a local ``temperature'' $T_{local}$ measures the degree of (de)coherence 
of the momentum transfer, then $ \Pi_e = (1/2) \Pi_0 
{\rm erfc}[(\Pi_0 - V_0)/\sqrt{2 k_B T_{local}}]$ and $\xi$ 
may be extracted from $V_0 - \Pi_e = V_0 \exp(-\alpha \omega^2 \xi)$.
Substituting for $\Pi_e$ we obtain the extra absorption due to the
interface,
\begin{equation}
\Delta^2 =  4.\alpha c_0^2 \xi = - a \log [ 1 - \frac{\Pi_0}{2 V_0}{\rm erfc}\{\frac{\Pi_0 - V_0}{\sqrt{2 k_B T_{local}}}\} ]
\label{ela}
\end{equation}
We use $a,\Pi_0$ and $T_{local}$ as fitting parameters. In Fig.~\ref{reson} we 
show a fit to Eq. \ref{ela} of our MD data and observe that it reproduces all
the features remarkably well. The larger error-bars near the peak 
in $\Delta^2(V_0)$ reflects the difficulty of fitting a Gaussian to the 
transmitted pulse when dissipation is large. Indeed, in this region the
pulse shape is systematically distorted away from Gaussian due to effects 
beyond the scope of our simple theory. Large fluctuations in $N_s$ 
(Fig.~\ref{nst}) lead to expected \cite{lanlif2,zener,isakovich,onuki} and  
detectable decrease in average pulse speed (Fig.~\ref{pdec}).
\begin{figure}[h]
\begin{center}
\includegraphics[width=7.0cm]{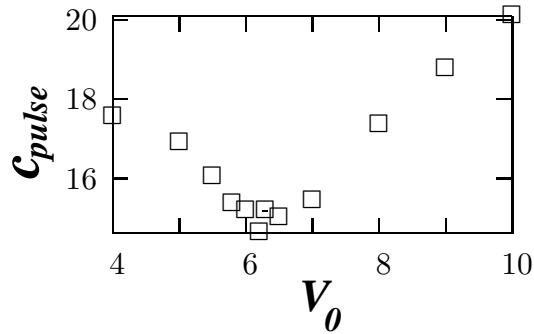}
\end{center}
\caption{Average pulse velocity $c_{pulse}(V_0)$ for $\mu = 4.8$; note the dip 
in $c_{pulse}$ where absorption is strongest.} 
\label{pdec}
\end{figure}

\section{Heat transport across the liquid - solid interface}
In the last section we studied the effect of the layering transition
on the transport of momentum across the liquid solid interface. In
this section we shall focus on the transport of energy. The kinetic
energy flux ${\bf j}_e$ is related to the temperature gradient ${\bf
\nabla {\cal T}}$ through the famous Fourier's law, viz.
\begin{eqnarray}
{\bf j}_e = -\lambda{\bf  \nabla {\cal T}}
\end{eqnarray}

where, the transport coefficient $\lambda$ is the thermal
conductivity. The transport of heat through small and low dimensional
system has enormous significance in the context of designing useful
nano-structures \cite{cahill}. Heat transport across a model liquid solid
interface has been studied in three dimensions with the interatomic potential
being Lennard-Jones \cite{barrat}. They have shown that the Kapitza resistance
can reach appreciable values when the liquid does not wet the solid. 
In two dimensions, strictly speaking,
Fourier's law needs to be modified by correction terms of the order of
the logarithm of the system size \cite{fourier}. For a small system,
however, we shall ignore these corrections and assume it to be valid.
  
The specific context in which we study thermal properties of the
liquid solid interface is the same as that we have used for our
studies of the acoustic properties in the last section. Again, a solid
region is created within a liquid using an external chemical potential
trap. The overall dimensions of the system and the size of the solid
region are also identical. The inter-atomic potential however is
chosen to be soft sphere with the potential given by Eqn. 4.4. The smooth 
potential allows us
to use standard MD simulations with a velocity Verlet algorithm. The
time step of $\delta t = 10^{-3}$ in our MD ensures that the total
energy is conserved (in equilibrium) to within $10^{-4}$. Periodic boundary
conditions are applied in the $x$-direction. A temperature gradient is
generated throughout the simulation cell by coupling the two sides of
the system in the $y$-direction to two reservoirs at different
temperatures, which drives the system out of equilibrium. Thus we have
a reservoir at $y = 0$ at a temperature ${\cal T}_1$ and another at $y
= L_y$ at temperature ${\cal T}_2$. The coupling to the two reservoirs
is implemented by the imposition of ``Maxwell boundary conditions'':
when a particle hits the left (right) boundary it get reflected such that
the velocity component parallel to the boundary ($v_x$)
derives a new velocity from normal Maxwell distribution at the given
temperature. The velocity component normal to the boundary ($v_y$) 
is taken from a Maxwell speed distribution given by
\begin{eqnarray}
f_{\alpha}(d{\bf v}) = \frac{m^2}{2\pi(k_B{\cal T}_{\alpha})^2}|v_y|
\exp\left[-\frac{mv^2}{2k_B{\cal T}_{\alpha}}\right]d{\bf v} \,\,\,\,\,\,\,\,\,\,\alpha = 1,2
\end{eqnarray}
corresponding to a temperature ${\cal T}_1 ({\cal T}_2)$ at the left (right) 
boundary. When the two temperatures are different, a net energy
flux $j_E$ in the $y$-direction results in the steady state condition,
which is computed by averaging the kinetic energy added (or removed)
by each reservoir per unit time and surface. Temperature profiles are
obtained from the local kinetic energy density. We use the standard
velocity Verlet scheme of molecular dynamics with equal time update,
except when the particles collide with the ``hard wall''
reservoirs. We treat the collision between the particles and the
reservoir as that between a hard disk of unit diameter colliding
against a hard structure less wall. If the time, $\tau_c$, of the next
collision with any of the two reservoirs at either end is smaller than
$\delta t$, the usual update time step of the MD simulation, we update
the system with $\tau_c$. During the collision with the walls Maxwell
boundary conditions are imposed to simulate the velocity of an atom
emerging from a reservoir at temperatures ${\cal T}_1$ or ${\cal T}_2$.

\begin{figure}[t]
\begin{center}
\includegraphics[width=16cm,height=5cm]{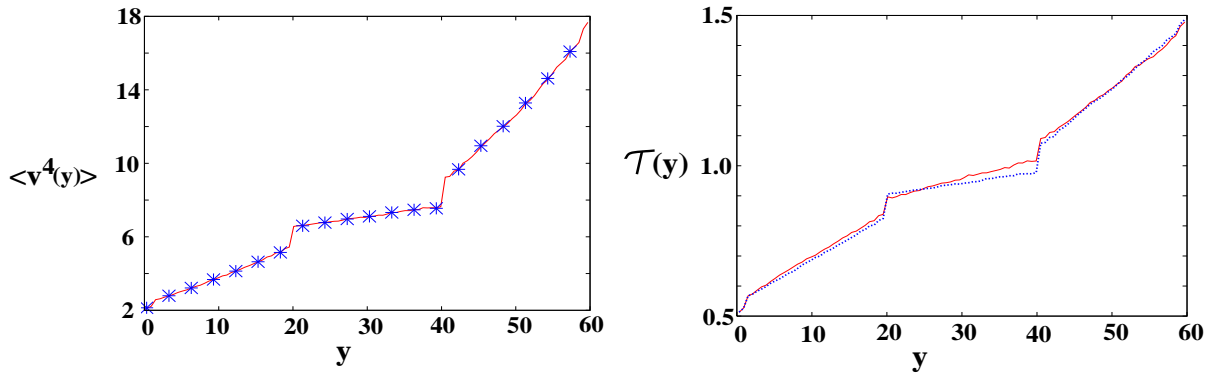}
\end{center}
\caption{(a) $\langle v^4(y) \rangle$ and $8{\cal T}^2(y)$ as a function of 
$y$, the system coordinate perpendicular to the reservoirs, for $\mu=8.0$.
(b) ${\cal T}(y)$ as a function of $y$, 
the system coordinate perpendicular to the reservoirs, for $\mu=7.0$ (Red line
) and $\mu=8.0$ (Blue line).
}
\label{lte}
\end{figure}

The system is allowed to reach the steady state where the current is
the same in the entire system. To check for local thermal equilibrium
, we first define the local temperature ${\cal T}(y) = \langle
1/2~mv^2(y)\rangle$, where the averaging is done locally over strips
parallel to the $x$-direction. This is called the kinematic
temperature. In Fig.~\ref{lte}(a) we plot, from the MD simulations with
$\mu=8.0$, $\langle v^4 \rangle$ and see that it overlaps with $8{\cal
T}^2$ locally. This is consistent with the thermodynamic relation
$\langle v^4 \rangle = 8{\cal T}^2$. Thus the system preserves local
thermal equilibrium. Note that this verification in turn justifies our
definition of the local temperature.
The temperature profile (Fig.~\ref{lte}(b)) shows that (1) the thermal 
conductivity of the solid region is larger than the liquid, which is expected 
because of the larger density of the former and (2) there is a significant jump
in the temperature as one crosses the two interface. Such a jump in
the temperature is also expected and is due to the Kapitza or contact
resistance ($R_K$) \cite{kapitza}. This is defined as,
\begin{eqnarray}
R_K = \frac{\Delta {\cal T}}{j_E}
\end{eqnarray}
where $\Delta {\cal T}$ is the difference in temperature across the
interface. We are next interested in determining this Kapitza
resistance across the solid liquid interface as a function of the
strength of the external potential $\mu$.

We have already seen that with increasing $\mu$, the system shows a
jump in the density of the solid region corresponding to the addition
of an entire layer of atoms. From the profile shown in
Fig.~\ref{lte}(b), the Kapitza resistance is easily obtained by
dividing the temperature jump by the energy flux. Note that a slight
dependence of $R_K$ on ${\cal T}$ is visible in Fig.~\ref{lte}(b) with
a larger temperature jump on the ``warm'' side. The results shown
correspond to the average values of $R_K$ between the ``warm'' and
``cold'' sides.
\begin{figure}[t]
\begin{center}
\includegraphics[width=8.6cm]{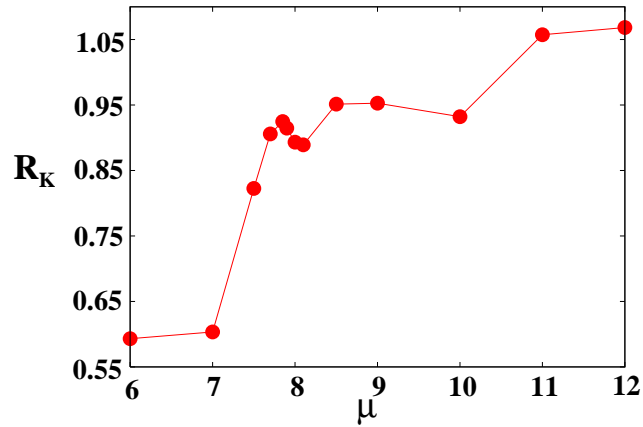}
\end{center}
\caption{Plot of the Kapitza resistance, $R_K$, as a function of $\mu$, shows
a jump at the layering transition 
}
\label{kapr}
\end{figure}
\begin{figure}[b]
\begin{center}
\includegraphics[width=8.6cm]{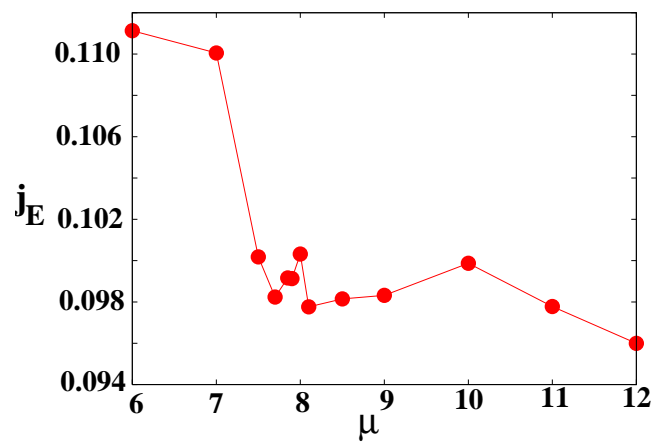}
\end{center}
\caption{Plot of the heat flux, $j_E$, as a function of the trap depth, $\mu$.
Note that the overall flux decreases as a function of $\mu$. 
}
\label{flux}
\end{figure}
\begin{figure}[t]
\begin{center}
\includegraphics[width=8.6cm]{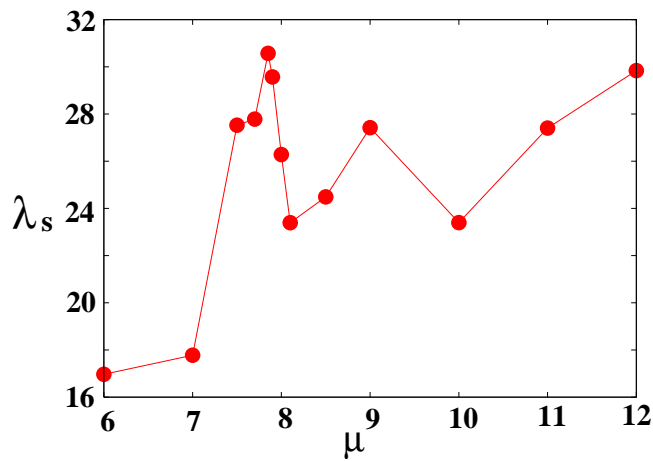}
\end{center}
\caption{Plot of the thermal conductivity of the solid, $\lambda_s$ as a 
function of $\mu$. $\lambda_s$ increases as the solid takes in a complete
layer of atoms.
}
\label{lambda}
\end{figure}
\begin{figure}[b]
\begin{center}
\includegraphics[width=8.6cm]{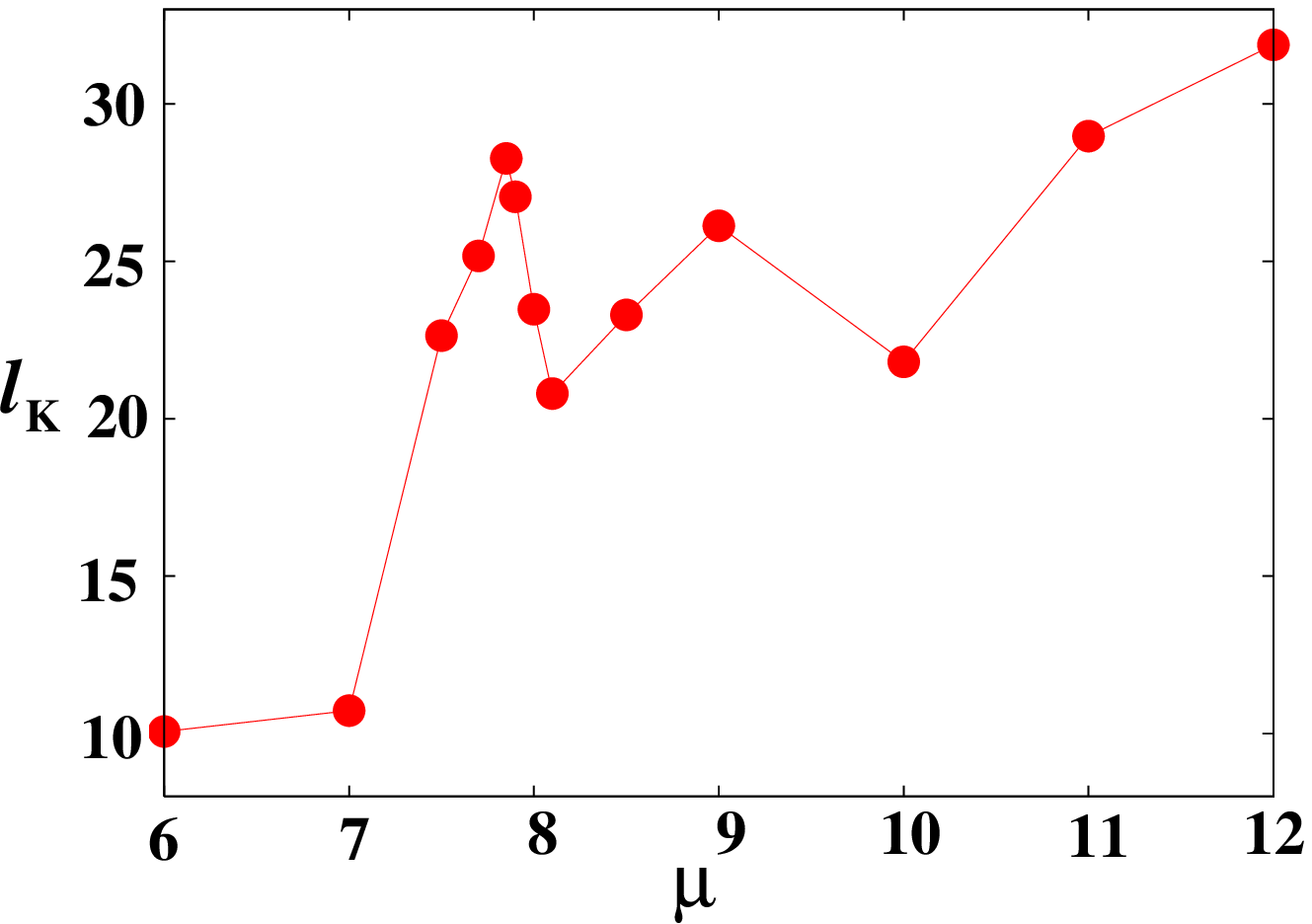}
\end{center}
\caption{Plot of Kapitza length, $l_K$ as a function of $\mu$ also shows a 
jump at the layering transition.
}
\label{kapl}
\end{figure}

The plot of $R_K$ as a function of $\mu$ shows a distinct jump as a
layer is included (for a value of $\mu$ close to $7$) in the solid
region as shown in Fig.~\ref{kapr}. The jump in $R_K$ is also
accompanied by a local peak at the transition. In Fig.~\ref{flux} we
have plotted the heat flux through the system as a function of
$\mu$. As $\mu$ increases, the atoms from the surrounding liquid get
attracted into the potential well and the density of the liquid
progressively becomes lower than that of the solid. This results in an
overall decrease in the heat flux and consequently the overall
conductivity. However, close to the layering transition at $\mu \sim
7-8$ there is a local peak in the value of the heat flux suggesting that
a significant amount of kinetic energy is exchanged between the liquid
and solid through the interface at the layering transition. The thermal 
conductivity of the solid $\lambda_s$ also shows a peak at the transition 
(Fig.~\ref{lambda}). The combined effect of the enhanced Kapitza resistance as
well as enhanced conductivity of the solid can be summarized by defining the
Kapitza length $l_K$, which is the length of solid which has a thermal
resistance equivalent to the liquid solid interface. This is
reminiscent of the "effective liquid" concept which has been used in
explaining some of the features of acoustic shock absorption in the
last section. The Kapitza length is given by,
\begin{eqnarray}
l_K = R_K \lambda_S
\end{eqnarray}
Note that substitution of the definition of the Kapitza length in the
equation for the Kapitza resistance (Eqn. 6.3) reproduces Fourier's law once the
thermal gradient is identified with $\Delta {\cal T}/l_K$. The Kapitza length
also shows a peak near the layering transition (Fig.~\ref{kapl}).

With the help of these results we may conclude that the layering
transition has a profound effect on the thermal properties of the
liquid solid interface. An important consequence of this study is the
possibility that the thermal resistance of interfaces may be altered
using external potential which cause layering transitions in a trapped
nano solid. We believe that this phenomenon has the potential for
useful applications for e.g. as tunable thermal switches or other nano
engineered devices.





\appendix



\printindex 

\bibliography{/nfs2/abhishek/Bibliography/abhi}
\end{document}